\documentclass[twocolumn,tighten]{aastex63}
\usepackage{natbib,graphicx,morefloats}
\usepackage{hyperref}
\usepackage{breakurl}
\newcommand{\kms}{km s$^{-1}$}
\newcommand{\msun}{$M_{\odot}$}
\newcommand{\lsun}{$L_{\odot}$}
\newcommand{\co}{$^{12}$CO}
\newcommand{\tco}{$^{13}$CO}

\newcommand{\jto}{${J=2\rightarrow1}$}
\newcommand{\jtt}{${J=3\rightarrow2}$}
\newcommand{\jsf}{${J=6\rightarrow5}$}
\newcommand{\jss}{${J=7\rightarrow6}$}

\shorttitle{Mid-$J$ CO observations of protostars}
\shortauthors{Kang et al.}
\begin{document}

\title{Mid-$J$ CO Line Observations of Protostellar Outflows
       in the Orion Molecular Clouds}

\author{Miju Kang}
\affiliation{Korea Astronomy and Space Science Institute,  
        776 Daedeokdae-ro, Yuseong-gu, Daejeon 34055, Republic of Korea}
\email{mjkang@kasi.re.kr}        
\author{Minho Choi}
\affiliation{Korea Astronomy and Space Science Institute, 776 Daedeokdae-ro, Yuseong-gu, Daejeon 34055, Republic of Korea}
\author{Friedrich Wyrowski}
\affiliation{Max-Planck-Institut f\"ur Radioastronomie, Auf dem H\"ugel 69, 53121, Bonn, Germany}
\author{Gwanjeong Kim}
\affiliation{Nobeyama Radio Observatory, National Astronomical Observatory of Japan, National Institutes of Natural Sciences, 462-2 Nobeyama, Minamimaki, Minamisaku, Nagano 384-1305, Japan}
\author{John H. Bieging}
\affiliation{Steward Observatory, The University of Arizona, Tucson, AZ 85721, USA}
\author{Mi-Ryang Kim}
\affiliation{Korea Astronomy and Space Science Institute, 776 Daedeokdae-ro, Yuseong-gu, Daejeon 34055, Republic of Korea}
\author{Geumsook Park}
\affiliation{Korea Astronomy and Space Science Institute, 776 Daedeokdae-ro, Yuseong-gu, Daejeon 34055, Republic of Korea}
 
\author{S. T. Megeath}
\affiliation{Ritter Astrophysical Research Center, University of Toledo, 2801 W. Bancroft Street, Toledo, OH 43606, USA} 

\author{Yunhee Choi}
\affiliation{Korea Astronomy and Space Science Institute, 776 Daedeokdae-ro, Yuseong-gu, Daejeon 34055, Republic of Korea} 

\author{Sung-Ju Kang}
\affiliation{Korea Astronomy and Space Science Institute, 776 Daedeokdae-ro, Yuseong-gu, Daejeon 34055, Republic of Korea} 
\affiliation{Korea Foundation for the Advancement of Science and Creativity, 602 Seollungno, Gangnam-Gu, Seoul 06097, Republic of Korea}

\author{Hyunju Yoo}
\affiliation{Korea Astronomy and Space Science Institute, 776 Daedeokdae-ro, Yuseong-gu, Daejeon 34055, Republic of Korea} 
\author{P. Manoj}
\affiliation{Department of Astronomy and Astrophysics, Tata Institute of Fundamental Research, Homi Bhabha Road, Colaba, Mumbai 400005, India} 

\begin{abstract}
Ten protostellar outflows in the Orion molecular clouds
were mapped in the \co/\tco\ \jsf\ and \co\ \jss\ lines.
The maps of these mid-$J$ CO lines
have an angular resolution of about 10\arcsec\
and a typical field size of about 100\arcsec.
Physical parameters of the molecular outflows were derived,
including mass transfer rates, kinetic luminosities, and outflow forces.
The outflow sample was expanded
by re-analyzing archival data of nearby low-luminosity protostars,
to cover a wide range of bolometric luminosities.
Outflow parameters derived from other transitions of CO were compared.
The mid-$J$ ($J_{\rm up} \approx 6$)
and low-$J$ ($J_{\rm up} \leq 3$) CO line wings
trace essentially the same outflow component.
By contrast, the high-$J$ (up to $J_{\rm up} \approx 50$) 
line-emission luminosity of CO
shows little correlation with the kinetic luminosity from the \jsf\ line,
which suggests that they trace distinct components.
The low/mid-$J$ CO line wings trace long-term outflow behaviors
while the high-$J$ CO lines are sensitive to short-term activities.
The correlations between the outflow parameters
and protostellar properties are presented,
which shows that the strengths of molecular outflows
increase with bolometric luminosity and envelope mass.
\\[3mm]
\end{abstract}

\section{Introduction}

Stars are formed in dense molecular clouds,
and the feedback from protostars in the form of radiation and outflows
is thought to play an important role
in mediating the collapse of cloud cores
and the accretion of material onto the protostars
\citep{Krumholz:2014ho,Offner:2014bx,Osorio:2017bw}.
Therefore, detailed studies of protostellar feedback are required
for understanding the physical processes
that control the ultimate masses of newly forming stars.
The rotational transitions of CO
provide a unique window into protostellar feedback.
In the last few decades,
many outflows were studied with low-$J$ CO lines ($J_{\rm up} \leq 3$)
tracing cold entrained gas
\citep{Bontemps:1996vb, Arce:2006dt, Dunham:2014gh}.
Relatively warm ($\ga$ 30 K) gas components
were studied with mid-$J$ CO lines ($J_{\rm up} \approx 6$)
\citep{vanKempen:2009jm, Yldz:2012cz, Yldz:2015ib, vanKempen:2016bv}.

High-$J$ CO lines  (up to $J_{\rm up} \approx 50$) in the far-IR range
were observed with instruments
such as the Photodetecting Array Camera and Spectrometer (PACS)
onboard the {\it Herschel} Space Observatory
\citep{vanKempen:2010jq, Visser:2012in, Herczeg:2012ej,
Manoj:2013ie, Manoj:2016et}.
The rotational diagrams of far-IR CO lines from protostars
can be fit with temperatures ranging from 200 to 1000 K
assuming local thermodynamic equilibrium \citep{vanKempen:2010jq}
or temperatures higher than 2000 K
assuming an isothermal, sub-thermally excited gas component
\citep{Manoj:2013ie}.
The origin of this hot gas is being debated.
The outflow pushes out the surrounding molecular gas,
and the outflow cavity inflates.
\cite{Visser:2012in} argued that the hot gas is heated
by a mixture of ultraviolet heating and C-type shocks,
while \cite{Manoj:2013ie} argued that the CO lines originate
in shock-heated, hot ($>$ 2000 K), sub-thermally excited molecular gas.
Therefore, it is important to investigate the relative
contributions of heating by ultraviolet radiation and by shocks.
In the case of shocks, it is needed to understand
whether the shocks are located
at the cavity walls or within the outflow itself. 

The {\it Herschel} Orion Protostar Survey (HOPS)
is a {\it Herschel} open-time key program
designed to study protostellar evolution
using a combination of {\it Herschel}/PACS imaging and spectroscopy
\citep{Fischer:2010bl,Furlan:2016df}.
Using the HOPS data, \cite{Manoj:2013ie} presented
the far-IR CO emission from 21 protostars in the Orion molecular clouds,
at a distance of $\sim$420 pc \citep{Menten:2007ew, Kim:2008gs}.
\cite{Manoj:2013ie} found that the total luminosity of the CO lines
in the range from $J = 14 \rightarrow 13$ to $46 \rightarrow 45$
exhibited a strong correlation
with the protostellar bolometric luminosity ($L_{\rm bol}$),
but not with the bolometric temperature ($T_{\rm bol}$)
or the envelope density,
estimated from model fits to the spectral energy distributions (SEDs).
\cite{Manoj:2013ie} argued
that the dominant component of the far-IR CO emission
is not caused by ultraviolet heating.
In addition, they found
that the rotational temperatures (which vary with $J$)
were remarkably independent of protostellar luminosity,
envelope density, or bolometric temperature.
They argued that the invariant rotational temperatures can be explained
by a gas component with high temperature ($>$ 2000 K)
and moderate density ($n(\rm{H_2)} < 10^6$ cm$^{-3}$).
In this regime, the rotational curve shows a weak dependence
on the gas kinetic temperature \citep[also see][]{Neufeld:2012fb}.
They proposed that the emission arises from the shock heated gas
in the protostellar wind that fills the outflow cavity.  

In this paper, we present the results of a survey of protostellar outflows
observed in the \co/\tco\ \jsf\ and \co\ \jss\ lines.
The outline of the paper is as follows.
Section 2 explains the observations.
The survey results and the physical parameters of the CO outflows
are given in Section 3.
The protostellar outflow activities are discussed in Section 4.
The observed sources are described in detail in Section 5.
A summary is given in Section 6.
This paper focuses on the presentation of the data,
and further analyses of the results will be presented in the future.

\section{Observations and Data}

\subsection{Observations}

Outflow activities of the protostars
in the far-IR CO study by \cite{Manoj:2013ie} were examined,
and ten outflows driven by relatively luminous protostars were selected.
(See Section 2 of \cite{Manoj:2013ie}
for the explanation on how the far-IR targets were selected.) 
A total of nine regions were observed,
because one of the regions contains two target outflows. 
The CHAMP$^+$ instrument on the APEX 12 m telescope in Chile
was used to observe the \co/\tco\ \jsf\ and \co\ \jss\ lines.
Most maps were obtained on 2012 August 28 and 29,
and two maps of the \tco\ \jsf\ line were made on 2014 November 6.
The CHAMP$^+$ instrument consists of two heterodyne receiver arrays,
each with seven-pixel detector elements
and a usable intermediate-frequency bandwidth of 2 GHz per pixel,
for simultaneous operations
in the 620--720 GHz and 780--950 GHz frequency ranges
\citep{Kasemann:2006jb,Gusten:2008ew}. 
Simultaneous observations were carried out
in the lower and higher frequency bands
with the settings of \co\ \jsf\ and \jss, respectively.
The observations were carried out under good weather conditions
(precipitable water vapor $\approx$ 0.5 mm).
The telescope pointing and focus checks were made
at typically one hour intervals on various planets and other strong sources.
The pointing was found to be accurate within 2\arcsec.
Table \ref{table_obs_lines} lists the parameters of the observed lines.

 \begin{deluxetable}{cccc}
 \tabletypesize{\small}
 \tablecaption{Observed CO Transitions \label{table_obs_lines}}
 \tablecolumns{5}
 \tablewidth{0pt}
 \tablehead{
 \colhead{Transition} & \colhead{Frequency} & \colhead{Beam}
 & \colhead{$T_{\rm sys}$\tablenotemark{a}} \\
 & \colhead{(GHz)} & \colhead{(\arcsec)} & \colhead{(K)}}
 \startdata
 \tco\ \jsf & 661.067280 & 9.4 & 1100 \\
 \co\  \jsf & 691.473076 & 9.0 & 1700 \\
 \co\  \jss & 806.651806 & 7.7 & 4000 \\
 \enddata
 \tablenotetext{a}{Mean system temperature.}
 \end{deluxetable}
 
The target protostars are listed in Table \ref{table_sources}.
For most of the target regions, except for HOPS 370,
each of the \co\ maps covered a $110\arcsec \times 110\arcsec$ area
with the on-the-fly observing mode.
The HOPS 370 maps cover a larger area to include the HOPS 368 outflow.
For each mapping field,
the background subtraction was done by observing a reference position
at an offset of 10$\arcmin$ in the right ascension.
Because the maps have elevated noise levels near the edge,
the outer areas were excluded by masking.
For each of the \co\ \jsf\ maps,
the area with the noise level larger by a factor of 7 
than that of the central area was masked out.
The effective size of the map is $\sim$90\arcsec\ in diameter
(about $100\arcsec \times 170\arcsec$ for the HOPS 370 region).
For the \tco\ \jsf\ and \co\ \jss\ maps,
the masking threshold was a factor of 3,
and the effective map size is somewhat smaller.

\begin{deluxetable*}{lccrrrrl}
\tabletypesize{\small}
\tablecaption{Target Protostars \label{table_sources}}
\tablewidth{0pt}
\tablehead{
\colhead{HOPS} & \colhead{R.A.} & \colhead{Decl.} & \colhead{$L_{\rm bol}$}
& \colhead{$T_{\rm bol}$} & \colhead{$M_{\rm env}$\tablenotemark{a}}
& \colhead{$i$\tablenotemark{b}} & \colhead{Association\tablenotemark{c}} \\
& \colhead{(J2000.0)} & \colhead{(J2000.0)} & \colhead{($L_{\odot}$)}
& \colhead{(K)} & \colhead{($M_{\odot}$)} & & } 
\startdata
310 & 05 42 27.67 & $-$01 20 01.0 &  13.8 &  51.8 &  6.3 & 70
    & L1630 HH 92 IRAS 05399--0121 \\
88  & 05 35 22.44 & $-$05 01 14.2 &  15.8 &  42.4 &  6.8 & 70
    & OMC 3 MMS 5 \\
68  & 05 35 24.31 & $-$05 08 30.5 &   5.7 & 100.6 &  7.5 & 50
    & OMC 2 FIR 2 \\
370 & 05 35 27.62 & $-$05 09 33.5 & 360.9 &  71.5 & 15.6 & 87
    & OMC 2 FIR 3, VLA 11 \\
368\tablenotemark{d}
    & 05 35 24.72 & $-$05 10 30.4 &  68.9 & 137.5 & \nodata & 18
    & OMC 2 VLA 13 \\
60  & 05 35 23.33 & $-$05 12 03.2 &  21.9 &  54.1 &  5.6 & 81
    & OMC 2 FIR 6b, CSO 25 \\
56  & 05 35 19.46 & $-$05 15 32.8 &  23.3 &  48.1 &  3.5 & 50
    & OMC 2 CSO 33 \\
182 & 05 36 18.84 & $-$06 22 10.2 &  71.1 &  51.9 & 27.5 & 76
    & L1641N MM1 \\
203 & 05 36 22.85 & $-$06 46 06.2 &  20.4 &  43.7 & 10.8 & 70
    & L1641 HH 1/2 VLA 1 \\
288 & 05 39 55.94 & $-$07 30 28.1 & 135.5 &  48.6 & 52.1 & 76
    & L1641 S3 MMS 1 \\
\enddata
\tablecomments{The HOPS source number and protostellar parameters
               are from \cite{Furlan:2016df}.
               Units of right ascension are hours, minutes, and seconds,
               and units of declination are degrees, arcminutes,
               and arcseconds.}
\tablenotetext{a}{Envelope mass calculated from 850 \micron\ flux densities.
                  See Section 4.3 for details.}
\tablenotetext{b}{Inclination to the line of sight in degrees,
                  from \cite{Furlan:2016df}.}
\tablenotetext{c}{Association from \cite{Mezger:1990vx}, \cite{Chini:1997hm},
                  \cite{Lis:1998dz}, \cite{Reipurth:1999jg},
                  \cite{Stanke:2007ba}, \cite{Rodriguez:1990du},
                  and \cite{Stanke:2000tv}.}
\tablenotetext{d}{HOPS 368 is located
                  in the southwestern corner of the HOPS 370 field.}
\end{deluxetable*}

The spectra were converted to the main beam temperature scale
with a forward efficiency of 0.95
and the beam efficiencies taken from the CHAMP$^+$ website%
\footnote{\burl{https://www.mpifr-bonn.mpg.de/4480868/efficiencies}}.
The original velocity resolution of the spectra is about 0.3 \kms.
The spectra were smoothed to a velocity resolution of 1 \kms.
All the maps were convolved to a resulting angular resolution of 10\arcsec.
The mean noise levels ($1\sigma$ for a channel width of 1 \kms)
at the target protostellar positions are about 0.24, 0.23, and 0.30 K
for the \tco\ \jsf, \co\ \jsf, and \co\ \jss\ lines, respectively.
The data were processed with the GILDAS/CLASS software
from Institut de Radioastronomie Millim\'etrique%
\footnote{\burl{http://www.iram.fr/IRAMFR/GILDAS}}.

\subsection{Archival Data of Nearby Protostars}

A survey of outflows driven by relatively low-mass protostars
in nearby molecular clouds was presented by \cite{Yldz:2015ib}.
These data will be incorporated in the statistical analysis in Section 4.
Instead of using the values given by \cite{Yldz:2015ib}, however,
we will use the outflow parameters
recalculated with the method described in Section 3.
There are two reasons for reprocessing the raw data.
First, there are some inconsistencies
between the observed data and the derived values in \cite{Yldz:2015ib}.
For example, the mass outflow rate, force, and luminosity
of the red lobes in Tables 2 and 3 of \cite{Yldz:2015ib}
are almost always larger than those of corresponding blue lobes,
even though the maps and spectra do not show such trends.
Second, some parameters such as maximum velocities
were taken from the \jtt\ line data,
which tends to produce larger values of outflow parameters
than those derived using the \jsf\ line data only.

Details for calculating the new outflow parameters
are given in the Appendix A.
In the discussions below, the outflow parameters of low-mass protostars
refer to the values presented in the Appendix A of this paper.
 
\section{Results}

\subsection{Overview}

\begin{figure*}[!t]
\includegraphics[width=0.5\textwidth]{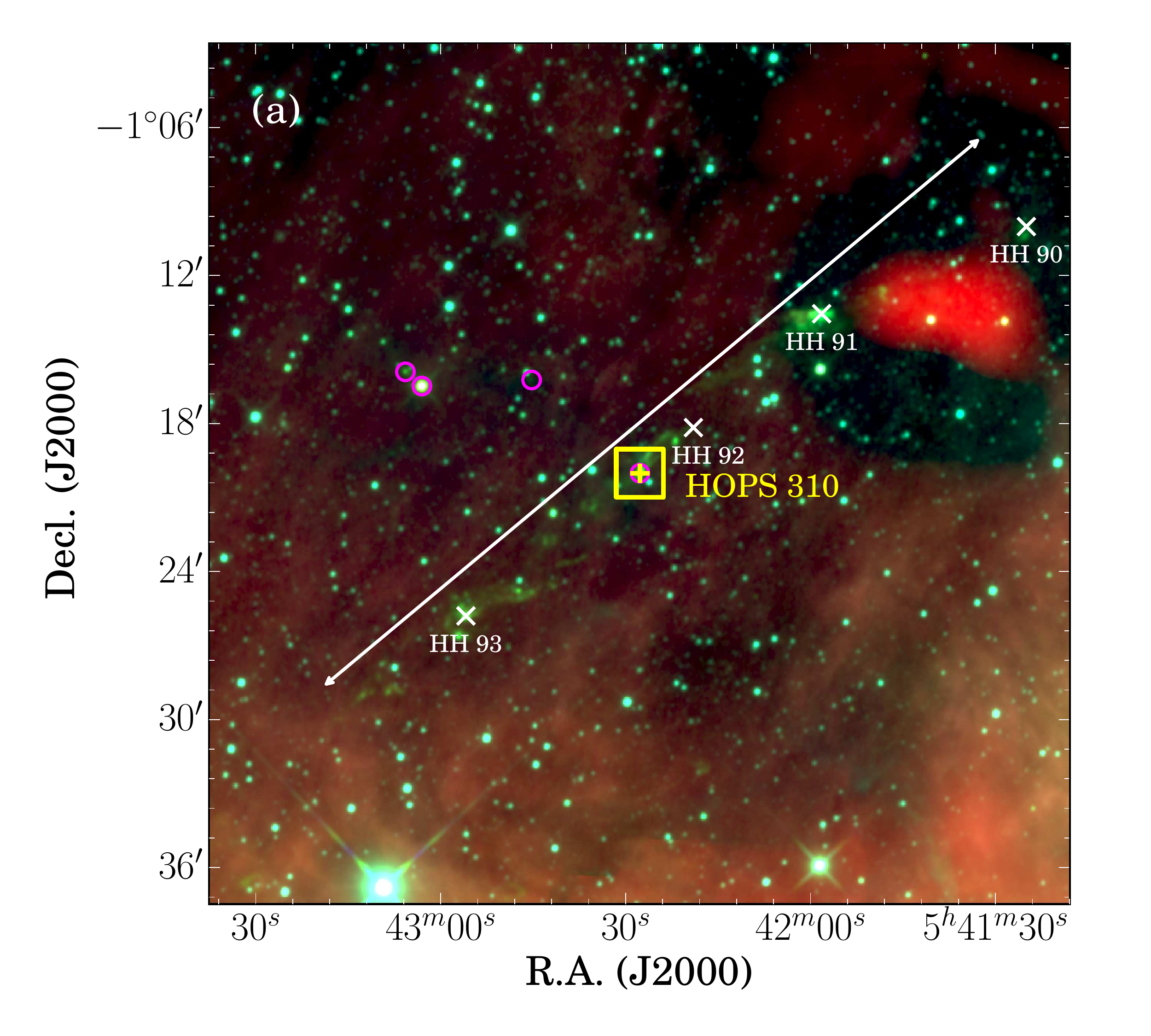}
\includegraphics[width=0.5\textwidth]{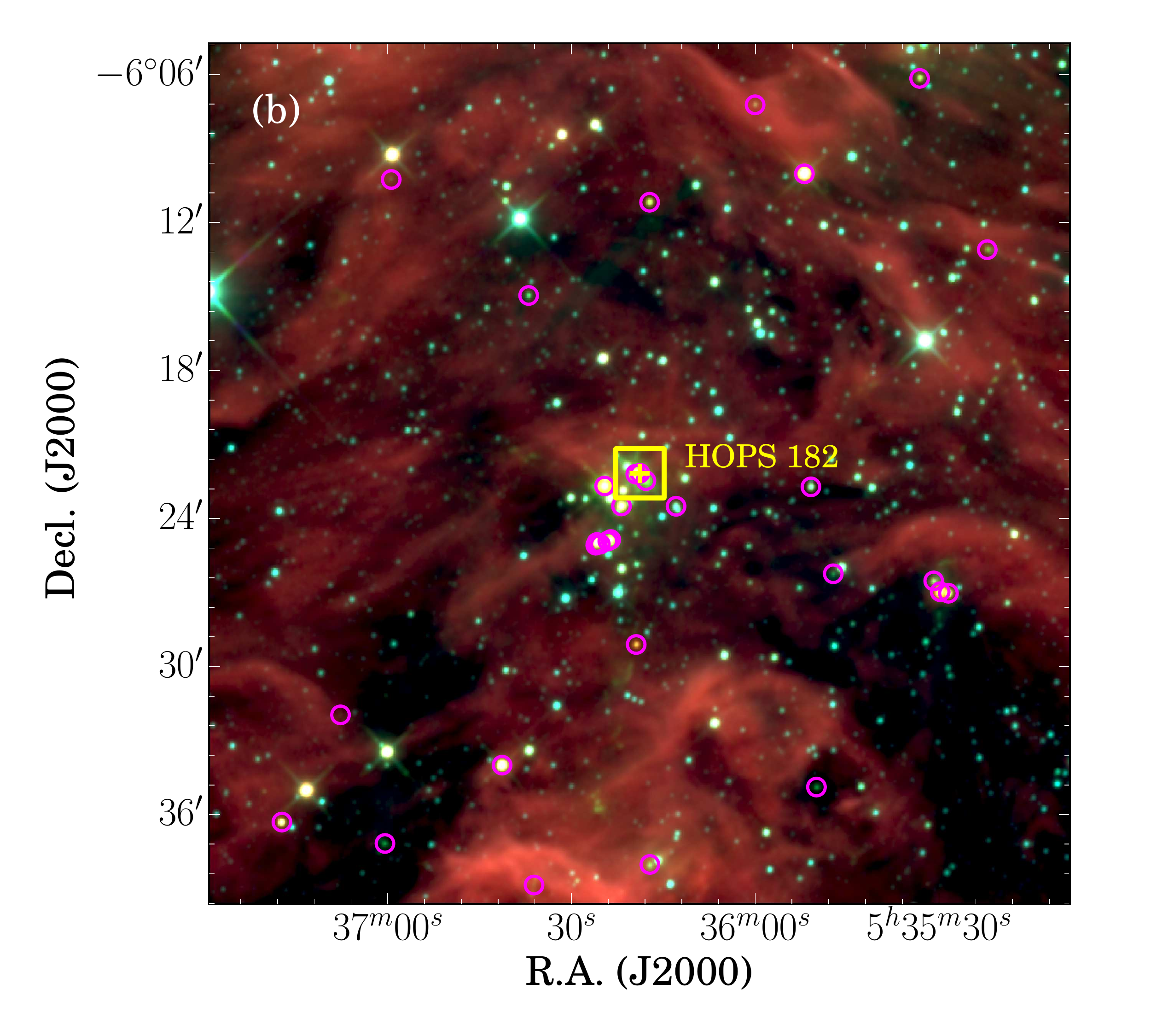}
\includegraphics[width=0.5\textwidth]{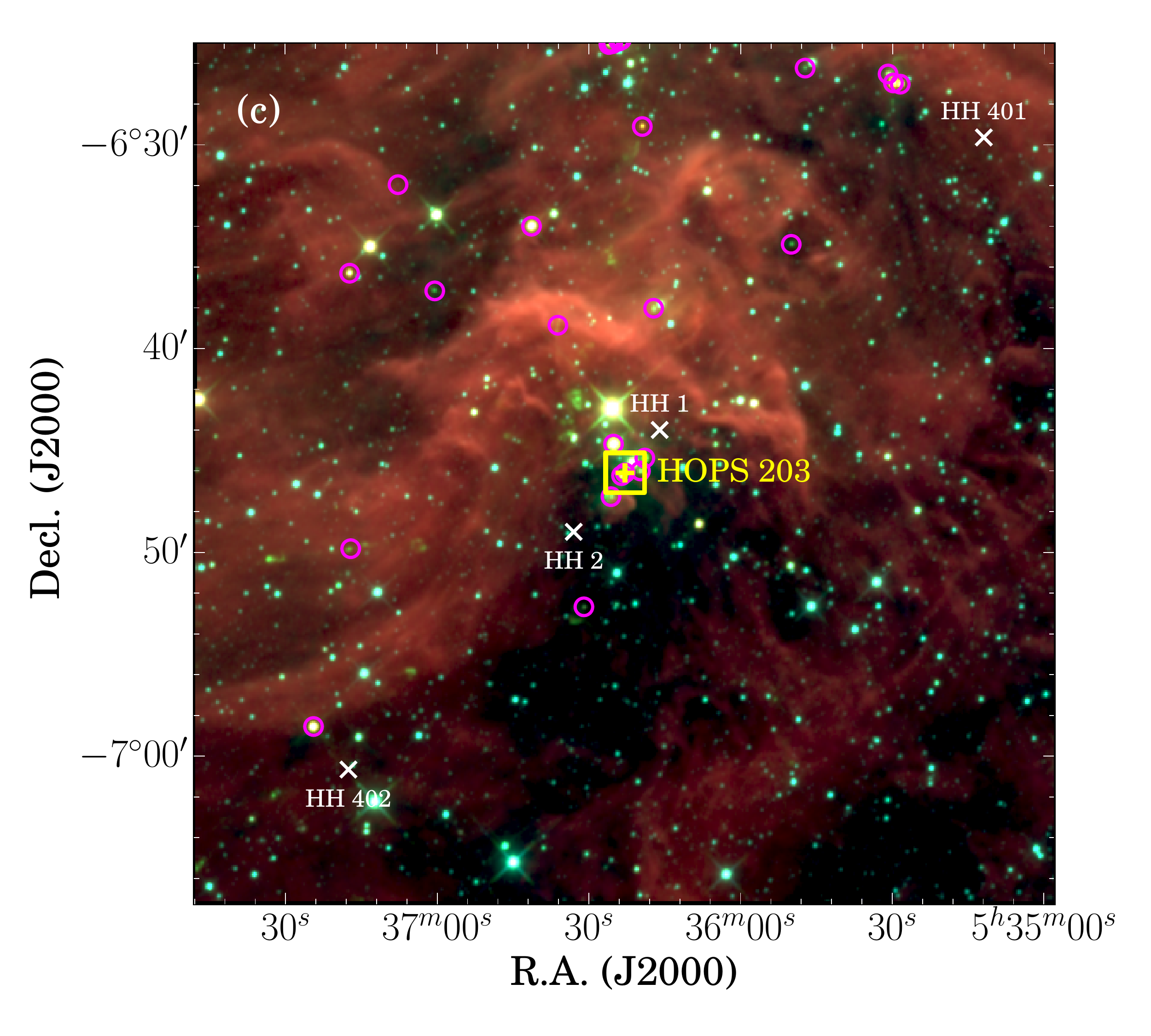}
\includegraphics[width=0.5\textwidth]{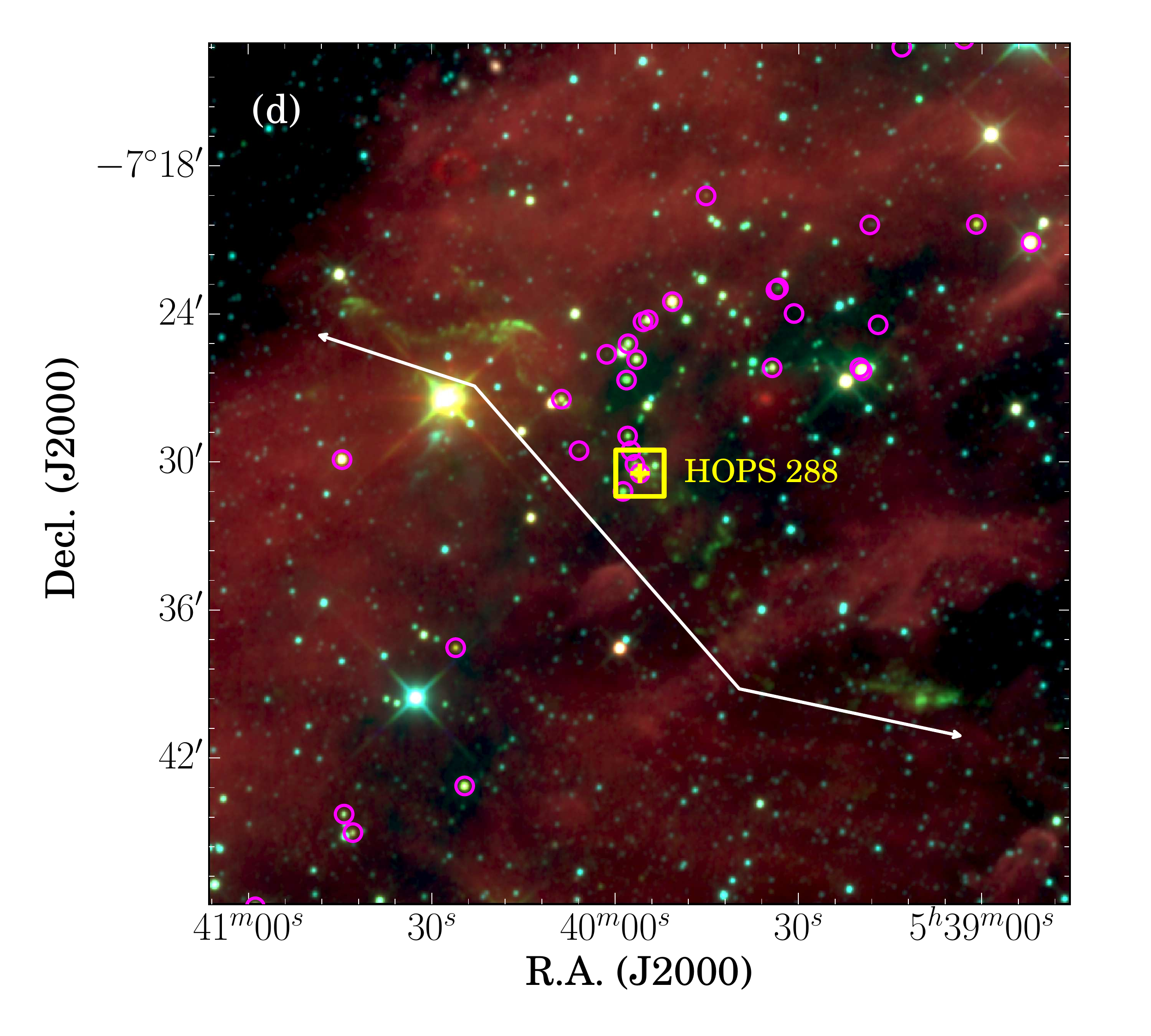}
\caption{
Color images composed of the {\it WISE} 12 $\micron$ (red),
4.6 $\micron$ (green), and 3.4 $\micron$ (blue) images
for (a) HOPS 310 with Herbig-Haro (HH) flows, (b) HOPS 182,
(c) HOPS 203 with HH flows, and (d) HOPS 288.
Yellow boxes show the areas covered in this work.
Open circles mark HOPS protostars,
and plus signs represent the driving sources of the target outflows.
Crosses mark HH objects associated with the target outflows.
White arrows indicate the large scale outflows
of HOPS 310 and HOPS 288.\label{fig_wisemap}}
\end{figure*}

\addtocounter{figure}{-1}
\begin{figure*} \centering
\includegraphics[width=0.8\textwidth]{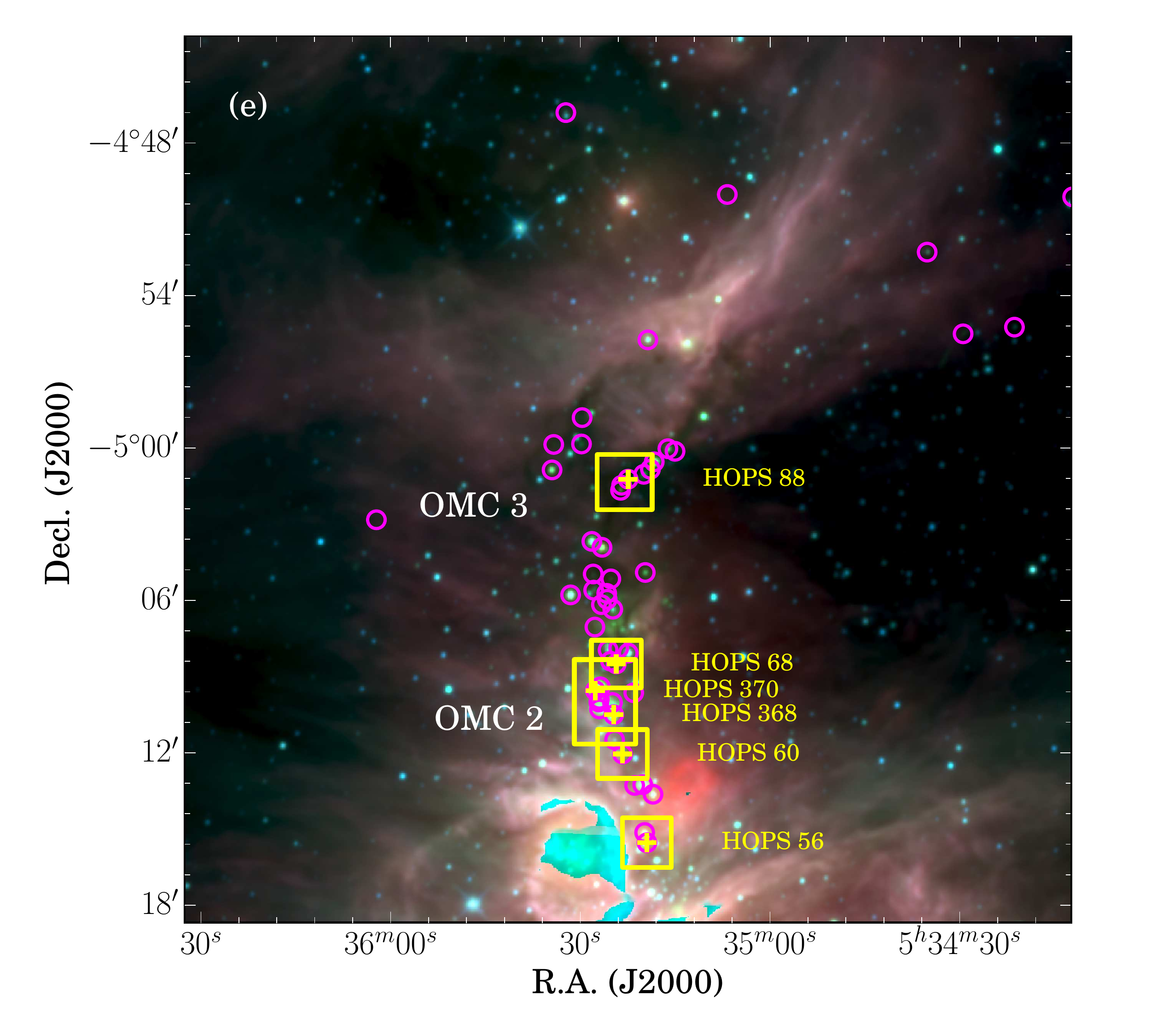}
\caption{(continued)
{\it WISE} color image for (e) HOPS 88, 68, 370, 368, 60, and 56.
The cyan color near the bottom of the map shows the area
where the 12 $\micron$ emission is saturated.}
\end{figure*}

Ten outflows were identified from the nine regions observed,
with the HOPS 370 map containing the HOPS 370 and HOPS 368 outflows.
Most of the outflows show clear blue-red bipolar outflow structures,
which is consistent with previous studies with low-$J$ CO lines
\citep{Stanke:2002bd,Aso:2000hx,Shimajiri:2008ih,
Takahashi:2012hh,Takahashi:2008ei}.
Table \ref{table_sources} lists the properties of the protostars
driving these outflows \citep{Furlan:2016df}.

Figure \ref{fig_wisemap} presents large-scale infrared images
composed with the Wide-field Infrared Survey Explorer {\it (WISE)} data.
The {\it WISE} 4.6 $\micron$ emission images
are useful for studying protostellar jets,
similar to the {\it Spitzer} IRAC 4.5 \micron\ images
showing shock-excited line features \citep{DeBuizer:2010dv}.
The outflow features are seen clearly
in the {\it WISE} three-color composite image as green features.
In particular, the {\it WISE} images show
that the HOPS 310 and HOPS 288 outflows are extended
over a length of $\sim$30\arcmin.

\begin{figure*}[t]
\includegraphics[width=0.85\textwidth]{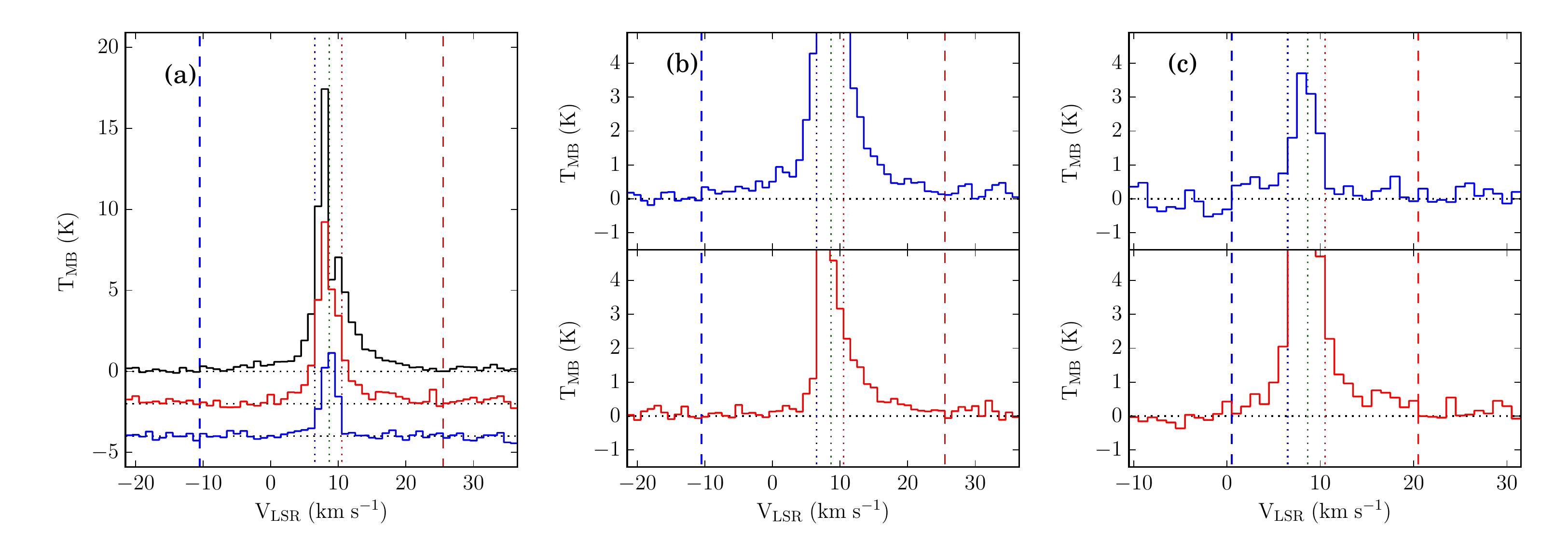}
\includegraphics[width=0.85\textwidth]{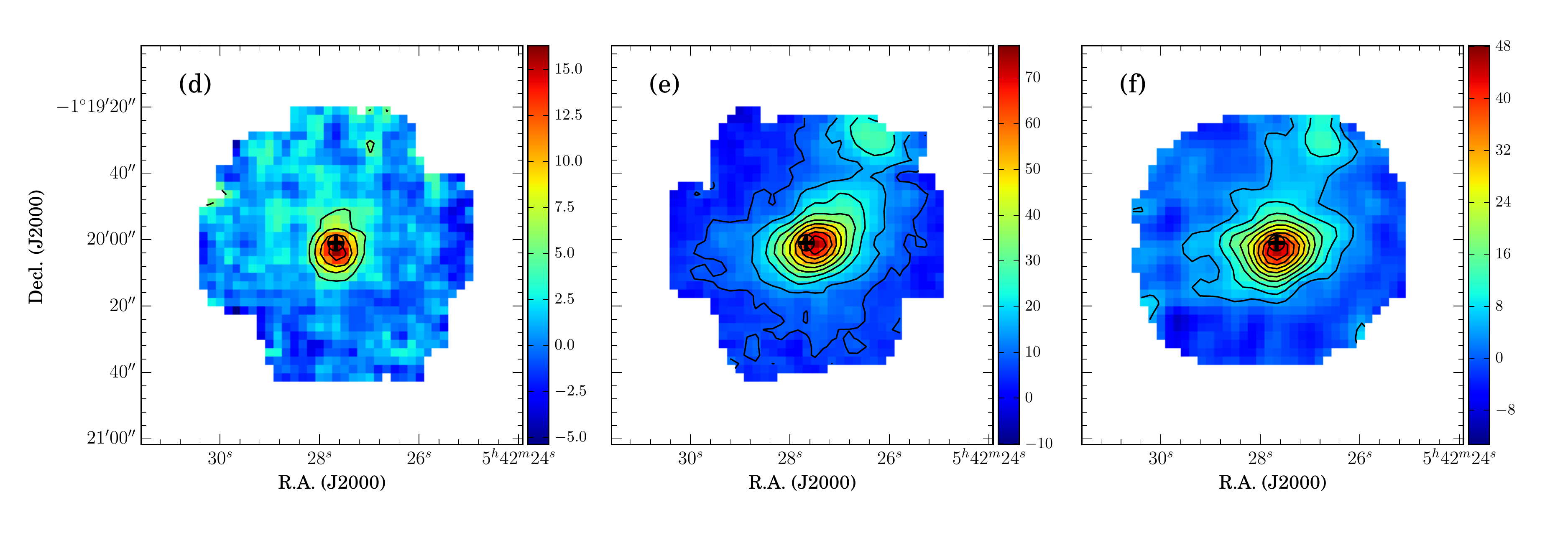}
\epsscale{0.75}
\plotone{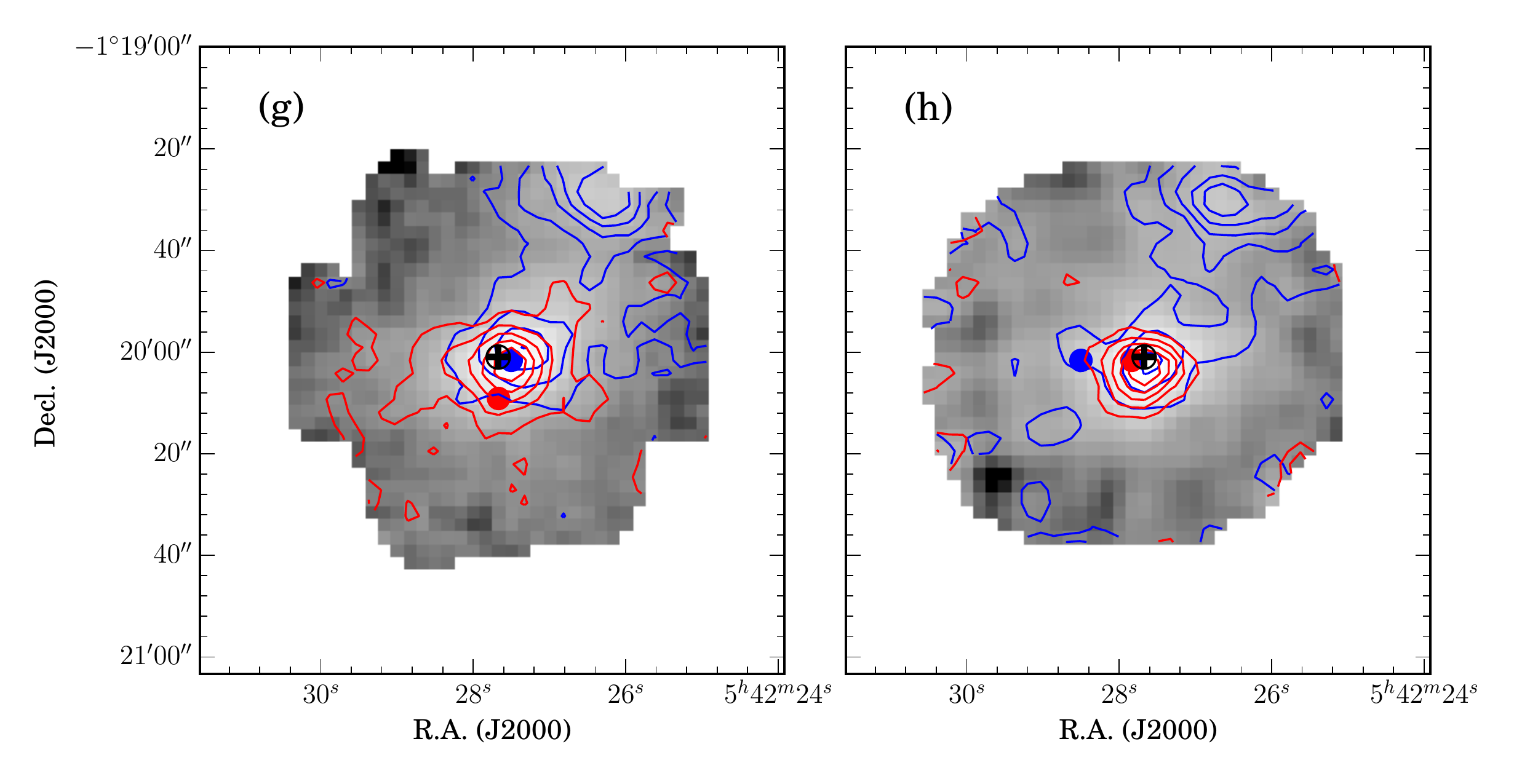}
\caption{
CO spectra and maps for HOPS 310 (IRAS 05399--0121).
(a) \co\ \jsf\ (black), \jss\ (red), and \tco\ \jsf\ (blue) spectra
at the protostellar position. 
(b) \co\ \jsf\ and (c) \co\ \jss\ spectra at the positions
where the line wings extend to the highest velocities,
for the blue (top) and red (bottom) wings.
The green dotted vertical line corresponds 
to the systemic velocity ($V_{\rm c}$)
determined with the \tco\ \jsf\ line.
The dotted and dashed vertical lines show
the inner and outer velocity limits of the line wings, respectively.
Color scale maps show the line intensities
integrated over the whole velocity range,
from $V_{\rm out,b}$ to  $V_{\rm out,r}$,
for the (d) \tco\ \jsf, (e) \co\ \jsf, and (f) \co\ \jss\ lines.
Contour levels on (d) start from 3$\sigma$ and in steps of 2$\sigma$, 
where $\sigma$ is the noise rms level of the map.
Contour levels on (e) and (f) are
10\% to 90\% in steps of 10\% of the peak intensity.
Plus signs mark the positions of the HOPS sources,
and the open circle represents
the driving source of the target outflow.
Contour maps show the line intensities
integrated over the blueshifted and redshifted wings
for the (g) \co\ \jsf\ and (h) \co\ \jss\ lines.
Contour levels are listed in Table \ref{table_contour}.
Blue and red dots show the positions
where the line wings are the widest, as shown in (b) and (c).
Grayscale maps are the same as shown in (e) and (f). 
CO spectra and maps for the other protostars in the survey
are shown in Appendix B.
\label{fig_310_co}}
\end{figure*}

Figures \ref{fig_310_co} and B2--B10 show
the spectra and maps of the target outflows.
In each figure,
the top panels show representative spectra.
The middle panels show the line intensity maps,
integrated over the whole velocity range.
The bottom panels show the maps of molecular outflows.

The \co\ \jsf\ and \jss\ line profiles at the protostellar positions 
show comparable emission strengths
between the line wings indicating strong outflow activities
and the line cores arising from the protostellar envelopes. 
The typical wing-to-core ratios of the integrated intensities
are 0.8 and 0.7 for the \co\ \jsf\ and \jss\ lines, respectively.
The wing-to-core ratios can be much larger
at outflow peak positions located away from the protostellar positions.
Table \ref{table_13co} lists the Gaussian-fit parameters
of the \tco\ \jsf\ spectra at the protostellar positions.
The \tco\ line width ranges from 1.7 \kms\ to 3.1 \kms.
Some \co\ spectra show self-absorption features,    
but the \tco\ spectra do not,
suggesting that the \tco\ line is mostly optically thin.

The velocity limits of the \co\ line wings
were determined carefully for each outflow.
The inner limit ($V_{\rm in}$) was determined
from the full-width at half-maximum (FWHM) of the averaged quiescent emission, 
which was extracted from the spatial region not associated with the outflow, 
i.e., several pixels in the region outside the first contours
in the outflow maps of Figures \ref{fig_310_co} and B2--B10.
This process needed a few iterations.
The outer velocity limit ($V_{\rm out}$) is at the velocity
where the wing intensity reaches the 1$\sigma$ level for the first time,
at the position where the wing is the widest, except HOPS 68 and 88.
For these two outflows,
the outer limits are set to values high enough
to include the emission from the extremely high velocity bullets
that were reported in previous studies
\citep{GomezRuiz:2019ka,Matsushita:2019kb}. 
Table \ref{table_vel} lists the velocity limits
for the blue and red line wings,
and they are also marked on the spectra
in the panels (b) and (c) of Figures \ref{fig_310_co} and B2--B10.

\begin{deluxetable*}{lRRRR}
\tabletypesize{\small}
\tablecaption{Parameters of the \tco\ \jsf\ Line \label{table_13co}}
\tablewidth{0pt}
\tablehead{
\colhead{HOPS} & \colhead{$\int T_{\rm MB} {\rm d}v$}
& \colhead{$V_{\rm c}$} & \colhead{FWHM} & \colhead{$T_{\rm peak}$} \\
& \colhead{(K km s$^{-1}$)} & \colhead{(km s$^{-1}$)}
& \colhead{(km s$^{-1}$)} & \colhead{(K)}} 
\startdata
310 & 14.2 \pm  0.6 & 8.66  \pm  0.04 & 2.52 \pm  0.09 & 5.29  \pm  0.16 \\
88  & 14.7 \pm  0.6 & 11.13 \pm  0.03 & 1.73 \pm  0.05 & 7.96  \pm  0.22 \\
68  & 25.8 \pm  0.8 & 11.27 \pm  0.02 & 1.76 \pm  0.04 & 13.75 \pm  0.29 \\
370 & 51.9 \pm  1.5 & 11.31 \pm  0.03 & 3.13 \pm  0.07 & 15.57 \pm  0.29 \\
368 & 25.2 \pm  1.1 & 11.06 \pm  0.03 & 2.38 \pm  0.08 & 9.97  \pm  0.29 \\
60  & 21.7 \pm  0.7 & 10.84 \pm  0.02 & 1.96 \pm  0.05 & 10.37 \pm  0.23 \\
56  & 25.1 \pm  0.6 & 10.22 \pm  0.02 & 2.11 \pm  0.04 & 11.17 \pm  0.18 \\
182 & 20.5 \pm  0.8 & 7.06  \pm  0.03 & 2.45 \pm  0.07 & 7.86  \pm  0.19 \\
203 & 10.1 \pm  0.8 & 10.31 \pm  0.06 & 2.26 \pm  0.14 & 4.20  \pm  0.22 \\
288 & 14.3 \pm  0.7 & 4.89  \pm  0.04 & 2.51 \pm  0.09 & 5.36  \pm  0.17 \\
\enddata
\end{deluxetable*}

\begin{deluxetable}{lRRRRRRR}
\tabletypesize{\small}
\tablecaption{Velocity Limits of Line Wings \label{table_vel}}
\tablewidth{0pt}
\tablehead{
\colhead{HOPS} & \multicolumn{3}{c}{Blue wing}
&& \multicolumn{3}{c}{Red wing} \\
\cline{2-4}\cline{6-8}
& \colhead{$V_{\rm max,b}$} & \colhead{$V_{\rm out,b}$}
& \colhead{$V_{\rm in,b}$}
&& \colhead{$V_{\rm in,r}$} & \colhead{$V_{\rm out,r}$}
& \colhead{$V_{\rm max,r}$}}
\startdata
\multicolumn{8}{c}{\co\ \jsf}\\
\hline
 310 &  19.2 & -10.5 & 6.5 && 10.5 & 25.5 & 16.8 \\
  88 & 103.6 & -92.5 & 7.5 && 14.5 & 75.5 & 64.4 \\
  68 &  66.8 & -55.5 & 7.5 && 15.5 & 83.5 & 72.2 \\
 370 &  50.8 & -39.5 & 7.5 && 14.5 & 39.5 & 28.2 \\
 368 &  13.6 &  -2.5 & 7.5 && 14.5 & 21.5 & 10.4 \\
  60 &  14.3 &  -3.5 & 7.5 && 13.5 & 52.5 & 41.7 \\
  56 &  19.7 &  -9.5 & 6.5 && 14.5 & 23.5 & 13.3 \\
 182 &  51.6 & -44.5 & 2.5 && 11.5 & 49.5 & 42.4 \\
 203 &  11.8 &  -1.5 & 6.5 && 13.5 & 27.5 & 17.2 \\
 288 &  19.4 & -14.5 & 2.5 &&  7.5 & 24.5 & 19.6 \\
\hline
\multicolumn{8}{c}{\co\ \jss}\\
\hline
 310 &   8.2 &   0.5 & 6.5 && 10.5 & 20.5 & 11.8 \\
  88 & 103.6 & -92.5 & 7.5 && 14.5 & 74.5 & 63.4 \\
  68 &  63.8 & -52.5 & 7.5 && 15.5 & 81.5 & 70.2 \\
 370 &  46.8 & -35.5 & 7.5 && 14.5 & 40.5 & 29.2 \\
 368 &  14.6 &  -3.5 & 7.5 && 14.5 & 21.5 & 10.4 \\
  60 &  16.3 &  -5.5 & 7.5 && 13.5 & 42.5 & 31.7 \\
  56 &  15.7 &  -5.5 & 6.5 && 13.5 & 22.5 & 12.3 \\
 182 &  49.6 & -42.5 & 2.5 && 11.5 & 48.5 & 41.4 \\
 203 &  11.8 &  -1.5 & 6.5 && 14.5 & 26.5 & 16.2 \\
 288 &  12.4 &  -7.5 & 1.5 &&  8.5 & 16.5 & 11.6 \\
\enddata
\tablecomments{All velocities are in \kms.
               $V_{\rm max} = |V_{\rm out} - V_{\rm c}|$.}
\end{deluxetable}

\begin{deluxetable}{lrrrrr}
\tabletypesize{\small}
\tablecaption{Contour Levels of the Outflow Maps \label{table_contour}}
\tablewidth{0pt}
\tablehead{
\colhead{HOPS} & \multicolumn{2}{c}{Blue wing}
&& \multicolumn{2}{c}{Red wing} \\
\cline{2-3}\cline{5-6}
& \colhead{Lowest level} & \colhead{Step}
&& \colhead{Lowest level} & \colhead{Step}}
\startdata
\multicolumn{6}{c}{\co\ \jsf}\\
\hline
 310 &  3.8 &  3.6 &&  2.9 &  3.1 \\
  88 & 15.9 & 15.3 && 10.7 & 10.2 \\
  68 & 22.8 & 34.5 && 28.1 & 22.6 \\
 370 & 28.4 & 51.3 && 22.6 & 35.7 \\
 368 & 14.4 & 10.1 && 11.2 &  7.3 \\
  60 &  5.9 &  3.3 && 11.7 & 11.4 \\
  56 &  4.9 &  4.0 &&  4.6 &  3.0 \\
 182 & 17.8 & 16.7 && 16.1 & 18.8 \\
 203 &  4.7 &  1.8 &&  4.4 &  2.4 \\
 288 &  7.2 &  5.6 &&  7.3 &  5.8 \\
\hline
\multicolumn{6}{c}{\co\ \jss} \\
\hline
 310 &  1.8 &  1.8 &&  2.0 &  1.5 \\
  88 & 20.1 &  9.3 && 14.5 &  6.4 \\
  68 & 27.3 & 14.9 && 25.6 &  8.0 \\
 370 & 25.4 & 31.7 && 19.7 & 23.5 \\
 368 & 13.9 &  4.7 && 10.6 &  3.9 \\
  60 &  6.3 &  1.4 && 10.9 &  5.6 \\
  56 &  4.2 &  1.7 &&  3.2 &  1.6 \\
 182 & 17.9 & 10.4 && 17.8 & 13.8 \\
 203 &  2.9 &  1.4 &&  3.8 &  0.8 \\
 288 &  6.0 &  1.6 &&  4.4 &  2.0 \\
\enddata
\tablecomments{All contour levels are in K \kms.}
\end{deluxetable}

Integrated intensity maps of the three lines are presented
in the middle panels of Figures \ref{fig_310_co} and B2--B10.
The line wing maps of the \co\ lines are presented in the bottom panels.
The contour maps show the line intensity
integrated over the velocity range listed in Table \ref{table_vel},
and the contour levels are given in Table \ref{table_contour}.
The lowest contour levels were chosen
to delineate the outflows as clearly as possible,
low enough to include the outflows
and high enough to exclude the ambient clouds.

\subsection{Outflow Properties}

Several quantities were measured for each outflow lobe
to describe the outflow properties.
The length of the outflow lobe, $R_{\rm lobe}$,
is the extent of the outflow in the line wing map,
from the protostellar position.
Some of the outflows are extended beyond the mapping fields,
and the corresponding $R_{\rm lobe}$
is the distance to the edge of the map.
For these outflows, the physical quantities described below
are relevant to the mapped portion of the outflow, not the whole outflow.
The maximum outflow velocity, $V_{\rm max}$,
is the maximum velocity of the line wing
relative to the centroid velocity, $|V_{\rm out}-V_{\rm c}|$,
as listed in Table \ref{table_vel}.
The dynamical time is calculated by
\begin{equation}
t_{\rm dyn} = \frac{R_{\rm lobe}}{V_{\rm max}}.
\end{equation}

The mass of molecular outflow is calculated by
\begin{equation}
M_{\rm CO} = \mu_{\rm H_2} m_{\rm H} A
               \sum\limits_{i}N_{\rm H_{2,{\it i}}},
\end{equation}
where $m_{\rm H}$ is the mass of the hydrogen atom,
$\mu_{\rm H_2}$ is the mean molecular weight per hydrogen molecule
\citep[$\mu_{\rm H_2}$ = 2.8;][]{Kauffmann:2008jj}, 
$A$ is the surface area of one pixel ($2\farcs5 \times 2\farcs5$),
$N_{\rm H_{2,{\it i}}}$ is the pixel-averaged ${\rm H_2}$ column density
over the selected velocity range,
and the sum is over spatial pixels encompassing the outflow
seen in the contour maps.
Assuming that the emission from the outflow is optically thin (see below),
the column density is proportional to the integrated intensity.
For the outflow lobes showing
predominantly blueshifted or redshifted emission,
the integration was done over the velocity interval
of ($V_{\rm out,b}$, $V_{\rm in,b}$) or
($V_{\rm in,r}$, $V_{\rm out,r}$), respectively.
For the outflow lobes showing both blueshifted and redshifted emission,
the integration was done over the velocity intervals
of ($V_{\rm out,b}$, $V_{\rm in,b}$) and ($V_{\rm in,r}$, $V_{\rm out,r}$).

The column density $N_{\rm H_2}$ was obtained
by adopting a typical abundance ratio
of [$^{12}$CO]/[H$_{2}$] = $10^{-4}$ \citep{Frerking:1982iq}.
An excitation temperature of $T_{\rm ex}$ = 75 K was assumed
for consistency with other studies such as \cite{Yldz:2012cz, Yldz:2015ib}.
In principle, the excitation temperature can be estimated
from the \co\ \jss\ / \jsf\ line ratio.
The excitation temperatures at the positions
of the strongest \co\ \jsf\ wing emission were estimated
using the integrated intensities of the line wings,
and the derived values range from 33 to 103 K.
Therefore, $T_{\rm ex}$ = 75 K is a reasonable assumption.

The kinetic energy within each velocity channel in each pixel
is given by
\begin{equation}
E_{\rm {{\it v},pixel}} = \frac{1}{2}M_{\rm {{\it v},pixel}}\ v^{2},
\end{equation}
where $v$ is the velocity of each channel
with respect to the systemic velocity.
The kinetic energy ($E_{\rm CO}$) of each molecular outflow lobe
is calculated by summing over the same velocities and pixels as for the mass.

The mass outflow rate of each lobe is given by
\begin{equation}
\dot{M}_{\rm CO} = \frac{M_{\rm CO}}{t_{\rm dyn}}.
\end{equation}
The molecular outflow kinetic luminosity and force of each lobe
are computed by
\begin{equation}
L_{\rm CO} = \frac{E_{\rm CO}}{t_{\rm dyn}},
\end{equation}
and
\begin{equation}
F_{\rm CO} = \sqrt{2\dot{M}_{\rm CO}L_{\rm CO}},
\end{equation}
respectively.
Here, CO as a subscript refers to 
the total mass, energy, and related quantities of molecular gas
as traced by the CO emission. 

In this study, the CO outflows are assumed to be optically thin.
Previous studies in the \jtt\ line showed
that outflows are optically thin
at velocities larger than $\sim$3 \kms\ away from the systemic velocity
\citep{vanderMarel:2013hy,Dunham:2014gh}.
Considering the uncertainties of outflow parameters,
the optical depth effect may be negligible \citep{vanderMarel:2013hy}.
Therefore, the outflows are expected to be optically thin
in the \co\ \jsf\ line.
In most cases of this survey,
there is no detectable wing feature in the \tco\ line profiles.
An exception is the blueshifted outflow of HOPS 288 (Figure B10(a)).
It shows a wing-like emission feature over a few velocity channels,
and the \co/\tco\ line ratio gives
an optical depth of $\sim$6 for the \co\ \jsf\ line,
using a \co/\tco\ abundance ratio of $\sim$60 \citep{Wilson:1994hu}.
This exceptional feature may be related with the fact
that the envelope mass of HOPS 288 is very large (Table \ref{table_sources}).
It can be seen only at the protostellar position,
and the correction factors for the mass and energy estimates of this lobe
are 1.6 and 1.1, respectively.
Therefore, the assumption of optically thin outflow is reasonable.

\begin{figure*}[!t]
\epsscale{0.8}
\plotone{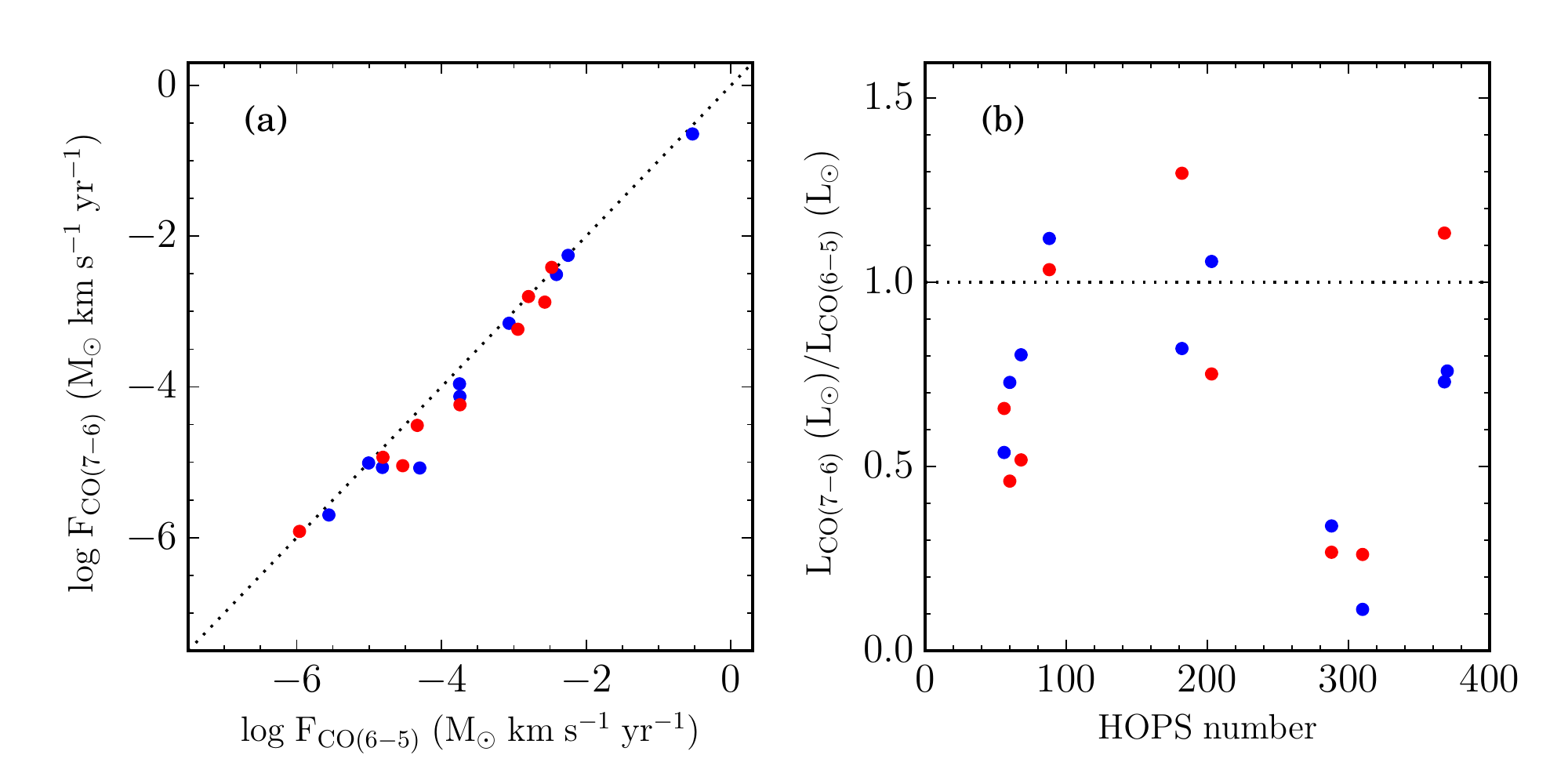}
\caption{
(a)
Scatter diagram of the outflow forces
$F_{\rm CO(6-5)}$ and $F_{\rm CO(7-6)}$. 
(b)
Ratio plot of the outflow kinetic luminosities of the two lines.
The dotted lines show
where the outflow parameters from the two lines are the same.
Blue and red circles represent
the outflow parameters of the blue and red lobes, respectively.
\label{fig_force76_force65}}
\end{figure*}

The line core usually contains
a blend of emission from the ambient cloud and the outflow.
The emission from the outflow
in the velocity interval ($V_{\rm in,b}$, $V_{\rm in,r}$)
is not included in the calculation of the integrated intensity.
Previous studies suggested correction factors for this ``missing'' mass,
ranging from $\sim$2 to $\sim$40,
in the case of low-$J$ CO line observations
(see Section 4.3 of \cite{vanderMarel:2013hy}
and Section 4.3 of \cite{Dunham:2014gh}).
The mass estimate from the \jsf\ line is less affected by this issue
because the typical kinetic temperature of the ambient gas
is much lower than the $J$ = 6 energy level (115 K above ground).
Considering that the integrated intensities of the line wing and core
are comparable (Section 3.1),
the correction factor for mass would be smaller than $\sim$2.
In regions near the protostars
where the emission from the ambient cloud is strong,
the innermost channels of line wings
may contain a blend of emission from the outflow and the ambient gas.
It is difficult to separate
this ``contaminating'' ambient emission from the outflow emission.
The effects of the missing and the contaminating emission components
may cancel out to a certain degree.
This issue barely affects the estimates of kinetic energy
because most of the contribution comes from high-velocity channels.

Corrections for the outflow inclination are necessary
because the measured quantities are
the velocity component along the line-of-sight
and the outflow lobe size projected on the plane of the sky.
In this paper, the inclination is defined to be the angle
between the outflow axis and the line of sight
(i.e., $i = 0\arcdeg$ corresponds to a pole-on system).
The inclination angles (listed in Table \ref{table_sources})
were taken from \cite{Furlan:2016df},
to treat the target sources consistently.
\cite{Furlan:2016df} derived the best-fit inclination angles 
from the continuum SED of protostars.
They used 30,400 different model SEDs 
to determine the best-fit model parameters 
for 330 young stellar objects (YSOs). 
There are 3040 models in grid covering 
eight values for the total luminosity, four disk radii, 
19 envelope infall rates which correspond to envelope densities, 
and five cavity opening angles.
They used the envelope densify profile
of a rotating, collapsing cloud core with constant infall rate
\citep{Terebey:1984hi}.
Each model is calculated for 10 different inclination angles.
The inclination angle can be degenerate
with the envelope density and other parameters.
(See the discussion in Section 6.3, Section 7.2,
and Appendix B of \cite{Furlan:2016df}.)
Inclination corrections to the outflow length, dynamical time,
energy, momentum, mass outflow rate, luminosity, and force
have been applied
in the way described by \cite{Dunham:2014gh} and \cite{Plunkett:2015jo}.

\begin{deluxetable*}{lcCCCCCLC}
\tabletypesize{\small}
\tablecaption{Properties of the CO Outflows \label{table_para}}
\tablewidth{0pt}
\tablehead{
\colhead{HOPS} & \colhead{Lobe} & \colhead{$R_{\rm lobe}$}
& \colhead{$t_{\rm dyn}$} & \colhead{$M_{\rm CO}$}
& \colhead{$E_{\rm CO}$} & \colhead{$\dot{M}_{\rm CO}$}
& \colhead{$L_{\rm CO}$} & \colhead{$F_{\rm CO}$} \\
& & \colhead{(au)} & \colhead{(year)} & \colhead{(\msun)}
& \colhead{(erg)} & \colhead{(\msun\ yr$^{-1}$)} & \colhead{(\lsun)}
& \colhead{(\msun\ yr$^{-1}$ \kms)}}
\startdata
\multicolumn{9}{c}{\co\ \jsf} \\
\hline
310 & Blue  & 2.1 \times 10^{ 4 } & 1.8 \times 10^{ 3 } & 4.2 \times 10^{ -3 } & 1.9 \times 10^{ 43 } & 2.4 \times 10^{ -6 } & 8.8 \times 10^{ -2 } & 5.0 \times 10^{ -5 } \\
    & Red   & 1.1 \times 10^{ 4 } & 1.0 \times 10^{ 3 } & 1.7 \times 10^{ -3 } & 5.4 \times 10^{ 42 } & 1.6 \times 10^{ -6 } & 4.3 \times 10^{ -2 } & 2.9 \times 10^{ -5 } \\
 88 & Blue  & 1.8 \times 10^{ 4 } & 2.8 \times 10^{ 2 } & 1.5 \times 10^{ -2 } & 1.6 \times 10^{ 45 } & 5.4 \times 10^{ -5 }  & 4.8 \times 10^{ 1 } & 5.6 \times 10^{ -3 } \\
    & Red   & 1.3 \times 10^{ 4 } & 3.2 \times 10^{ 2 } & 5.8 \times 10^{ -3 } & 4.5 \times 10^{ 44 } & 1.8 \times 10^{ -5 }  & 1.2 \times 10^{ 1 } & 1.6 \times 10^{ -3 } \\
 68 & Blue  & 1.6 \times 10^{ 4 } & 7.2 \times 10^{ 2 } & 1.6 \times 10^{ -2 } & 2.4 \times 10^{ 44 } & 2.2 \times 10^{ -5 }  & 2.8 \times 10^{ 0 } & 8.6 \times 10^{ -4 } \\
    & Red   & 1.1 \times 10^{ 4 } & 4.5 \times 10^{ 2 } & 1.1 \times 10^{ -2 } & 2.4 \times 10^{ 44 } & 2.4 \times 10^{ -5 }  & 4.3 \times 10^{ 0 } & 1.1 \times 10^{ -3 } \\
370\tablenotemark{a} & Blue  & 3.3 \times 10^{ 4 } & 1.6 \times 10^{ 2 } & 1.8 \times 10^{ -1 } & 1.3 \times 10^{ 47 } & 1.1 \times 10^{ -3 }  & 6.6 \times 10^{ 3 } & 2.9 \times 10^{ -1 } \\
368 & Blue  & 4.2 \times 10^{ 4 } & 1.4 \times 10^{ 4 } & 9.5 \times 10^{ -3 } & 3.6 \times 10^{ 42 } & 6.8 \times 10^{ -7 } & 2.1 \times 10^{ -3 } & 4.2 \times 10^{ -6 } \\
    & Red   & 2.8 \times 10^{ 4 } & 1.2 \times 10^{ 4 } & 3.9 \times 10^{ -3 } & 1.2 \times 10^{ 42 } & 3.2 \times 10^{ -7 } & 8.5 \times 10^{ -4 } & 1.8 \times 10^{ -6 } \\
 60 & Blue  & 1.9 \times 10^{ 4 } & 1.0 \times 10^{ 3 } & 4.4 \times 10^{ -3 } & 7.1 \times 10^{ 43 } & 4.4 \times 10^{ -6 } & 5.9 \times 10^{ -1 } & 1.8 \times 10^{ -4 } \\
    & Red   & 1.8 \times 10^{ 4 } & 3.1 \times 10^{ 2 } & 9.5 \times 10^{ -3 } & 7.4 \times 10^{ 44 } & 3.0 \times 10^{ -5 }  & 2.0 \times 10^{ 1 } & 2.7 \times 10^{ -3 } \\
 56 & Blue  & 1.8 \times 10^{ 4 } & 2.7 \times 10^{ 3 } & 3.8 \times 10^{ -3 } & 4.7 \times 10^{ 42 } & 1.4 \times 10^{ -6 } & 1.4 \times 10^{ -2 } & 1.5 \times 10^{ -5 } \\
    & Red   & 2.3 \times 10^{ 4 } & 5.4 \times 10^{ 3 } & 8.8 \times 10^{ -3 } & 7.8 \times 10^{ 42 } & 1.6 \times 10^{ -6 } & 1.2 \times 10^{ -2 } & 1.6 \times 10^{ -5 } \\
182 & Blue  & 1.2 \times 10^{ 4 } & 2.6 \times 10^{ 2 } & 1.3 \times 10^{ -2 } & 7.7 \times 10^{ 44 } & 5.0 \times 10^{ -5 }  & 2.5 \times 10^{ 1 } & 3.9 \times 10^{ -3 } \\
    & Red   & 1.4 \times 10^{ 4 } & 3.7 \times 10^{ 2 } & 2.0 \times 10^{ -2 } & 7.7 \times 10^{ 44 } & 5.3 \times 10^{ -5 }  & 1.7 \times 10^{ 1 } & 3.3 \times 10^{ -3 } \\
203 & Blue  & 1.9 \times 10^{ 4 } & 2.7 \times 10^{ 3 } & 1.5 \times 10^{ -3 } & 4.6 \times 10^{ 42 } & 5.6 \times 10^{ -7 } & 1.4 \times 10^{ -2 } & 9.9 \times 10^{ -6 } \\
    & Red   & 8.2 \times 10^{ 3 } & 7.7 \times 10^{ 2 } & 1.5 \times 10^{ -3 } & 8.3 \times 10^{ 42 } & 2.0 \times 10^{ -6 } & 8.9 \times 10^{ -2 } & 4.6 \times 10^{ -5 } \\
288\tablenotemark{b} & Blue  & 1.0 \times 10^{ 4 } & 5.9 \times 10^{ 2 } & 5.3 \times 10^{ -3 } & 3.7 \times 10^{ 43 } & 9.0 \times 10^{ -6 } & 5.2 \times 10^{ -1 } & 2.4 \times 10^{ -4 } \\
    & Red   & 1.8 \times 10^{ 4 } & 1.1 \times 10^{ 3 } & 6.0 \times 10^{ -3 } & 6.2 \times 10^{ 43 } & 5.6 \times 10^{ -6 } & 4.8 \times 10^{ -1 } & 1.8 \times 10^{ -4 } \\
\hline
\multicolumn{9}{c}{\co\ \jss} \\
\hline
310 & Blue  & 1.9 \times 10^{ 4 } & 3.7 \times 10^{ 3 } & 2.2 \times 10^{ -3 } & 4.4 \times 10^{ 42 } & 5.9 \times 10^{ -7 } & 9.8 \times 10^{ -3 } & 8.4 \times 10^{ -6 } \\
    & Red   & 7.9 \times 10^{ 3 } & 1.1 \times 10^{ 3 } & 6.4 \times 10^{ -4 } & 1.5 \times 10^{ 42 } & 5.9 \times 10^{ -7 } & 1.1 \times 10^{ -2 } & 9.0 \times 10^{ -6 } \\
 88 & Blue  & 1.3 \times 10^{ 4 } & 2.0 \times 10^{ 2 } & 9.5 \times 10^{ -3 } & 1.3 \times 10^{ 45 } & 4.7 \times 10^{ -5 }  & 5.4 \times 10^{ 1 } & 5.6 \times 10^{ -3 } \\
    & Red   & 9.5 \times 10^{ 3 } & 2.4 \times 10^{ 2 } & 4.2 \times 10^{ -3 } & 3.5 \times 10^{ 44 } & 1.7 \times 10^{ -5 }  & 1.2 \times 10^{ 1 } & 1.6 \times 10^{ -3 } \\
 68 & Blue  & 1.1 \times 10^{ 4 } & 5.3 \times 10^{ 2 } & 9.4 \times 10^{ -3 } & 1.5 \times 10^{ 44 } & 1.8 \times 10^{ -5 }  & 2.3 \times 10^{ 0 } & 7.0 \times 10^{ -4 } \\
    & Red   & 1.1 \times 10^{ 4 } & 4.7 \times 10^{ 2 } & 5.8 \times 10^{ -3 } & 1.3 \times 10^{ 44 } & 1.2 \times 10^{ -5 }  & 2.2 \times 10^{ 0 } & 5.8 \times 10^{ -4 } \\
370\tablenotemark{a} & Blue  & 3.0 \times 10^{ 4 } & 1.6 \times 10^{ 2 } & 1.3 \times 10^{ -1 } & 9.5 \times 10^{ 46 } & 8.5 \times 10^{ -4 }  & 5.0 \times 10^{ 3 } & 2.3 \times 10^{ -1 } \\
368 & Blue  & 4.4 \times 10^{ 4 } & 1.4 \times 10^{ 4 } & 5.8 \times 10^{ -3 } & 2.4 \times 10^{ 42 } & 4.2 \times 10^{ -7 } & 1.5 \times 10^{ -3 } & 2.7 \times 10^{ -6 } \\
    & Red   & 2.5 \times 10^{ 4 } & 1.1 \times 10^{ 4 } & 3.3 \times 10^{ -3 } & 1.1 \times 10^{ 42 } & 3.1 \times 10^{ -7 } & 8.6 \times 10^{ -4 } & 1.8 \times 10^{ -6 } \\
 60 & Blue  & 1.6 \times 10^{ 4 } & 7.2 \times 10^{ 2 } & 1.7 \times 10^{ -3 } & 3.7 \times 10^{ 43 } & 2.3 \times 10^{ -6 } & 4.3 \times 10^{ -1 } & 1.1 \times 10^{ -4 } \\
    & Red   & 1.6 \times 10^{ 4 } & 3.9 \times 10^{ 2 } & 6.2 \times 10^{ -3 } & 4.2 \times 10^{ 44 } & 1.6 \times 10^{ -5 }  & 9.0 \times 10^{ 0 } & 1.3 \times 10^{ -3 } \\
 56 & Blue  & 1.6 \times 10^{ 4 } & 3.0 \times 10^{ 3 } & 2.4 \times 10^{ -3 } & 2.8 \times 10^{ 42 } & 7.9 \times 10^{ -7 } & 7.5 \times 10^{ -3 } & 8.5 \times 10^{ -6 } \\
    & Red   & 2.1 \times 10^{ 4 } & 5.2 \times 10^{ 3 } & 7.3 \times 10^{ -3 } & 5.0 \times 10^{ 42 } & 1.4 \times 10^{ -6 } & 7.9 \times 10^{ -3 } & 1.2 \times 10^{ -5 } \\
182 & Blue  & 1.1 \times 10^{ 4 } & 2.5 \times 10^{ 2 } & 9.9 \times 10^{ -3 } & 6.2 \times 10^{ 44 } & 3.9 \times 10^{ -5 }  & 2.0 \times 10^{ 1 } & 3.1 \times 10^{ -3 } \\
    & Red   & 1.3 \times 10^{ 4 } & 3.6 \times 10^{ 2 } & 2.0 \times 10^{ -2 } & 9.8 \times 10^{ 44 } & 5.4 \times 10^{ -5 }  & 2.2 \times 10^{ 1 } & 3.8 \times 10^{ -3 } \\
203 & Blue  & 1.9 \times 10^{ 4 } & 2.6 \times 10^{ 3 } & 1.3 \times 10^{ -3 } & 4.8 \times 10^{ 42 } & 5.2 \times 10^{ -7 } & 1.5 \times 10^{ -2 } & 9.8 \times 10^{ -6 } \\
    & Red   & 9.5 \times 10^{ 3 } & 9.6 \times 10^{ 2 } & 1.1 \times 10^{ -3 } & 7.7 \times 10^{ 42 } & 1.2 \times 10^{ -6 } & 6.7 \times 10^{ -2 } & 3.1 \times 10^{ -5 } \\
288\tablenotemark{b} & Blue  & 5.9 \times 10^{ 3 } & 5.5 \times 10^{ 2 } & 2.5 \times 10^{ -3 } & 1.2 \times 10^{ 43 } & 4.6 \times 10^{ -6 } & 1.8 \times 10^{ -1 } & 9.9 \times 10^{ -5 } \\
    & Red   & 8.0 \times 10^{ 3 } & 7.9 \times 10^{ 2 } & 1.7 \times 10^{ -3 } & 1.2 \times 10^{ 43 } & 2.2 \times 10^{ -6 } & 1.3 \times 10^{ -1 } & 5.8 \times 10^{ -5 } \\
\enddata
\tablenotetext{a}{The blue lobe refers to the northeastern lobe.}
\tablenotetext{b}{The blue and red lobes refer to the northeastern
                  and southwestern lobes, respectively.}
\end{deluxetable*}

The derived outflow parameters are listed in Table \ref{table_para}.
The outflow parameters presented in this paper
are corrected for the inclination.
Corrections for the optical depth are applied
to the blueshifted lobe of HOPS 288 only.
No correction is applied
for the missing or contaminating low-velocity emission components.
The southwestern outflow lobe of HOPS 370 is omitted
because this region is overly complicated (Section \ref{hops370}).
Figure \ref{fig_force76_force65} shows
a comparison of the outflow parameters derived from the two \co\ lines.
The overall correlation is good,
but the estimates from the \jss\ line tend to be smaller.
This difference is mainly owing
to the relatively low signal-to-noise ratios of the \jss\ data.
For HOPS 288 and HOPS 310,
the kinetic luminosities from the \jss\ line
are much smaller than those from the \jsf\ line,
which is mainly owing to the differences 
in the maximum velocities measured with these lines.
In the discussion below,
the estimates from the \jsf\ line are mainly used
to discuss the outflow properties and star formation activities.

The outflow parameters of HOPS 370 are exceptionally large.
Its molecular outflow luminosity is
even larger than the bolometric luminosity, which is unusual.
This anomaly is probably owing to the large inclination angle.
The inclination correction factors \citep{Dunham:2014gh} are valid
when all the gas motion is parallel to the outflow axis,
but the real outflow must have non-parallel velocity components.
Because of the large inclination of the HOPS 370 outflow,
its $F_{\rm CO}$ and $L_{\rm CO}$ are probably overestimated.
Therefore, in the discussion below,
HOPS 370 is excluded from the statistical analysis of outflow parameters.

It is difficult to estimate
the uncertainties in individual outflow parameters
because major contributions come
from certain assumptions made in the calculations.
By trying some reasonable alternative values,
it seems that the largest contribution to the uncertainties
comes from the inclination angle.
\cite{Furlan:2016df} used a grid of inclination angles, 
equally spaced in $\cos i$. 
The grid size may be considered as an estimate of the uncertainty.
Scaling the grid size with the correction factors,
the uncertainties in $\log\dot{M}_{\rm CO}$,
$\log F_{\rm CO}$, and $\log L_{\rm CO}$
are $\sim$0.08, $\sim$0.15, and $\sim$0.21, respectively.
These values may be optimistic
because the inclination of the outflow cavity near the protostar
is not necessarily exactly same as that of the large-scale outflow axis.
Considering this issue and other sources of uncertainties,
we will use uncertainties of 0.1 for $\log\dot{M}_{\rm CO}$,
0.2 for $\log F_{\rm CO}$, and 0.3 for $\log L_{\rm CO}$.
Note that our data will be combined with the data from other surveys
for the statistical analysis.
Therefore, these uncertainties are only nominal values
as a rough guide to the interpretation of statistics.

\section{Discussion}

\subsection{Comparison with Low-J CO Transitions}
\label{LowJCO}

The mid-$J$ CO lines ($J_{\rm up} \approx 6$)
trace warm components of outflows
while the low-$J$ lines ($J_{\rm up} \leq 3$)
trace relatively cold components,
as mentioned in Section 1.
It is therefore interesting to compare the outflow parameters
derived from these lines.
\cite{Takahashi:2008ei} observed some HOPS protostars in the OMC 2/3 region
in the \co\ \jtt\ line.
Comparison of the spectra shows
that the line wings are more pronounced, relative to the line core,
in the \jsf\ line than the \jtt\ line.

\begin{figure*}[!pt]
\epsscale{1.1}
\plotone{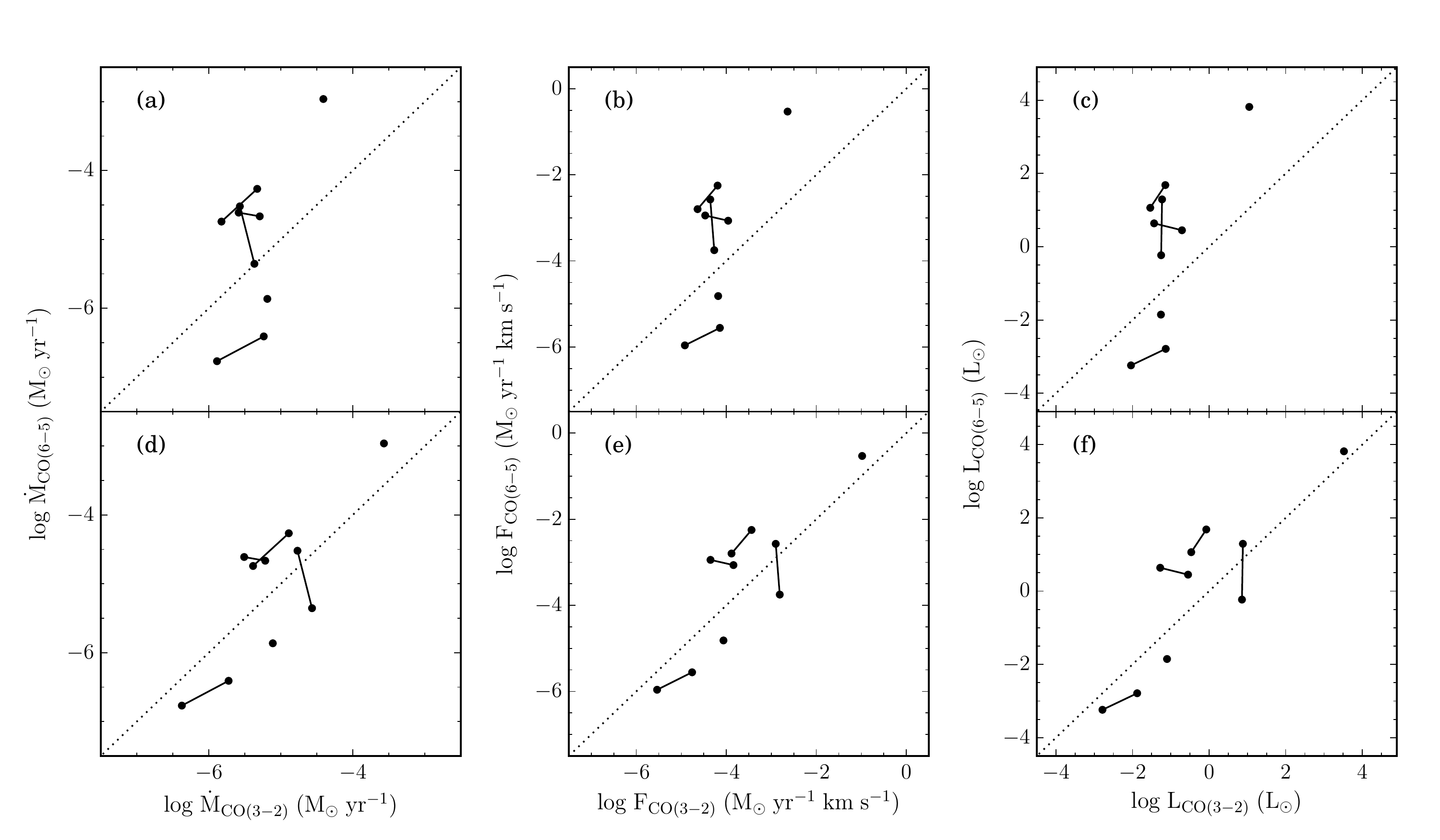}
\caption{
Scatter diagrams of (a) the mass outflow rate, (b) outflow force,
and (c) outflow kinetic luminosity
from the \co\ \jsf\ and \jtt\ lines
for the protostars in the OMC 2 and 3 regions
(HOPS 56, 60, 68, 88, 368, and 370).
The \co\ \jtt\ data are from \cite{Takahashi:2008ei}.
Short solid lines connect the bipolar outflow pairs of each protostar.
HOPS 56 and HOPS 370 have data for the blueshifted outflow lobes only.
The dotted lines show where the parameters from the two lines are the same.
Bottom panels show the scatter diagrams
with the \co\ \jtt\ parameters scaled for the same inclination angles
used in the calculation of the \co\ \jsf\ parameters.
\label{fig_co3265}}
\end{figure*}

The correlations between the outflow parameters derived from the two lines
are shown in the top panels of Figure \ref{fig_co3265}.
When the outflow parameters of \cite{Takahashi:2008ei}
are used without modification,
the correlations are moderately good,
and the linear correlation coefficients (Pearson's $r$) are
$r$ = 0.55, 0.64, and 0.68
for $\dot{M}_{\rm CO}$, $F_{\rm CO}$, and $L_{\rm CO}$, respectively.
However, the ranges of the \jtt\ parameters are
narrower than those of the \jsf\ parameters.
This difference is mostly owing to the inclination angle.
\cite{Takahashi:2008ei} used either $i$ = 45$\arcdeg$ or 70$\arcdeg$
depending on the outflow morphology.
When the outflow parameters are corrected for the inclination angles
used in this paper (Figure \ref{fig_co3265} bottom panels),
the correlation becomes even stronger ($r$ = 0.81, 0.84, and 0.85),
and the distributions of data points show
that the outflow parameters from the two lines are nearly equivalent.
These strong correlations suggest
that the two lines trace essentially the same outflow component.
The advantages in observing outflows in the \jsf\ line over the \jtt\ line
include higher angular resolution,
higher contrast to the cold ambient gas, and smaller optical depth.

Though the outflow strengths from the two lines are similar,
there are significant differences in the outflow timescales.
The timescales of the CO outflows traced by the \jtt\ line
in the study of \cite{Takahashi:2008ei}
are longer than those reported in this paper
typically by a factor of $\sim$5.
This contrast suggests
that the molecular outflows may be steady
over the range of timescales covered by the two studies
(from $\sim$200 to $\sim$30,000 years),
which implies that the outflow kinetic energy
is not easily converted to other forms of energy in these timescales.
Eventually, over a longer timescale,
molecular outflows may have disruptive effects on the dense cores,
develop to parsec scales, feed the interstellar turbulence,
and even disperse the parent clouds
\citep{Arce:2007wx,Arce:2010iq,Plunkett:2013cc}.

\subsection{Comparison with High-J CO Transitions}
\label{HighJCO}

The bolometric luminosity is
a good proxy of the accretion luminosity at the current epoch.
The high-$J$ CO lines in the far-IR range trace hot gas components,
and the far-IR CO luminosity ($L_{\rm CO}^{\rm FIR}$) is a good tracer
of the shocked gas near the base of molecular outflows
at the current epoch or in the recent past,
within the last $\sim$100 years \citep{Manoj:2016et}.
By contrast, the kinetic luminosities
from single-dish observations in low and mid-$J$ CO lines
trace the outflow power smoothed over a longer timescale,
300--20000 years for the outflows studied in this paper.
Therefore, it would be interesting to see
how $L_{\rm CO}^{\rm FIR}$ is related with the other two luminosities,
$L_{\rm bol}$, tracing short-term accretion activities,
and $L_{\rm CO}$, tracing long-term outflow activities.

\begin{figure*}[!t]
\epsscale{0.7}
\plotone{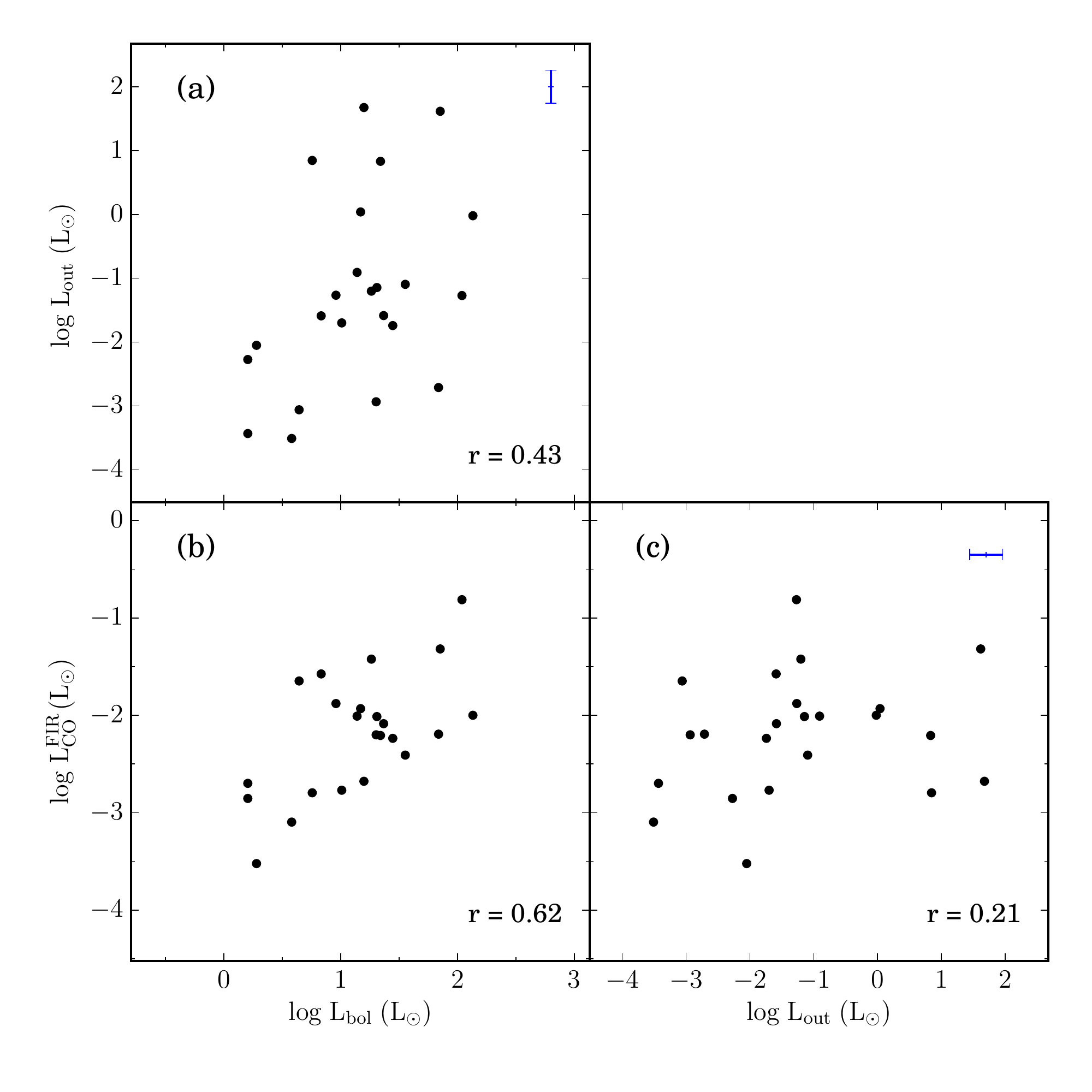}
\caption{
Scatter diagrams
of the outflow kinetic luminosities from the \co\ \jsf\ line,
the luminosities of the far-IR CO line emission,
and the bolometric luminosities.
The \co\ \jsf\ data are from this work
(Tables \ref{table_para} and \ref{a_table_para}).
The far-IR data are from \cite{Manoj:2016et} and \cite{Karska:2013ij}.
The bolometric luminosities are from \cite{Furlan:2016df},
\cite{Kristensen:2012kw}, and \cite{Green:2013dq}.
HOPS 370 is not shown
because of the lack of data for the redshifted outflow lobe.
The linear correlation coefficient
is written in each panel.
The bars in the upper right corners of panels (a) and (c)
represent nominal uncertainties of $\log L_{\rm out}$}.
\label{fig_co65fir}
\end{figure*}

The scatter diagrams of the three forms of luminosities
are shown in Figure \ref{fig_co65fir}.
The kinetic luminosity of molecular outflow is defined by
\begin{equation}
L_{\rm out} = 2 (L_{\rm CO}^{\rm blue}\ L_{\rm CO}^{\rm red})^{{1}\over{2}},
\end{equation}
where $L_{\rm CO}^{\rm blue}$ and $L_{\rm CO}^{\rm red}$ are
the kinetic luminosities from the \co\ \jsf\ line observations
of the blueshifted and redshifted outflow lobes, respectively.
Geometric mean is used
because the fits to the outflow data in the next section
are made in the logarithmic scale.
The factor 2 is needed
because $L_{\rm CO}$ is the power per outflow lobe
while $L_{\rm out}$ is the power per protostar.
Note that $L_{\rm out}$ is the kinetic luminosity of molecular outflow 
while $L_{\rm CO}^{\rm FIR}$ is the line-emission luminosity
of CO gas ($J_{\rm up}$ = 14--46).

Though both $L_{\rm out}$ and $L_{\rm CO}^{\rm FIR}$ trace outflow activities,
they show almost no correlation (Figure \ref{fig_co65fir}(c)),
which suggests that the high-$J$ lines and the low/mid-$J$ lines
trace very different components of molecular outflows.
The reason may be that they have different timescales
and their energetics are not directly related.
While $L_{\rm out}$ may depend
on relatively global properties of protostellar envelopes,
$L_{\rm CO}^{\rm FIR}$ may be
sensitive to the local condition of shocked regions.
Despite the different timescales,
the bolometric luminosity shows a moderate correlation with $L_{\rm out}$,
because the accretion process
is the underlying driving mechanism of outflows.
The bolometric luminosity shows a relatively good correlation
with $L_{\rm CO}^{\rm FIR}$,
because they have similar timescales and are directly related in energetics.
(See Sections 4.1 and 5.3 of \cite{Manoj:2016et}
for more discussions on these luminosities.)
The comparisons presented above show
that the low/mid-$J$ and high-$J$ CO lines are complementary,
and they can be useful in distinguishing
the long-term (evolutionary) outflow behaviors
from the short-term (episodic or variable) activities.

Results of a survey of protostars observed in the CO lines
from $J$ = 4 $\rightarrow$ 3 to 13 $\rightarrow$ 12
were presented by \cite{Yang:2018hj}.
They found that multiple excitation mechanisms affect
a wide range of CO energy levels with significant overlaps.
They also found that the line emission luminosity of the \jsf\ line
shows a good correlation with the total CO luminosity.
These findings suggest
that resolving the CO emission structure both spectrally and spatially
is critical to the understanding of the underlying physical processes. 

\subsection{Comparison with Protostellar Properties}

\begin{figure*}
\epsscale{0.9}
\plotone{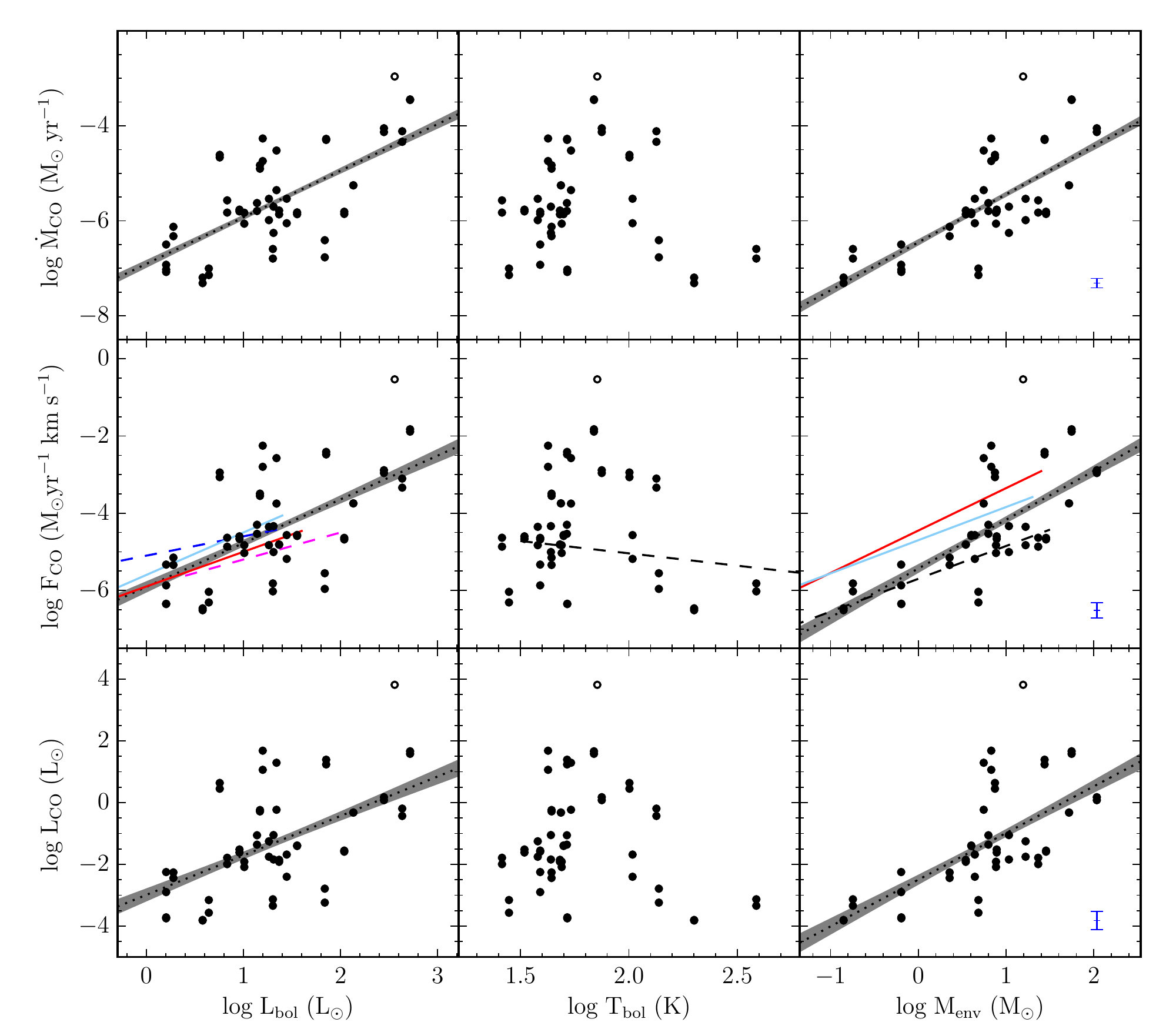}
\caption{
Scatter diagrams of the mass outflow rate,
outflow force, and outflow kinetic luminosity
with the protostellar bolometric luminosity, 
bolometric temperature, and envelope mass.
Each marker represents an outflow lobe.
The data are from this work and \cite{vanKempen:2016bv}.
The dotted lines show the best-fit lines to all data points
except for HOPS 370 (open dots).
The red solid line shows the correlation for Class I protostars,
proposed by \cite{Bontemps:1996vb} using the \co\ \jto\ line.
The dashed lines show the correlations
for protostars in the Perseus molecular cloud,
proposed by \cite{Curtis:2010ia} using the \co\ \jtt\ line
(in the left panel, the blue line is for Class 0,
and the magenta line is for Class I).
The light-blue solid line shows the correlation for protostars in Ophiuchus,
proposed by \cite{vanderMarel:2013hy}.
The shaded areas show the 95\% confidence region of the fits.
The vertical bars in the lower right corner represent nominal uncertainties.
\label{fig_ltm-mfl}}
\end{figure*}

The correlation between the outflow force
and bolometric luminosity of protostars
was revealed by \cite{Bontemps:1996vb}.
Since then, many studies corroborated
the $L_{\rm bol}$-$F_{\rm CO}$ correlation
\citep{Hatchell:2007cz,Curtis:2010ia,vanderMarel:2013hy,vanKempen:2016bv}.
Scatter diagrams
between the outflow parameters and protostellar properties 
are shown in Figure \ref{fig_ltm-mfl}.
Such diagrams are often made
by summing the quantities of the blueshifted and redshifted outflow lobes.
The meaning of simple sum, however, is unclear
when the dynamic times of the two lobes are different.
Therefore, the outflow parameters of each lobe
are displayed separately in these figures.
To cover a wide range of the parameter space,
the data from this work (Tables \ref{table_para} and \ref{a_table_para})
are combined with the \co\ \jsf\ data of three intermediate-mass protostars
(NGC 2071, Vela IRS 17, and IRAS 20050+2720) from \cite{vanKempen:2016bv}.
Vela IRS 19 in the sample of \cite{vanKempen:2016bv} is omitted
because its $T_{\rm bol}$ is unknown.
The protostellar parameters are from \cite{Furlan:2016df},
\cite{Kristensen:2012kw}, \cite{Froebrich:2005es},
and \cite{Strafella:2015}.
(Some parameters are updated and listed in the Appendix A.)
As a whole, the sample covers
a luminosity range of $L_{\rm bol}$ = 1.6--520 \lsun\
and a temperature range of $T_{\rm bol}$ = 26--387 K.

The scatter diagrams
in the left panels of Figure \ref{fig_ltm-mfl}
show clear trends of $\dot{M}_{\rm CO}$, $F_{\rm CO}$, and $L_{\rm CO}$
increasing with $L_{\rm bol}$.
The linear correlation coefficients are
$r$ = 0.68, 0.61, and 0.56, respectively.
Performing a linear least-squares fit to each diagram,
the best-fit lines are
\begin{equation}
\log \dot{M}_{\rm CO} = (-6.91 \pm 0.03) + (0.98 \pm 0.02) \log L_{\rm bol},
\end{equation}
\begin{equation}
\log F_{\rm CO} = (-5.91 \pm 0.06) + (1.13 \pm 0.04) \log L_{\rm bol},
\end{equation}
\begin{equation}
\log L_{\rm CO} = (-2.99 \pm 0.09) + (1.28 \pm 0.06) \log L_{\rm bol},
\label{eq_lco-lbol}
\end{equation}
where $\dot{M}_{\rm CO}$ is in \msun\ yr$^{-1}$,
$F_{\rm CO}$ is in \msun\ yr$^{-1}$ \kms,
and $L_{\rm CO}$ and $L_{\rm bol}$ are in  \lsun.

The slope of the $L_{\rm bol}$-$F_{\rm CO}$ relation in this work
is consistent with that given by \cite{vanderMarel:2013hy}
but steeper than those of \cite{Bontemps:1996vb} and \cite{Curtis:2010ia}.
This difference probably comes from the coverage of $L_{\rm bol}$.
The outflow sample of \cite{Bontemps:1996vb} covers
a range of $L_{\rm bol}$ = 0.2--41 \lsun,
and outflows were undetected for some protostars
at the lower part of the range (see Figure 5 of \cite{Bontemps:1996vb}).
Therefore, their $F_{\rm CO}$ data may be biased positively
at the lower end of their $L_{\rm bol}$ range,
and the slope may appear shallower.

The scatter diagrams between the outflow parameters and $T_{\rm bol}$ 
are shown in the central panels of Figure \ref{fig_ltm-mfl}.
They show trends of $\dot{M}_{\rm CO}$, $F_{\rm CO}$, and $L_{\rm CO}$
decreasing with $T_{\rm bol}$.
However, the linear correlation coefficients are small:
$r$ = $-0.06$ , $-0.12$, and $-0.16$, respectively.
The main reason is the large dispersions
in the middle range of $T_{\rm bol}$ (40--100 K).
Figure 6(a) of \cite{Curtis:2010ia} shows a similar feature.
This trend may be an evolutionary effect.
At the low end of $T_{\rm bol}$,
the protostellar mass is small, the accretion luminosity is low,
and consequently the outflow is relatively weak.
The outflow becomes stronger as the protostellar mass increases.
On the high side of $T_{\rm bol}$,
the envelope mass decreases with evolution,
and the outflow gradually weakens.
The evolution of outflow parameters with $T_{\rm bol}$
may be complicated
and cannot be described with simple power-law fits.
Despite the low degree of correlation,
the $T_{\rm bol}$-$F_{\rm CO}$ relation in this work
is consistent with that given by \cite{Curtis:2010ia} 
($F_{\rm CO} \propto T_{\rm bol}^{-0.64}$,
as shown by the dashed line
in the central panel of Figure \ref{fig_ltm-mfl}),
which suggests that the overall trend may be real.
This complicated evolutionary trend shows
that $L_{\rm bol}$ and $T_{\rm bol}$ should be considered together
for proper analyses of outflow statistics.

In addition to $L_{\rm bol}$ and $T_{\rm bol}$,
it has been well known
that the mass of protostellar envelope shows
a positive correlation with the outflow force
\citep{Bontemps:1996vb,Curtis:2010ia,vanderMarel:2013hy}.
However, the envelope mass derived from continuum SED 
is highly model dependent. For the outflow sample in this work,
the masses derived by \cite{Kristensen:2012kw} are usually much larger
than those of \cite{Furlan:2016df}
for protostars with similar $L_{\rm bol}$ and $T_{\rm bol}$.
This discrepancy may be caused by the differences
in the model density profiles and the definitions of envelope mass,
such as the inner and outer radii for integration.

For a simple comparison with other studies, 
the envelope masses ($M_{\rm env}$) were derived
from submillimeter flux densities
using Equation (1) of \cite{Nutter:2007fp}.
They assumed optically thin conditions, a dust temperature of 20 K,
and a mass emissivity of 0.01 cm$^2$ g$^{-1}$
at the wavelength of 850 \micron.
The 850 \micron\ flux densities were obtained
from the catalogues in \cite{Nutter:2007fp} and \cite{DiFrancesco:2008di}.
Four objects in the outflow sample are not included in these catalogues.
The flux density of HOPS 288 was obtained from \cite{vanKempen:2012fb},
and HOPS 368, Vela IRS 17, and BHR 71 are omitted.

The scatter diagrams
in the right panels of Figure \ref{fig_ltm-mfl}
show clear trends of $\dot{M}_{\rm CO}$, $F_{\rm CO}$, and $L_{\rm CO}$
increasing with $M_{\rm env}$.
The linear correlation coefficients are
$r$ = 0.75, 0.72, and 0.69, respectively.
(These coefficients are essentially the same as those of $L_{\rm bol}$,
calculated with HOPS 368, Vela IRS 17, and BHR 71 excluded.)
The best-fit lines are
\begin{equation}
\log \dot{M}_{\rm CO} = (-6.45 \pm 0.02) + (1.01 \pm 0.02) \log M_{\rm env},
\end{equation}
\begin{equation}
\log F_{\rm CO} = (-5.43 \pm 0.04) + (1.27 \pm 0.04) \log M_{\rm env},
\end{equation}
\begin{equation}
\log L_{\rm CO} = (-2.50 \pm 0.06) + (1.51 \pm 0.06) \log M_{\rm env},
\end{equation}
where $M_{\rm env}$ are in \msun.

The slope of the $M_{\rm env}$-$F_{\rm CO}$ relation in this work
is steeper than those of \cite{Bontemps:1996vb},
\cite{Curtis:2010ia}, and \cite{vanderMarel:2013hy}.
This difference may be owing to the same reason mentioned above,
in the discussion on the $L_{\rm bol}$-$F_{\rm CO}$ relation.

\section{Individual Sources}

\subsection{HOPS 310}

HOPS 310 is a Class 0 protostar associated with IRAS 05399--0121
in the LBS 30 core of the L1630 cloud \citep{Furlan:2016df}.
It drives a giant HH flow \citep{Bally:2002gt}.
The infrared image (Figure \ref{fig_wisemap}(a)) shows
several HH objects along the outflow.

The APEX CO maps (Figure \ref{fig_310_co}) cover
the $110\arcsec\ \times\ 110\arcsec$ region centered at HOPS 310.
The blue CO outflow peak
in the northwestern corner of Figure \ref{fig_310_co}(g) and (h)
corresponds to HH 92.
At the position of HH 92,
the \co\ spectra show emission from the outflow only,
with little emission from the ambient cloud.

\subsection{HOPS 88/87}

Several target outflows are located in the OMC 2/3 region
(Figure \ref{fig_wisemap}(e)).
The HOPS 88 mapping field
was originally selected for both HOPS 87 and 88.
Initial examination of the mapping data revealed
that the HOPS 87 outflow is unsuitable for a detailed study,
and it was subsequently removed from the target list.

HOPS 88 is a Class 0 protostar associated with OMC 3 MMS 5
\citep{Chini:1997hm,Furlan:2016df}
and has the lowest $T_{\rm bol}$ among the survey sample.
It drives an outflow in the east-west direction
\citep{Williams:2003dj,Takahashi:2008ei}.
Several infrared knots were detected along the western flow
\citep{Yu:1997do,Takahashi:2008ei}. 

In the central region of the CO mapping field,
there are three protostars: HOPS 86, 87, and 88 (Figure \ref{fig_88_co}).
Only HOPS 88 shows a prominent CO outflow.
This outflow shows clearly separated blue and red lobes. 
It has the largest $V_{\rm max}$ among the survey sample,
$\sim$100 \kms\ for the blue wing and $\sim$70 \kms\ for the red wing.
HOPS 88 also shows the narrowest \tco\ \jsf\ line profile
among the survey sample.
The blue outflow lobe is much stronger than the red one.
In the red lobe, there is a blueshifted emission structure
elongated in the north-south direction,
which is aligned with HOPS 87. 
This structure was not considered
for calculations of the red outflow parameters. 

HOPS 87 is stronger than HOPS 88
in the high-$J$ CO emission \citep{Manoj:2013ie}.
It is also brighter than HOPS 88
in the \tco\ line (Figure \ref{fig_88_co}(d)).
HOPS 87 is associated with OMC 3 MMS 6,
the brightest 1.3 mm continuum source
in the OMC 2/3 region \citep{Chini:1997hm}.
However, the \co\ maps (Figure \ref{fig_88_co}(g) and (h))
may be showing only a hint of an outflow around HOPS 87,
and it is too weak to measure outflow parameters.
No clear outflow of HOPS 87 was found
in previous studies of single-dish low-$J$ CO observations
\citep{Williams:2003dj, Takahashi:2008ei}.
\cite{Takahashi:2012hh} discovered a compact outflow
with interferometric observations.

\subsection{HOPS 68}

HOPS 68 is a Class I protostar associated with OMC 2 FIR 2
\citep{Mezger:1990vx,Furlan:2016df}.
It drives a bipolar outflow previously studied in low-$J$ CO lines
\citep{Aso:2000hx,Williams:2003dj,Takahashi:2008ei}.
HOPS 68 has the lowest $L_{\rm bol}$ among the survey sample.
It was detected in fewer than eight far-IR CO lines with PACS
and has the lowest rotational temperature
among the protostars studied by \cite{Manoj:2013ie}.

The \co\ maps of HOPS 68 (Figure \ref{fig_68_co}(g) and (h))
show a prominent bipolar outflow in the north-south direction.
It has the second largest $V_{\rm max}$
among the target outflows ($\sim$70 \kms).
The kinetic luminosity and outflow force of HOPS 68 are
on the high side of the distributions.

\subsection{HOPS 370}
\label{hops370}

HOPS 370 is a Class I protostar associated with OMC 2 FIR 3, 
with a $T_{\rm bol}$ close to the Class 0-I boundary
\citep{Mezger:1990vx,Furlan:2016df}.
It drives a spectacular bipolar outflow
\citep{Aso:2000hx,Williams:2003dj,Takahashi:2008ei}.
HOPS 370 has the highest $L_{\rm bol}$
and the highest far-IR CO luminosity among the survey sample.
The outflow from HOPS 370 shows the brightest far-IR emission 
of all the Orion low-to-intermediate mass protostars observed 
by Herschel \citep{Manoj:2013ie,GonzalezGarcia:2016cc}.

The \co\ maps (Figure \ref{fig_370_co}(g) and (h)) show
the bipolar outflow of HOPS 370
flowing in the northeast-southwest direction.
Both outflow lobes display redshifted and blueshifted line wings
in comparable strengths.
This emission pattern suggests
that the outflow axis is very close to the plane of the sky,
which is consistent with the high inclination angle
derived by \cite{Furlan:2016df}.
The kinetic luminosity and outflow force of the northeastern (blue) lobe
are the largest among the target outflows.

It is difficult to analyze the southwestern outflow lobe of HOPS 370
because of the confusion with other outflows.
There are several YSOs in this region
(HOPS 64, 108, and 369).
The star formation activity of HOPS 64,
located at 23$\arcsec$ southeast of HOPS 370,
is not well known,
and it exhibits optically thick free-free emission \citep{Osorio:2017bw}.
HOPS 108 (FIR 4) is associated with a thermal radio jet
and a near-IR nebulosity
\citep{Reipurth:1999jg,Takahashi:2008ei,Osorio:2017bw}.
Previous studies suggested
that the formation of HOPS 108 may have been triggered
by the HOPS 370 outflow
\citep{Shimajiri:2008ih,GonzalezGarcia:2016cc,Osorio:2017bw}.
Though the details are unclear,
the velocity structure in this part of the cloud is complicated
\citep{Osorio:2017bw}.
To the southwest of HOPS 108,
there are two radio sources, VLA 15 and VLA 16,
indicating other outflow activities of YSOs
\citep{Osorio:2017bw,Tobin:2019dm}.
While the redshifted \co\ emission structure
(Figure \ref{fig_370_co}(g) and (h)) seem to show the HOPS 370 outflow,
the blueshifted emission may be more closely related
to the cluster of YSOs in this region.

Near the northeastern outflow lobe of HOPS 370,
there is HOPS 66 (MIR 20, \cite{Nielbock:2003gl})
located 15$\arcsec$ northwest of HOPS 370.
\cite{Takahashi:2008ei} reported a jet-like near-IR feature
and extended blueshifted \co\ \jtt\ emission
to the west of this object.
This outflow can be seen in the \co\ \jsf\ map (Figure \ref{fig_370_co}(g)).
Though this extended blueshifted emission may be associated with HOPS 66,
it is hard to distinguish it
from the dominant blueshifted emission of HOPS 370.
In calculating the outflow parameters of HOPS 370,
this emission structure of HOPS 66 was excluded.

\subsection{HOPS 368}
\label{hops368}

HOPS 368 is a Class I protostar associated with OMC 2 VLA 13
\citep{Reipurth:1999jg,Furlan:2016df}.
It drives a relatively compact bipolar outflow \citep{Takahashi:2008ei}.

The \co\ maps (Figure \ref{fig_368_co}(g) and (h)) show
the bipolar outflow of HOPS 368 flowing in the north-south direction.
The outflow velocity is relatively low,
and the dynamical time is relatively long.
Despite the relatively high bolometric luminosity,
the kinetic luminosity and outflow force of HOPS 368
are the smallest among the target outflows.

\subsection{HOPS 60}

HOPS 60 is a Class 0 protostar associated with OMC 2 FIR 6b
\citep{Chini:1997hm,Furlan:2016df}.
It drives a prominent bipolar outflow \citep{Takahashi:2008ei}.

The \co\ maps (Figure \ref{fig_60_co}(g) and (h)) show
the bipolar outflow of HOPS 60 flowing in the northeast-southwest direction.
In addition, there is a blueshifted emission structure
near the southern edge of the maps,
which belongs to the outflow driven by FIR 6c (HOPS 409)
\citep{Takahashi:2008ei}.
The FIR 6c outflow lobe was excluded
from the analysis of the HOPS 60 outflow parameters.

In the \tco\ \jsf\ map (Figure \ref{fig_60_co}(d)),
there is yet another emission component
at $\sim$30\arcsec\ south of HOPS 60.
This component is associated with FIR 6a.
\cite{Shimajiri:2009kn} proposed
that the FIR 6a core may be interacting with the FIR 6c outflow.

\subsection{HOPS 56}

HOPS 56 is a Class 0 protostar associated with OMC 2 CSO 33
\citep{Lis:1998dz,Furlan:2016df}.
\cite{Takahashi:2008ei} categorized the outflow of HOPS 56 as ``probable''
because there is a directional discrepancy
between a near-IR nebulosity
and the redshifted \co\ ${J=3\rightarrow2}$ emission peak.
The redshifted emission structure subtends a large angle from HOPS 56.

The \co\ \jsf\ and \jss\ maps (Figure \ref{fig_56_co}(g) and (h)) show
outflow structures similar to those seen in the ${J=3\rightarrow2}$ line,
which suggests that the redshifted outflow lobe
may have a relatively large opening angle.
HOPS 57 is located near the redshifted outflow lobe,
but there is no known outflow activity of HOPS 57.

\subsection{HOPS 182}

HOPS 182 is a Class 0 protostar associated with L1641N MM1
\citep{Stanke:2007ba,Furlan:2016df}.
The L1641N cluster region contains many YSOs and multiple outflows
\citep{Fukui:1986fj,Wilking:1990gi,Stanke:2007ba,Galfalk:2008db}.
\cite{Stanke:2007ba} reported that
there are two deeply embedded objects, MM1 and MM3, in the cluster center,
each of them driving an outflow along a northeast-southwest direction. 
They suggested that the CO outflow seen in single-dish maps
may be driven by MM1,
while MM3 drives a well-collimated faint jet.
HOPS 181 (mid-IR source 18 of \cite{Ali:2004gy})
drives yet another outflow in the north-south direction
\citep{Stanke:2007ba}.
There is no known outflow activity of HOPS 183.

The \co\ \jsf\ and \jss\ maps (Figure \ref{fig_182_co}(g) and (h)) show
the outflow structures in the HOPS 182 region.
The well-defined bipolar lobes in the central region
belong to the HOPS 182 outflow.
The emission structures near the northern and southern edges of the maps
belong to the HOPS 181 giant bipolar outflow
and are excluded from the calculations of outflow parameters.
The kinetic luminosity and outflow force of HOPS 182
are the second largest among the target outflows.

\subsection{HOPS 203}

HOPS 203 is a Class 0 protostar associated with HH 1/2 VLA 1
\citep{Pravdo:1985bw, Rodriguez:1990du, Reipurth:1993hu,Furlan:2016df}
and has the second lowest $T_{\rm bol}$ among the survey sample.
It drives the HH 1/2/401/402 giant outflow (Figure \ref{fig_wisemap}(c))
that extends more than 20\arcmin\ on either side of the driving source
\citep{Ogura:1995ek, Reipurth:2013gw}.
The total projected size of this HH complex is 5.9 pc,
which is the second-largest HH flow known \citep{Reipurth:1997hk}.
The HH 1/2 flow axis may be
very close ($< 20\arcdeg$) to the plane of the sky
\citep{NoriegaCrespo:1991ck, Eisloffel:1994hc, Correia:1997wn},
which is consistent with the inclination angle of 70\arcdeg\
derived from the SED fitting by \cite{Furlan:2016df}.
This outflow has been studied extensively in low-$J$ CO lines
\citep{Choi:1997ke, Correia:1997wn,MoroMartin:1999cu}.

The \co\ \jsf\ and \jss\ maps (Figure \ref{fig_203_co}(g) and (h))
clearly show the bipolar outflow of HOPS 203
in the northwest-southeast direction.
Both outflow lobes display redshifted and blueshifted line wings.

A faint centimeter continuum source, VLA 2,
is located at 3\arcsec\ from VLA 1
and drives the HH 144 flow to the west \citep{Reipurth:1993hu}.
The \co\ maps (Figure \ref{fig_203_co}(g) and (h)) show a hint of this flow,
but it is too weak to derive outflow parameters.
HOPS 165 probably drives the HH 146 flow
to the south \citep{Reipurth:1993hu},
but it is unclear if this flow produces any detectable CO emission.

\subsection{HOPS 288}

HOPS 288 is a Class 0 protostar
associated with L1641 S3 MMS 1 (= L1641 S3 IRS)
\citep{Stanke:2000tv,Furlan:2016df}.
The bolometric luminosity of HOPS 288 is
the second highest among the survey sample.
It drives a giant H$_2$/CO outflow
\citep{Morgan:1991et,Wilking:1990gi,Stanke:2000tv,vanKempen:2016bv}.
The infrared H$_2$ features are distributed
along the bright 4.6 \micron\ emission,
showing a point-symmetric variability of flow direction
(Figure \ref{fig_wisemap}(d)).

Figure \ref{fig_288_co} shows
the mid-$J$ \co\ spectra and maps of the HOPS 288 region.
Both northeastern and southwestern outflow lobes display
redshifted and blueshifted line wings in comparable strengths,
which is consistent with the \co\ ${J=2\rightarrow1}$ maps
presented by \cite{Wilking:1990gi}.
This emission pattern indicates
that the outflow opening angle is larger
than the angle between the outflow axis and the plane of the sky.
This outflow may have a large opening angle
owing to the directional variability mentioned in the previous paragraph
(also see the discussion in \cite{Stanke:2000tv}).
Among the target outflows in this work,
the kinetic luminosity and outflow force of HOPS 288 are moderate,
despite the large bolometric luminosity.

\section{Summary}

Ten protostellar outflows in the Orion molecular clouds
were observed in the \co/\tco\ \jsf\ and \co\ \jss\ lines
with the APEX/CHAMP$^+$ instrument.
The target protostars were selected based on the bright far-IR CO lines
\citep{Manoj:2013ie, Manoj:2016et}.
Maps were made in the three lines
and have an angular resolution of 10\arcsec.

The target protostars show outflows
traced by the \co\ \jsf\ and \jss\ line wings.
For each outflow lobe,
parameters such as the maximum velocities, lobe lengths,
dynamical timescales, and masses were measured.
Properties of the molecular outflows were derived,
including the mass transfer rates,
kinetic luminosities, and outflow forces.

In addition to the data from the observations presented in this paper,
the data of low-luminosity protostars presented by \cite{Yldz:2015ib}
were re-analyzed.
The outflow parameters of \cite{vanKempen:2016bv}
were also included in the analysis.
By combining these data sets,
the outflow sample covers a wide range of luminosities,
$L_{\rm bol}$ = 1.6--520 \lsun.

The outflow parameters derived from the \co\ \jsf\ line
show strong correlations with those from the \jtt\ line,
which suggests
that the two lines trace essentially the same outflow component.
The outflow kinetic luminosities from the \jsf\ line
and the high-$J$ (far-IR) line-emission luminosities of CO
show little correlation,
which implies that they trace distinct components of molecular outflows.
The low/mid-$J$ CO line wings and the high-$J$ CO lines
are sensitive to the long-term outflow behaviors
and the short-term accretion activities, respectively.
Nevertheless, the underlying energy source may be related
because $L_{\rm bol}$ shows moderate correlations
with both $L_{\rm out}$ and $L_{\rm CO}^{\rm FIR}$.

The correlations between the properties of molecular outflows and protostars
were investigated.
Similarly to the findings of previous works,
the mass outflow rate, outflow force, and kinetic luminosity
increase with bolometric luminosity and envelope mass.
Power-law fits to the outflow parameters
as functions of these protostellar parameters are presented.
The strengths of molecular outflows show
little correlation with bolometric temperature.
The evolution of outflow parameters with $T_{\rm bol}$ may be complicated.

\acknowledgments

M. K. was supported by Basic Science Research Program
through the National Research Foundation of Korea (NRF)
funded by the Ministry of Science, ICT \& Future Planning
(NRF-2015R1C1A1A01052160). 
G. P. was supported by Basic Science Research Program through 
the National Research Foundation of Korea funded by 
the Ministry of Education (NRF-2020R1A6A3A01100208).

\clearpage
\appendix
\section{Outflow Properties of Low-mass Protostars}
\restartappendixnumbering 

The outflow parameters of low-mass protostars were recalculated
with the data presented by \cite{Yldz:2015ib}
using the method described in Section 3 of this paper.
Velocity limits were determined
using the \co\ \jsf\ spectra with a 1 \kms\ velocity resolution. 
Table \ref{a_table_contour} lists the newly determined limits.
The systemic velocities and distances
were taken from Table 4 of \cite{Yldz:2013bt}.
For several sources,
the protostellar parameters and distances were updated for various reasons,
and Table \ref{a_table_dist} lists the values used in this paper.
The outflow parameters were calculated using the \co\ \jsf\ data,
corrected for the inclination angles
listed in Tables 2 and 3 of \cite{Yldz:2015ib}.
The outflow properties of 14 protostars
are listed in Table \ref{a_table_para}.
There are 23 protostars in the \jsf\ line sample of \cite{Yldz:2015ib},
but 9 of them were omitted.
The line-wing emission intensities of 7 protostars
(Ced 110 IRS 4, L723MM, TMC 1A, TMC 1, DK Cha, Oph IRS 63, and RNO 91)
are not strong enough to calculate the outflow parameters.
TMR 1 and GSS 30 IRS 1 were excluded
because the regions around these protostars
are too complicated and contain multiple outflows.

\begin{figure*}
\epsscale{0.65}
\plotone{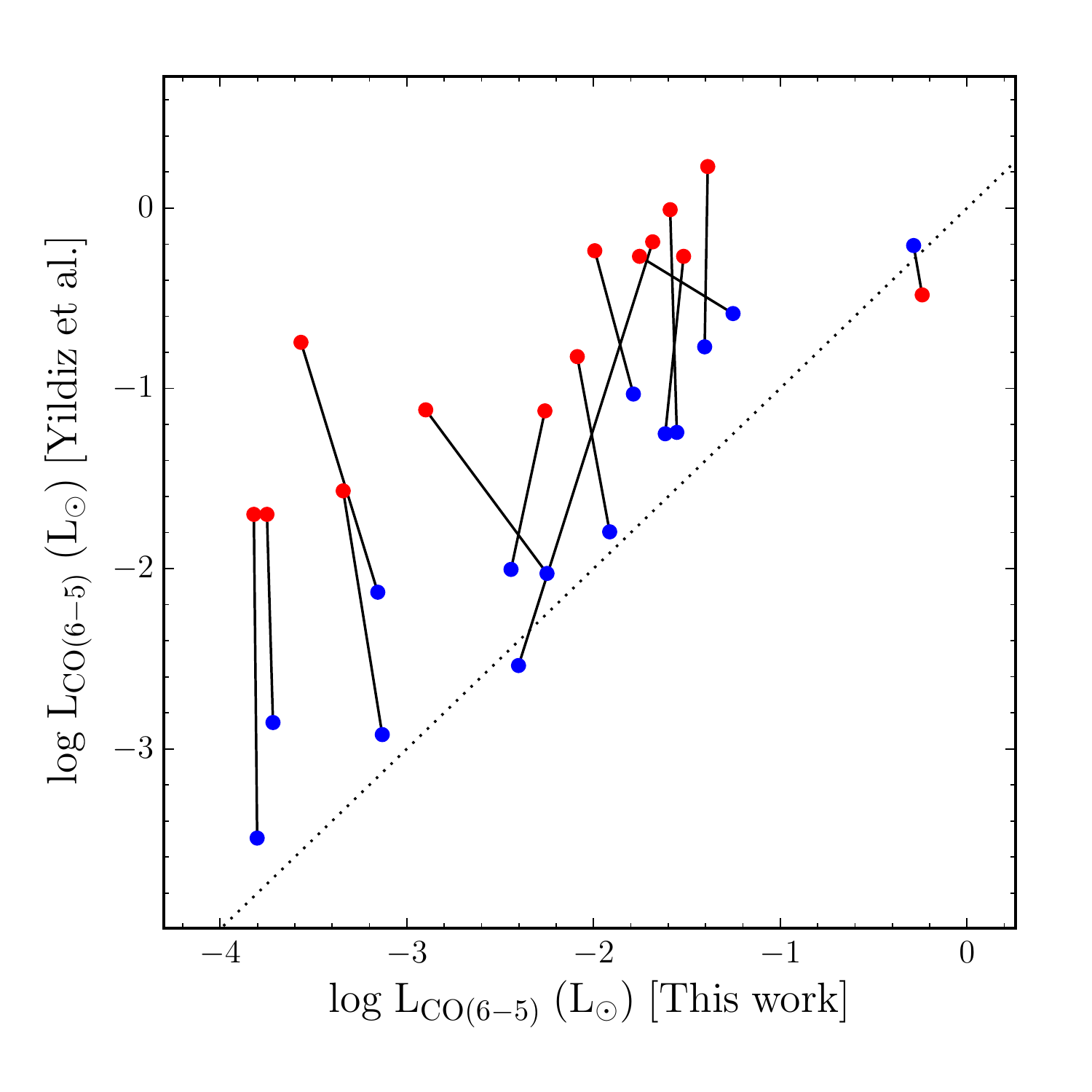}
\caption{
Comparison of the \co\ \jsf\ outflow kinetic luminosities
between \cite{Yldz:2015ib} and this work.
Blue and red circles represent the blue and red lobes, respectively.
Short solid lines connect the bipolar outflow pairs of each protostar.
The dotted line shows where the two luminosities are the same.
\label{fig_kang_yildiz}}
\end{figure*}

The outflow properties reported by \cite{Yldz:2015ib}
have a strong bias toward the redshifted outflow.
The red-to-blue luminosity ratio is
$\log(L_{\rm CO}^{\rm red}/L_{\rm CO}^{\rm blue})$ = 1.1 on average,
with a standard deviation of 0.6.
The reason for this unnatural bias is unclear.
The newly derived values in Table \ref{a_table_para} do not show such a bias:
$\log(L_{\rm CO}^{\rm red}/L_{\rm CO}^{\rm blue})$ = --0.1 on average,
with a standard deviation of 0.3.
Figure \ref{fig_kang_yildiz} shows
a lobe-by-lobe comparison between the two works.

\begin{deluxetable}{lRRRRRRR}[b]
\tabletypesize{\small}
\tablecaption{Velocity Limits of Line Wings \label{a_table_contour}}
\tablewidth{0pt}
\tablehead{
\colhead{Source}
& \multicolumn{3}{c}{Blue lobe} && \multicolumn{3}{c}{Red lobe} \\
\cline{2-4} \cline{6-8}
& \colhead{$V_{\rm max,b}$} & \colhead{$V_{\rm out,b}$}
& \colhead{$V_{\rm in,b}$}
&& \colhead{$V_{\rm in,r}$} & \colhead{$V_{\rm out,r}$}
& \colhead{$V_{\rm max,r}$}}
\startdata
 NGC 1333 IRAS 2A & 11.2 &  -3.5 &  4.5 && 10.5 & 20.5 & 12.8 \\
 NGC 1333 IRAS 4A & 19.5 & -12.5 &  4.5 &&  9.5 & 29.5 & 22.5 \\
 NGC 1333 IRAS 4B & 17.6 & -10.5 &  4.5 &&  9.5 & 19.5 & 12.4 \\
 L1527            &  7.4 &  -1.5 &  4.5 &&  7.5 & 13.5 &  7.6 \\
 BHR71            & 16.1 & -20.5 & -5.5 && -2.5 & 15.5 & 19.9 \\
 IRAS 15398       & 10.6 &  -5.5 &  3.5 &&  6.5 & 14.5 &  9.4 \\
 L483MM           &  8.7 &  -3.5 &  4.5 &&  6.5 & 12.5 &  7.3 \\
 Ser SMM1         & 18.0 &  -9.5 &  3.5 && 13.5 & 27.5 & 19.0 \\
 Ser SMM4         & 20.5 & -12.5 &  4.5 && 11.5 & 20.5 & 12.5 \\
 Ser SMM3         & 21.1 & -13.5 &  3.5 && 12.5 & 23.5 & 15.9 \\
 B335             & 10.9 &  -2.5 &  6.5 && 10.5 & 15.5 &  7.1 \\
 L1489            &  6.7 &   0.5 &  5.5 &&  8.5 & 14.5 &  7.3 \\
 HH46 IRS         &  9.7 &  -4.5 &  4.5 &&  6.5 & 19.5 & 14.3 \\
 Elias 29         & 11.8 &  -7.5 &  2.5 &&  6.5 & 15.5 & 11.2 \\
\enddata
\tablecomments{All velocities are in \kms.}
\end{deluxetable}


\begin{deluxetable}{llRRR}
\tabletypesize{\small}
\tablecaption{Source Parameters \label{a_table_dist}}
\tablewidth{0pt}
\tablehead{
\colhead{Source} & \colhead{$D$} & \colhead{$L_{\rm bol}$} 
& \colhead{$T_{\rm bol}$} & \colhead{$L_{\rm CO}^{\rm FIR}$} \\
& \colhead{(pc)} & \colhead{($L_\odot$)} & \colhead{(K)}
& \colhead{($\times 10^{-3}~L_\odot$)}}
\startdata
BHR 71\tablenotemark{a}        & 200   &  14.8 &  44 &  11.7 \\
Serpens SMM 1\tablenotemark{b} & 436   & 109.2 &  39 & 153.8 \\
Serpens SMM 4\tablenotemark{b} & 436   &   6.8 &  26 &  26.6 \\
Serpens SMM 3\tablenotemark{b} & 436   &  18.3 &  38 &  37.7 \\
B335\tablenotemark{c}          & 164.5 &   1.6 &  39 &   1.4 \\
Elias 29\tablenotemark{d}      & 125   &  20.1 & 387 &   6.3 \\
\enddata
\tablenotetext{a}{For consistency,
                  the $L_{\rm CO}^{\rm FIR}$ \citep{Manoj:2016et}
                  is scaled for the distance
                  listed in \cite{Kristensen:2012kw}.}
\tablenotetext{b}{The $L_{\rm bol}$ and $L_{\rm CO}^{\rm FIR}$
                  \citep{Karska:2013ij}
                  are scaled for the revised distance
                  from \cite{OrtizLeon:2017ey}.}
\tablenotetext{c}{The $L_{\rm bol}$ \citep{Green:2013dq}
                  and $L_{\rm CO}^{\rm FIR}$ \citep{Manoj:2016et}
                  are scaled for the revised distance
                  from \cite{Watson:2020cy}.
                  The $T_{\rm bol}$ is from \cite{Green:2013dq}.}
\tablenotetext{d}{The $L_{\rm bol}$ and $T_{\rm bol}$
                  are from \cite{Green:2013dq}.
                  The values in \cite{Kristensen:2012kw}
                  have an unclear origin.}
\end{deluxetable}


\begin{deluxetable}{lcCCCCCCC}
\tabletypesize{\small}
\tablecaption{Properties of the CO Outflows \label{a_table_para}}
\tablewidth{0pt}
\tablehead{
\colhead{Source} & \colhead{Lobe} & \colhead{$R_{\rm lobe}$}
& \colhead{$t_{\rm dyn}$} & \colhead{$M_{\rm CO}$}
& \colhead{$E_{\rm CO}$} & \colhead{$\dot{M}_{\rm CO}$}
& \colhead{$L_{\rm CO}$} & \colhead{$F_{\rm CO}$} \\
& & \colhead{(au)} & \colhead{(year)} & \colhead{(\msun)}
& \colhead{(erg)} & \colhead{(\msun\ yr$^{-1}$)} & \colhead{(\lsun)}
& \colhead{(\msun\ yr$^{-1}$ \kms)}}
\startdata
NGC 1333 IRAS 2A & Blue & 3.9 \times 10^{ 3 }  & 5.6 \times 10^{ 2 }  & 7.8 \times 10^{ -4 }  & 2.7 \times 10^{ 42 }  & 1.4 \times 10^{ -6 }  & 3.9 \times 10^{ -2 }  & 2.6 \times 10^{ -5 }\\ 
                 & Red  & 6.7 \times 10^{ 3 }  & 8.5 \times 10^{ 2 }  & 1.3 \times 10^{ -3 }  & 4.2 \times 10^{ 42 }  & 1.5 \times 10^{ -6 }  & 4.1 \times 10^{ -2 }  & 2.7 \times 10^{ -5 }\\ 
NGC 1333 IRAS 4A & Blue & 4.1 \times 10^{ 4 }  & 6.5 \times 10^{ 3 }  & 1.0 \times 10^{ -2 }  & 1.9 \times 10^{ 43 }  & 1.6 \times 10^{ -6 }  & 2.4 \times 10^{ -2 }  & 2.2 \times 10^{ -5 }\\ 
                 & Red  & 4.7 \times 10^{ 4 }  & 6.3 \times 10^{ 3 }  & 1.1 \times 10^{ -2 }  & 2.3 \times 10^{ 43 }  & 1.7 \times 10^{ -6 }  & 3.0 \times 10^{ -2 }  & 2.5 \times 10^{ -5 }\\ 
NGC 1333 IRAS 4B & Blue & 2.8 \times 10^{ 4 }  & 7.3 \times 10^{ 3 }  & 7.3 \times 10^{ -4 }  & 6.2 \times 10^{ 41 }  & 9.9 \times 10^{ -8 }  & 7.0 \times 10^{ -4 }  & 9.2 \times 10^{ -7 }\\ 
                 & Red  & 1.6 \times 10^{ 4 }  & 6.1 \times 10^{ 3 }  & 4.4 \times 10^{ -4 }  & 2.0 \times 10^{ 41 }  & 7.2 \times 10^{ -8 }  & 2.7 \times 10^{ -4 }  & 4.9 \times 10^{ -7 }\\ 
L1527            & Blue & 1.1 \times 10^{ 4 }  & 2.4 \times 10^{ 3 }  & 1.1 \times 10^{ -3 }  & 1.0 \times 10^{ 42 }  & 4.7 \times 10^{ -7 }  & 3.6 \times 10^{ -3 }  & 4.6 \times 10^{ -6 }\\ 
                 & Red  & 1.0 \times 10^{ 4 }  & 2.2 \times 10^{ 3 }  & 1.6 \times 10^{ -3 }  & 1.4 \times 10^{ 42 }  & 7.5 \times 10^{ -7 }  & 5.5 \times 10^{ -3 }  & 7.1 \times 10^{ -6 }\\ 
BHR 71           & Blue & 5.7 \times 10^{ 4 }  & 5.7 \times 10^{ 3 }  & 7.2 \times 10^{ -2 }  & 3.6 \times 10^{ 44 }  & 1.3 \times 10^{ -5 }  & 5.2 \times 10^{ -1 }  & 2.8 \times 10^{ -4 }\\ 
                 & Red  & 5.6 \times 10^{ 4 }  & 4.5 \times 10^{ 3 }  & 6.7 \times 10^{ -2 }  & 3.2 \times 10^{ 44 }  & 1.5 \times 10^{ -5 }  & 5.8 \times 10^{ -1 }  & 3.2 \times 10^{ -4 }\\ 
IRAS 15398       & Blue & 6.2 \times 10^{ 3 }  & 2.4 \times 10^{ 3 }  & 2.0 \times 10^{ -4 }  & 5.6 \times 10^{ 40 }  & 8.4 \times 10^{ -8 }  & 1.9 \times 10^{ -4 }  & 4.4 \times 10^{ -7 }\\ 
                 & Red  & 4.1 \times 10^{ 3 }  & 1.8 \times 10^{ 3 }  & 1.7 \times 10^{ -4 }  & 3.9 \times 10^{ 40 }  & 9.4 \times 10^{ -8 }  & 1.8 \times 10^{ -4 }  & 4.5 \times 10^{ -7 }\\ 
L483MM           & Blue & 1.5 \times 10^{ 4 }  & 2.8 \times 10^{ 3 }  & 4.2 \times 10^{ -3 }  & 4.2 \times 10^{ 42 }  & 1.5 \times 10^{ -6 }  & 1.2 \times 10^{ -2 }  & 1.5 \times 10^{ -5 }\\ 
                 & Red  & 1.5 \times 10^{ 4 }  & 3.3 \times 10^{ 3 }  & 2.9 \times 10^{ -3 }  & 3.3 \times 10^{ 42 }  & 8.7 \times 10^{ -7 }  & 8.2 \times 10^{ -3 }  & 9.3 \times 10^{ -6 }\\ 
Serpens SMM 1    & Blue & 4.6 \times 10^{ 4 }  & 7.9 \times 10^{ 3 }  & 1.2 \times 10^{ -2 }  & 2.6 \times 10^{ 43 }  & 1.6 \times 10^{ -6 }  & 2.8 \times 10^{ -2 }  & 2.3 \times 10^{ -5 }\\ 
                 & Red  & 6.2 \times 10^{ 4 }  & 1.0 \times 10^{ 4 }  & 1.4 \times 10^{ -2 }  & 3.1 \times 10^{ 43 }  & 1.4 \times 10^{ -6 }  & 2.6 \times 10^{ -2 }  & 2.1 \times 10^{ -5 }\\ 
Serpens SMM 4    & Blue & 6.0 \times 10^{ 4 }  & 1.2 \times 10^{ 4 }  & 3.3 \times 10^{ -2 }  & 2.4 \times 10^{ 43 }  & 2.7 \times 10^{ -6 }  & 1.6 \times 10^{ -2 }  & 2.3 \times 10^{ -5 }\\ 
                 & Red  & 6.3 \times 10^{ 4 }  & 2.1 \times 10^{ 4 }  & 3.1 \times 10^{ -2 }  & 2.5 \times 10^{ 43 }  & 1.5 \times 10^{ -6 }  & 1.0 \times 10^{ -2 }  & 1.4 \times 10^{ -5 }\\ 
Serpens SMM 3    & Blue & 1.2 \times 10^{ 4 }  & 1.8 \times 10^{ 3 }  & 5.2 \times 10^{ -3 }  & 1.2 \times 10^{ 43 }  & 2.9 \times 10^{ -6 }  & 5.6 \times 10^{ -2 }  & 4.4 \times 10^{ -5 }\\ 
                 & Red  & 7.3 \times 10^{ 3 }  & 1.4 \times 10^{ 3 }  & 1.4 \times 10^{ -3 }  & 3.0 \times 10^{ 42 }  & 1.0 \times 10^{ -6 }  & 1.8 \times 10^{ -2 }  & 1.5 \times 10^{ -5 }\\ 
B335             & Blue & 4.4 \times 10^{ 3 }  & 6.5 \times 10^{ 2 }  & 2.1 \times 10^{ -4 }  & 4.4 \times 10^{ 41 }  & 3.2 \times 10^{ -7 }  & 5.6 \times 10^{ -3 }  & 4.7 \times 10^{ -6 }\\ 
                 & Red  & 6.3 \times 10^{ 3 }  & 1.4 \times 10^{ 3 }  & 1.7 \times 10^{ -4 }  & 2.2 \times 10^{ 41 }  & 1.2 \times 10^{ -7 }  & 1.3 \times 10^{ -3 }  & 1.3 \times 10^{ -6 }\\ 
L1489            & Blue & 3.7 \times 10^{ 3 }  & 1.7 \times 10^{ 3 }  & 8.1 \times 10^{ -5 }  & 3.2 \times 10^{ 40 }  & 4.9 \times 10^{ -8 }  & 1.6 \times 10^{ -4 }  & 3.1 \times 10^{ -7 }\\ 
                 & Red  & 4.3 \times 10^{ 3 }  & 1.8 \times 10^{ 3 }  & 1.2 \times 10^{ -4 }  & 3.3 \times 10^{ 40 }  & 6.4 \times 10^{ -8 }  & 1.5 \times 10^{ -4 }  & 3.4 \times 10^{ -7 }\\ 
HH 46 IRS        & Blue & 1.2 \times 10^{ 4 }  & 3.8 \times 10^{ 3 }  & 3.4 \times 10^{ -3 }  & 1.8 \times 10^{ 42 }  & 8.9 \times 10^{ -7 }  & 4.0 \times 10^{ -3 }  & 6.5 \times 10^{ -6 }\\ 
                 & Red  & 1.8 \times 10^{ 4 }  & 3.9 \times 10^{ 3 }  & 1.1 \times 10^{ -2 }  & 9.8 \times 10^{ 42 }  & 2.9 \times 10^{ -6 }  & 2.1 \times 10^{ -2 }  & 2.7 \times 10^{ -5 }\\ 
Elias 29         & Blue & 1.7 \times 10^{ 4 }  & 6.1 \times 10^{ 3 }  & 1.6 \times 10^{ -3 }  & 5.4 \times 10^{ 41 }  & 2.6 \times 10^{ -7 }  & 7.4 \times 10^{ -4 }  & 1.5 \times 10^{ -6 }\\ 
                 & Red  & 1.6 \times 10^{ 4 }  & 5.9 \times 10^{ 3 }  & 9.5 \times 10^{ -4 }  & 3.3 \times 10^{ 41 }  & 1.6 \times 10^{ -7 }  & 4.6 \times 10^{ -4 }  & 9.4 \times 10^{ -7 }\\ 
\enddata
\end{deluxetable}

\clearpage
\newpage
\section{CO spectra and maps of the target outflows}
\label{COspandmaps}

The spectra and maps of HOPS 310 are shown in Figure \ref{fig_310_co},
and those of the other protostars are presented here.
Details are explained in the caption of Figure \ref{fig_310_co}.

\begin{figure}[h]
\includegraphics[width=1.0\textwidth]{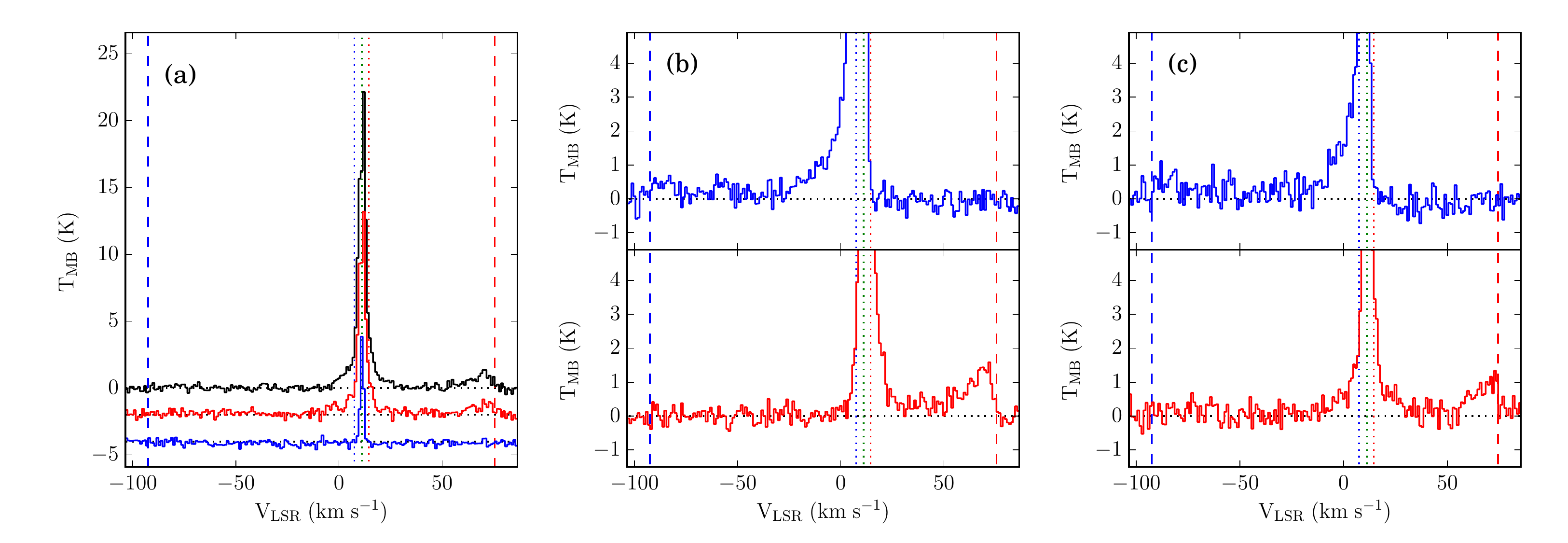}
\includegraphics[width=1.0\textwidth]{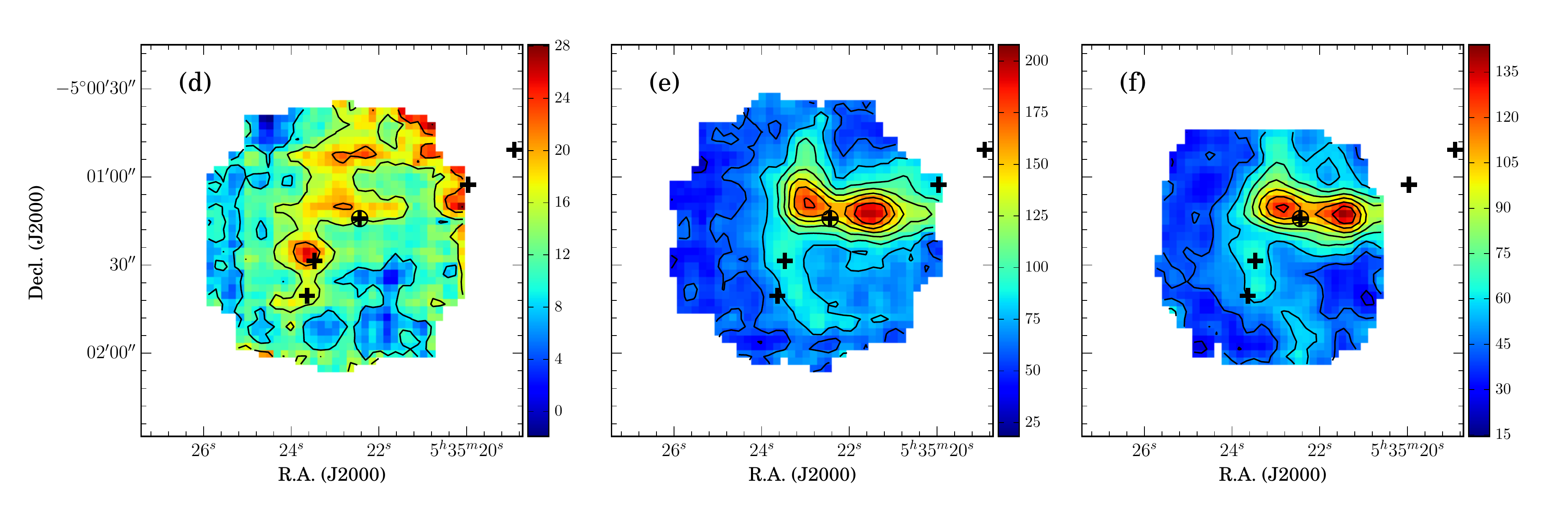}
\epsscale{0.9}
\plotone{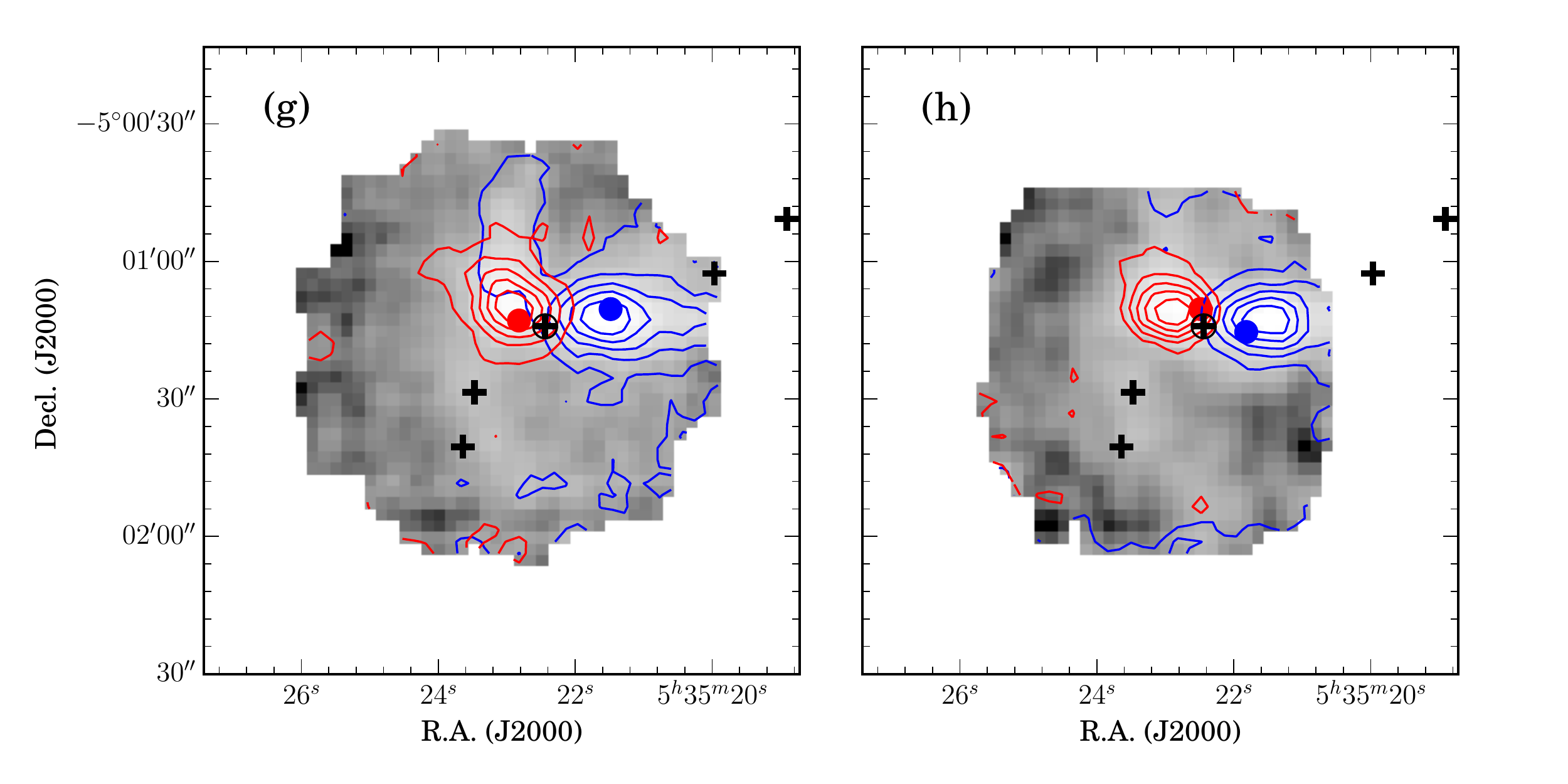}
\caption{
CO spectra and maps for HOPS 88 (OMC 3 MMS 5).
\label{fig_88_co}}
\end{figure}

\begin{figure*}
\includegraphics[width=1.0\textwidth]{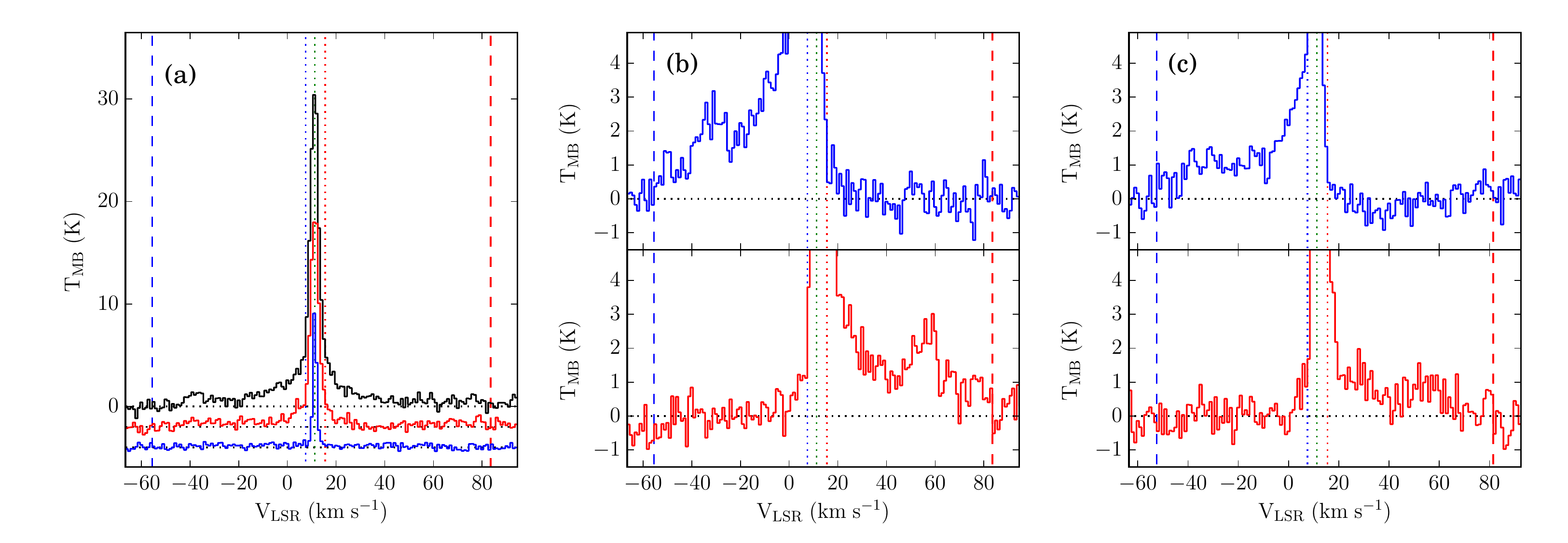}
\includegraphics[width=1.0\textwidth]{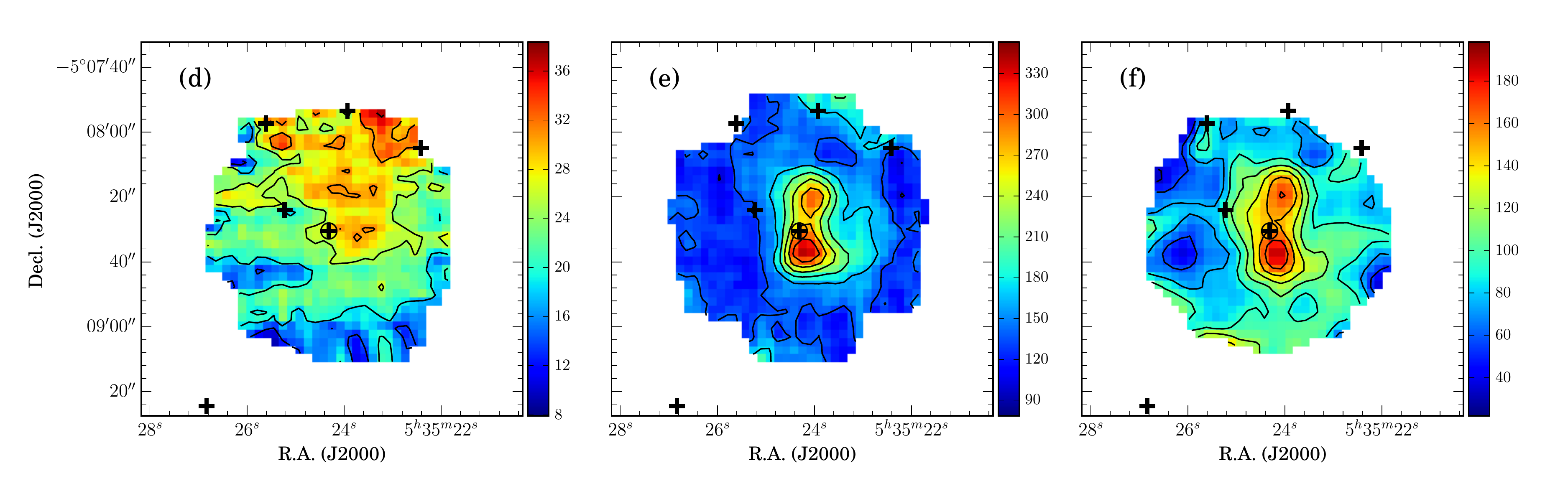}
\epsscale{0.9}
\plotone{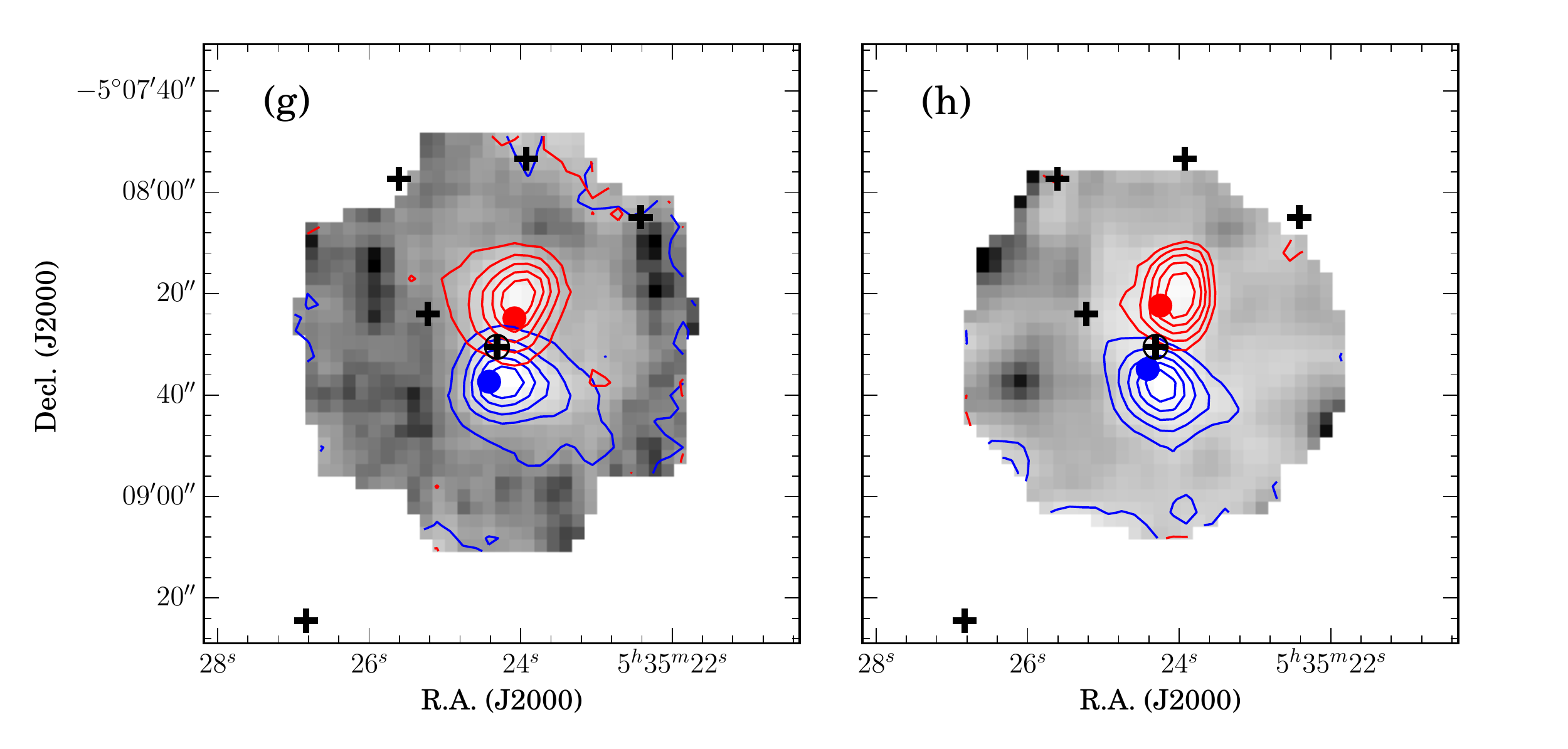}
\caption{
CO spectra and maps for HOPS 68 (OMC 2 FIR 2).
\label{fig_68_co}}
\end{figure*}

\begin{figure*}
\includegraphics[width=0.9\textwidth]{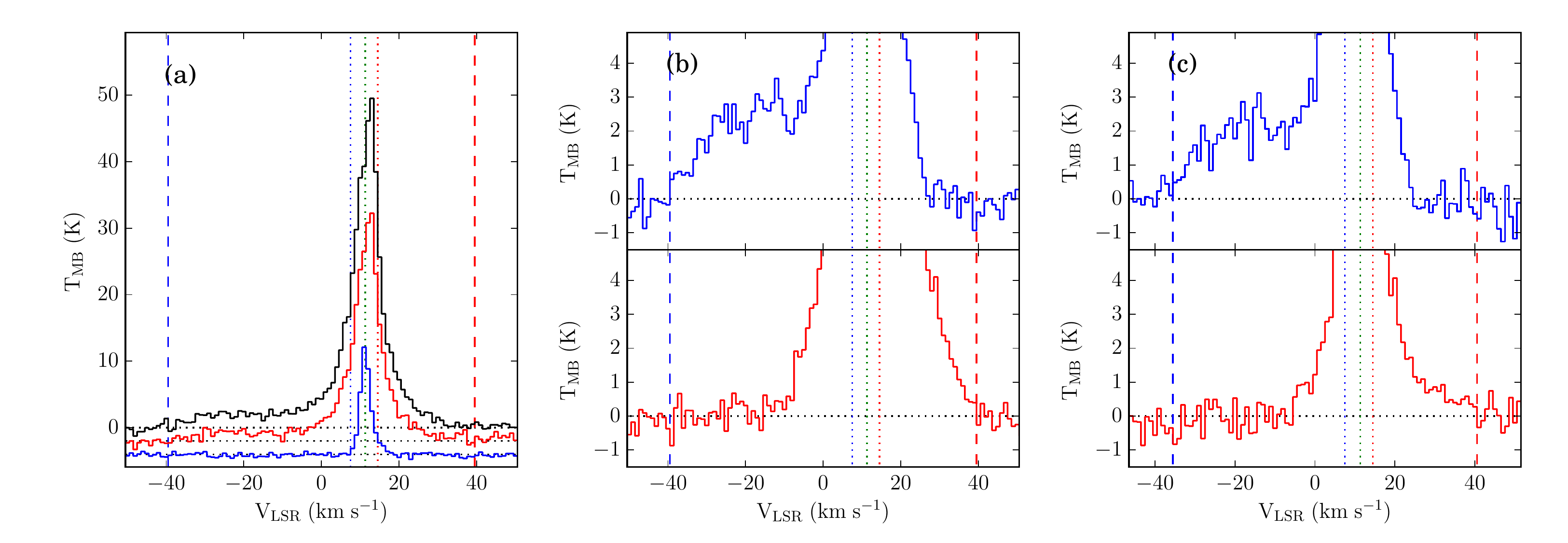}
\includegraphics[width=0.9\textwidth]{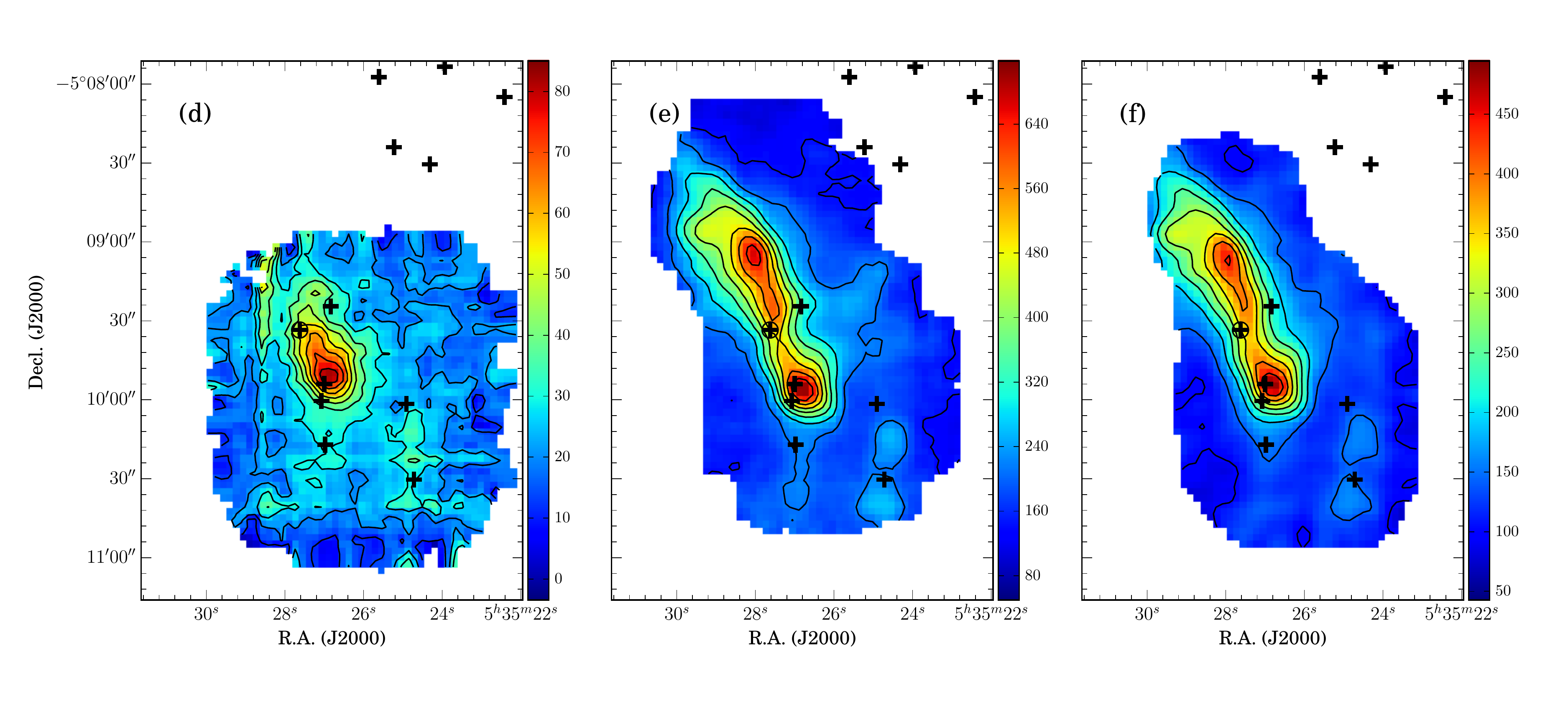}
\epsscale{0.8}
\plotone{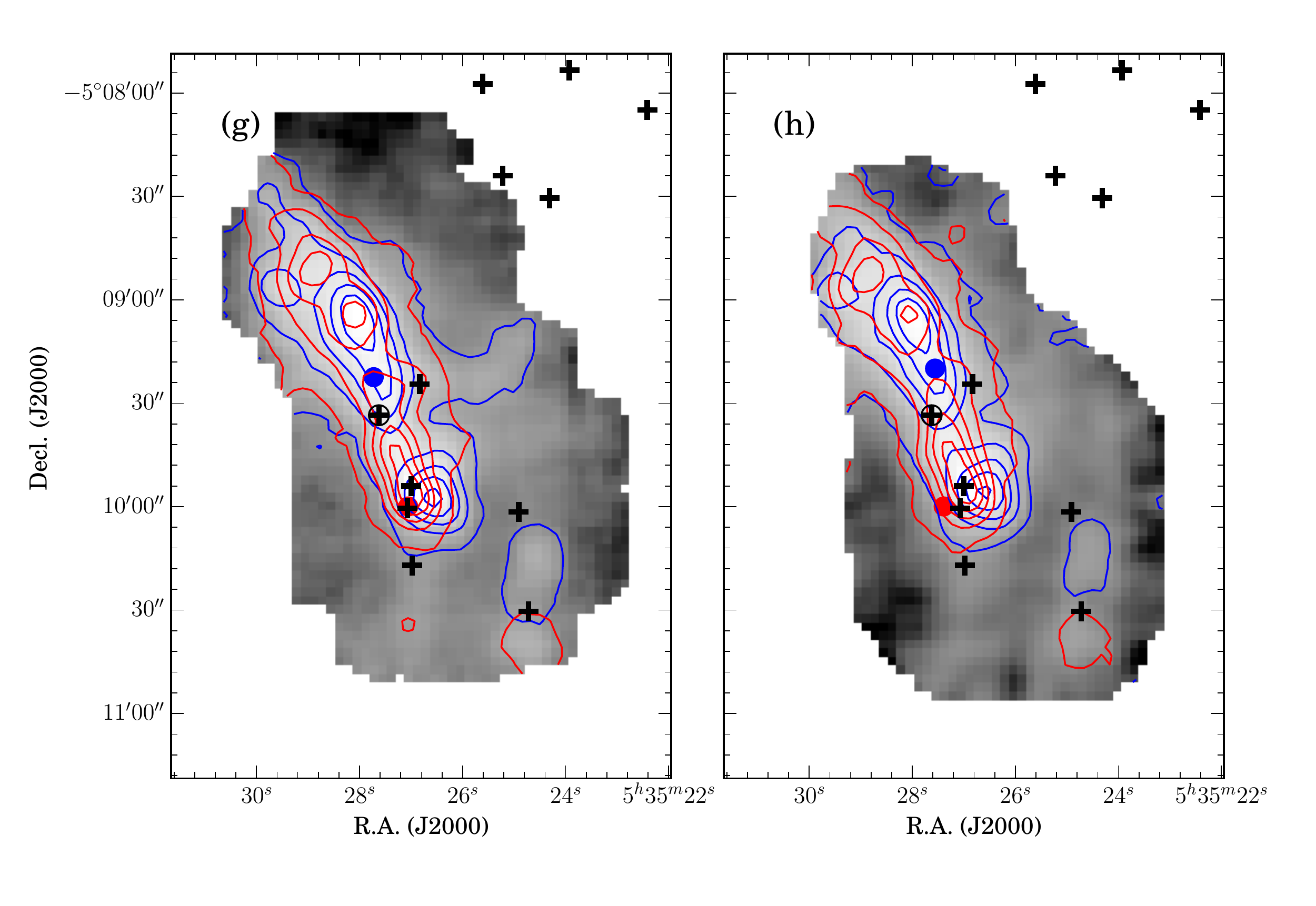}
\caption{
CO spectra and maps for HOPS 370 (OMC 2 FIR 3).
\label{fig_370_co}}
\end{figure*}

\begin{figure*}
\includegraphics[width=0.95\textwidth]{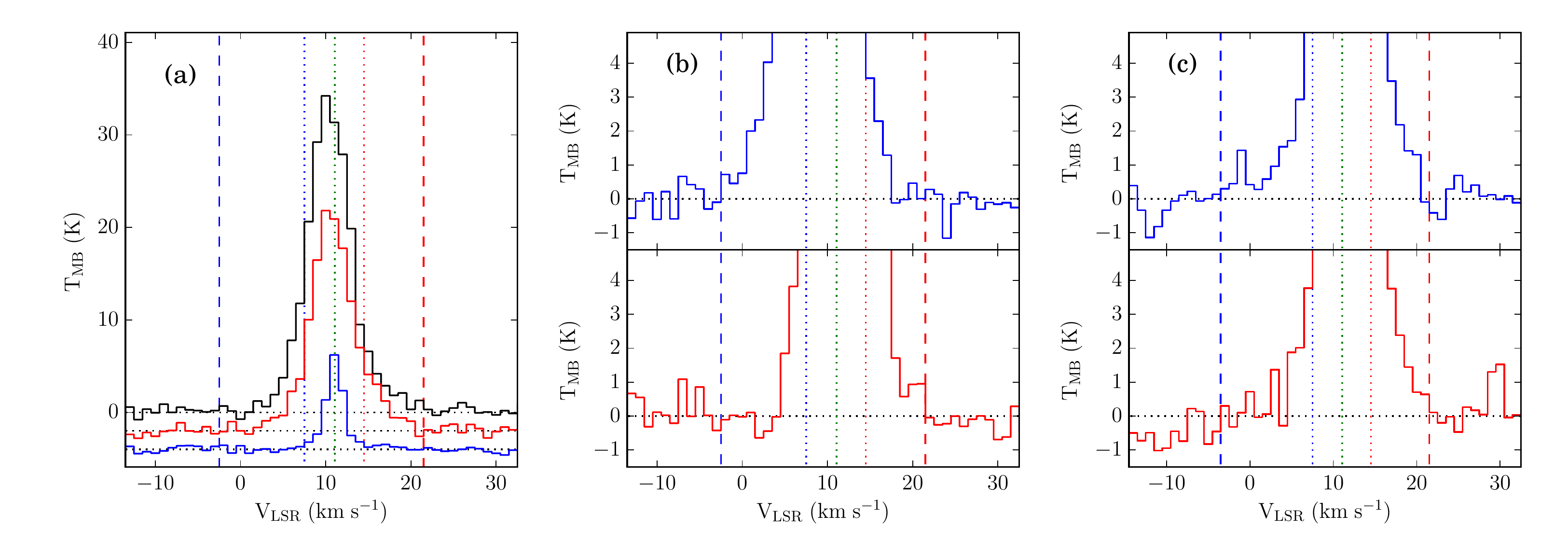}
\includegraphics[width=0.95\textwidth]{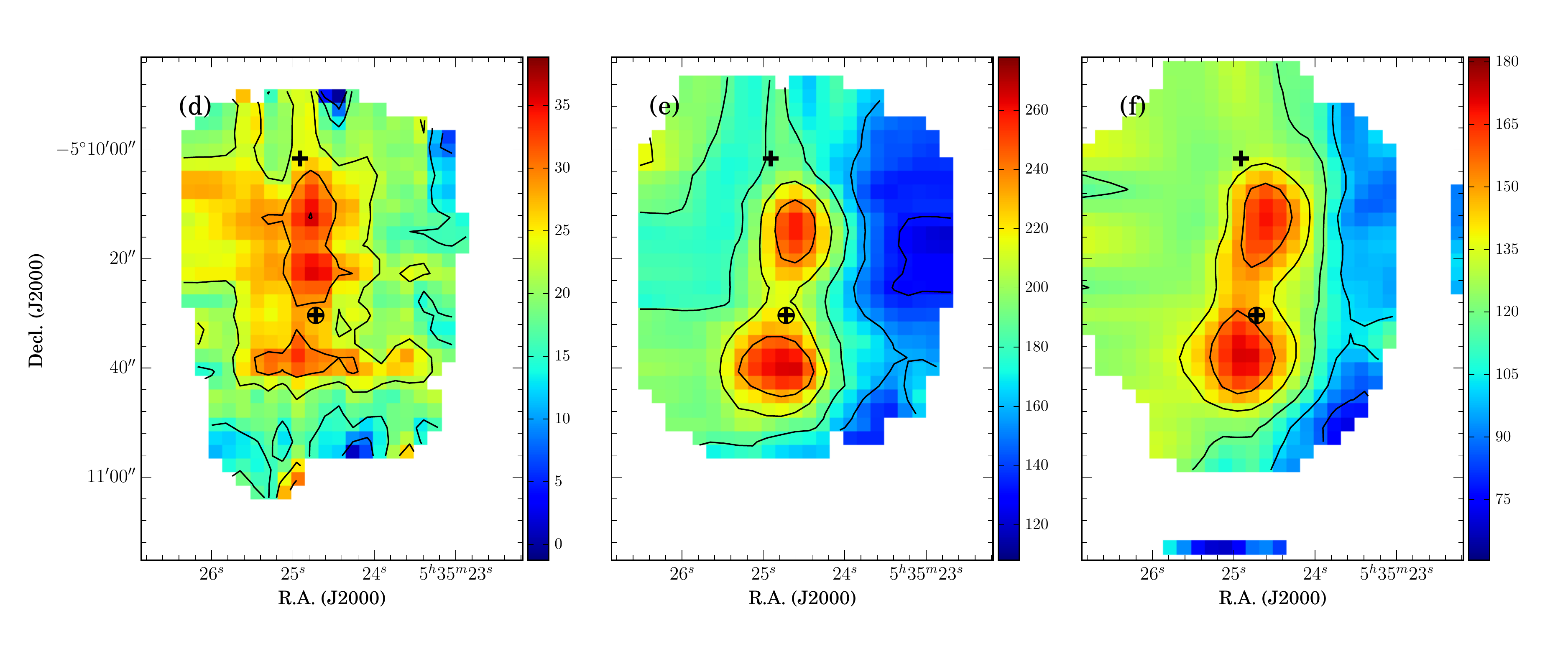}
\epsscale{0.85}
\plotone{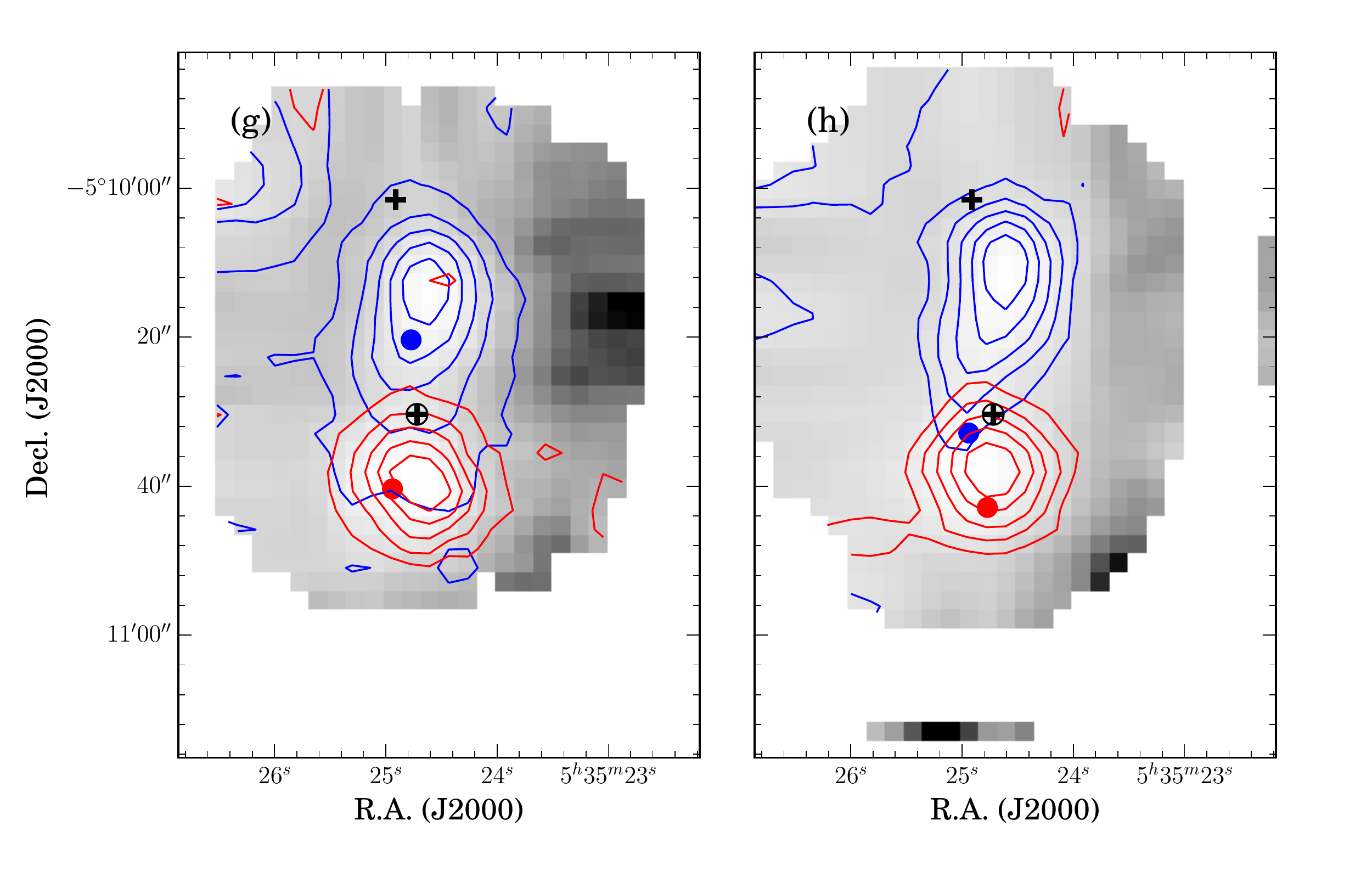}
\caption{
CO spectra and maps for HOPS 368 (OMC 2 VLA 13).
\label{fig_368_co}}
\end{figure*}

\begin{figure*}
\includegraphics[width=1.0\textwidth]{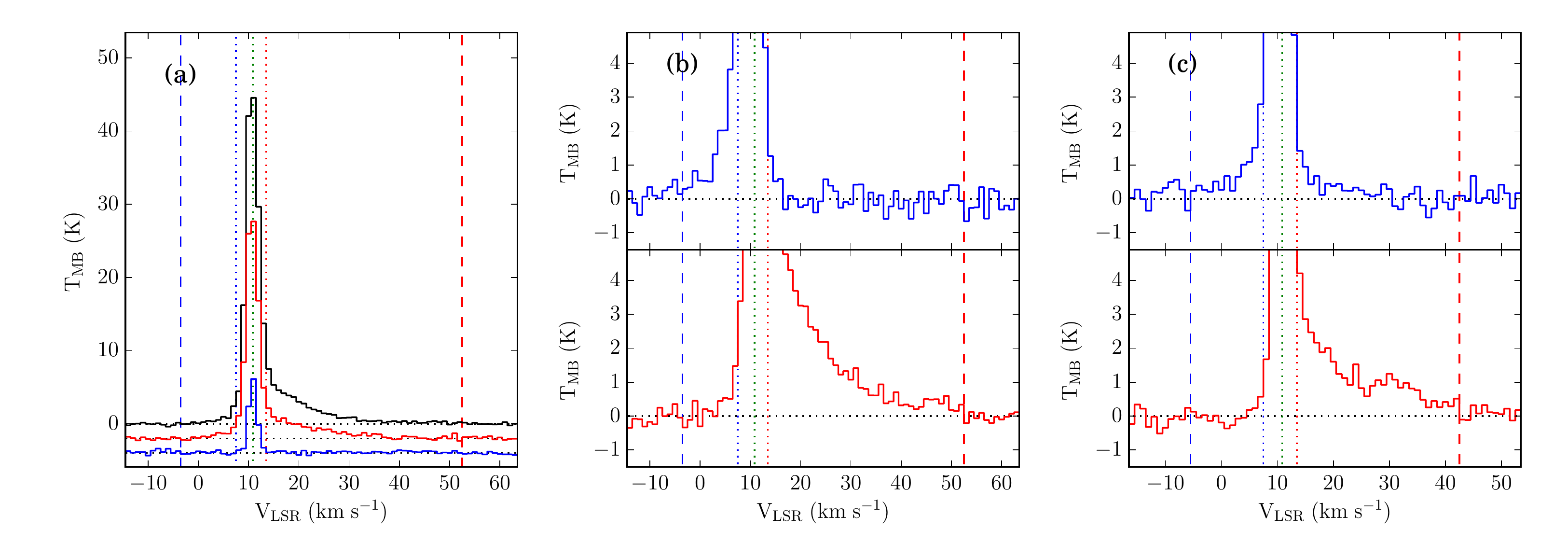}
\includegraphics[width=1.0\textwidth]{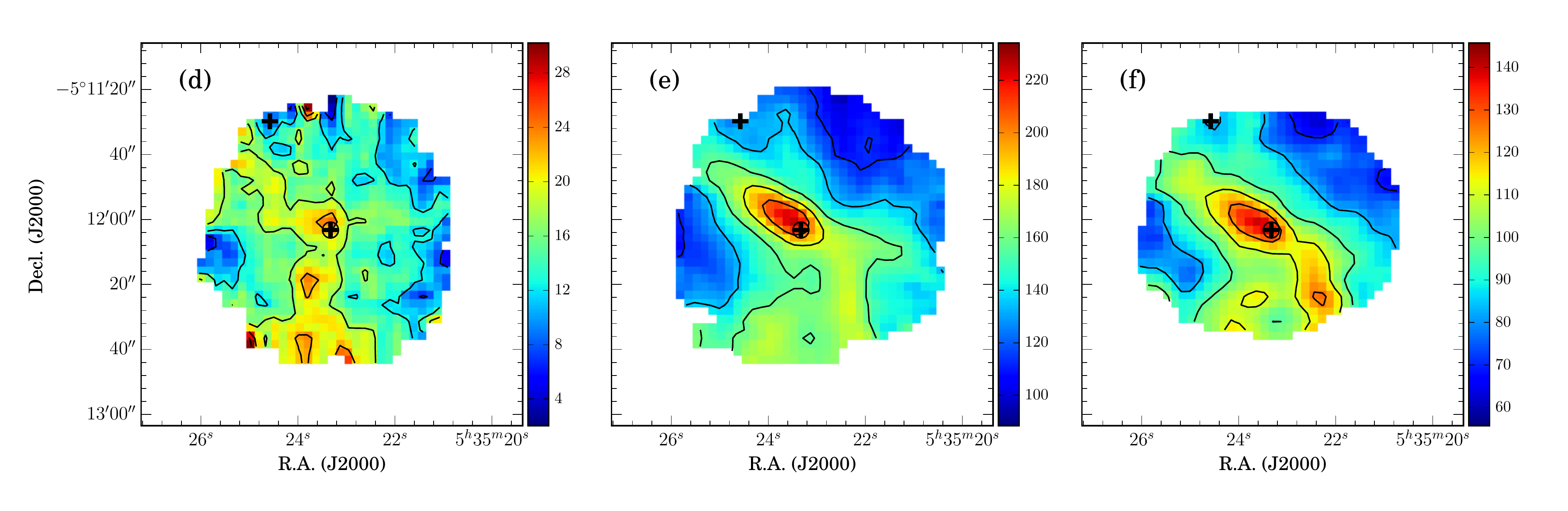}
\epsscale{0.9}
\plotone{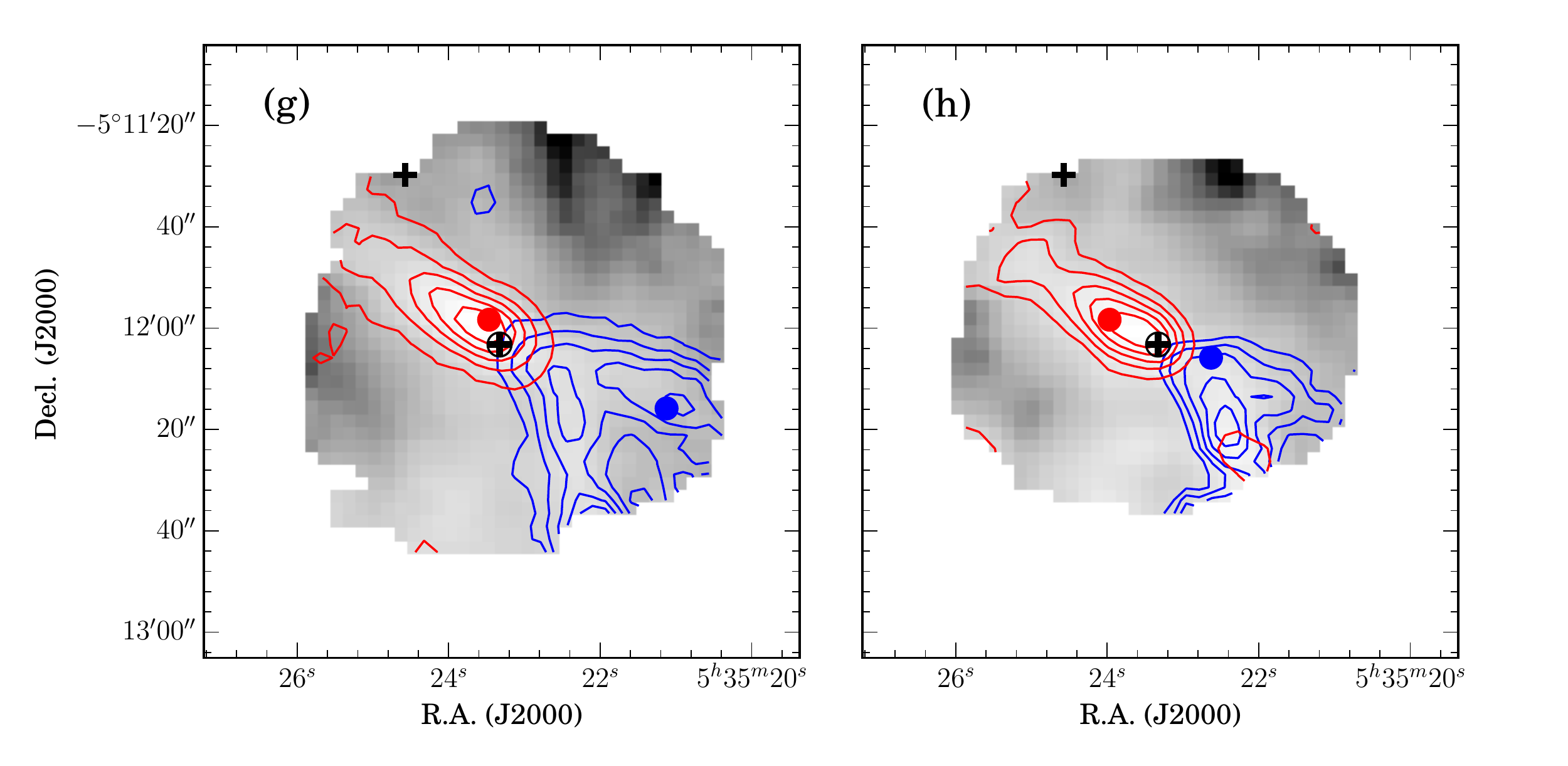}
\caption{
CO spectra and maps for HOPS 60 (OMC 2 FIR 6b).
\label{fig_60_co}}
\end{figure*}

\begin{figure*}
\includegraphics[width=1.0\textwidth]{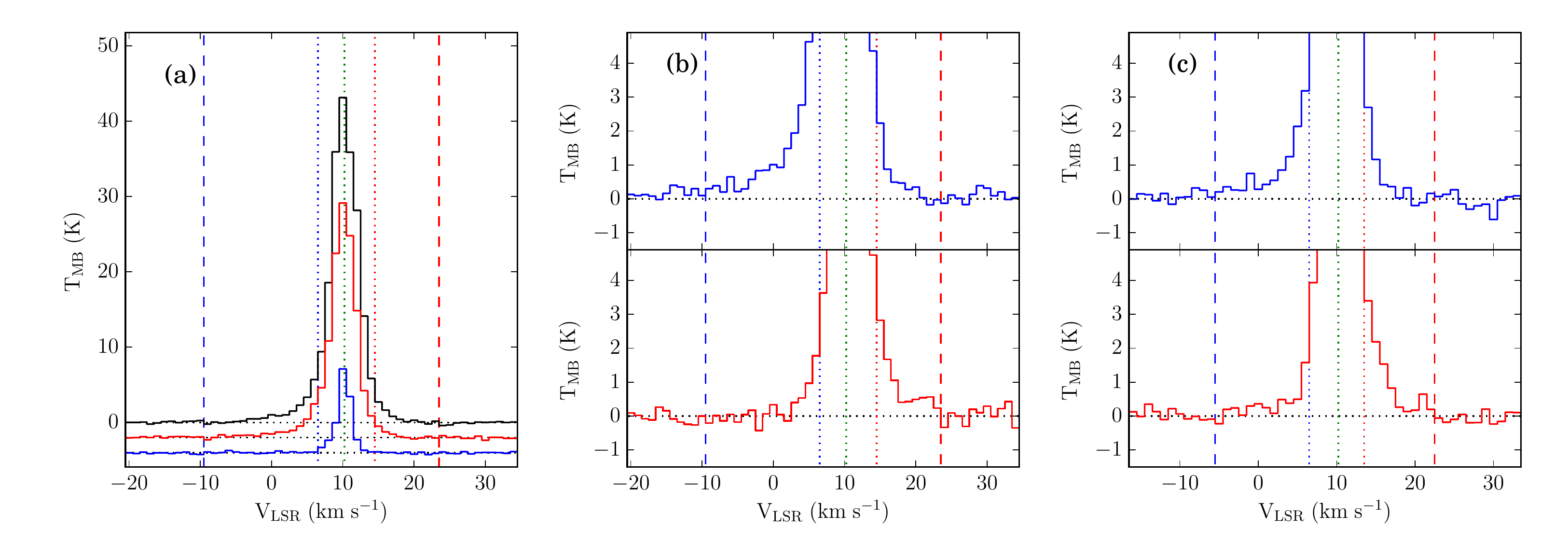}
\includegraphics[width=1.0\textwidth]{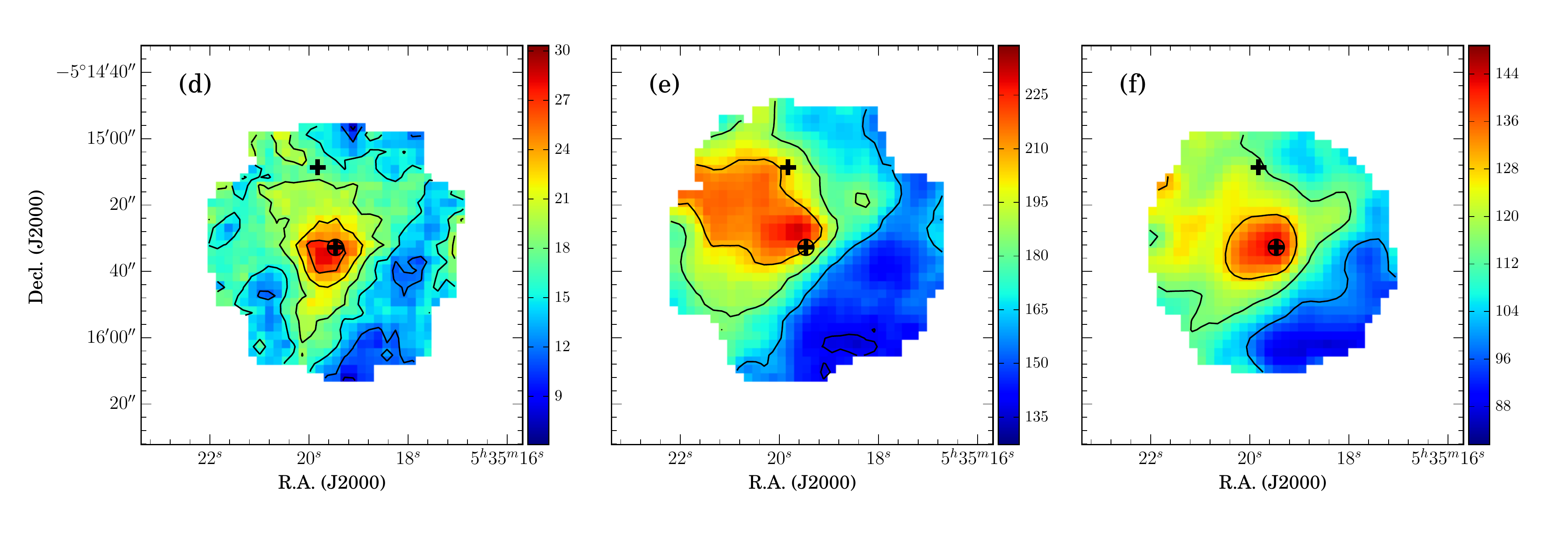}
\epsscale{0.9}
\plotone{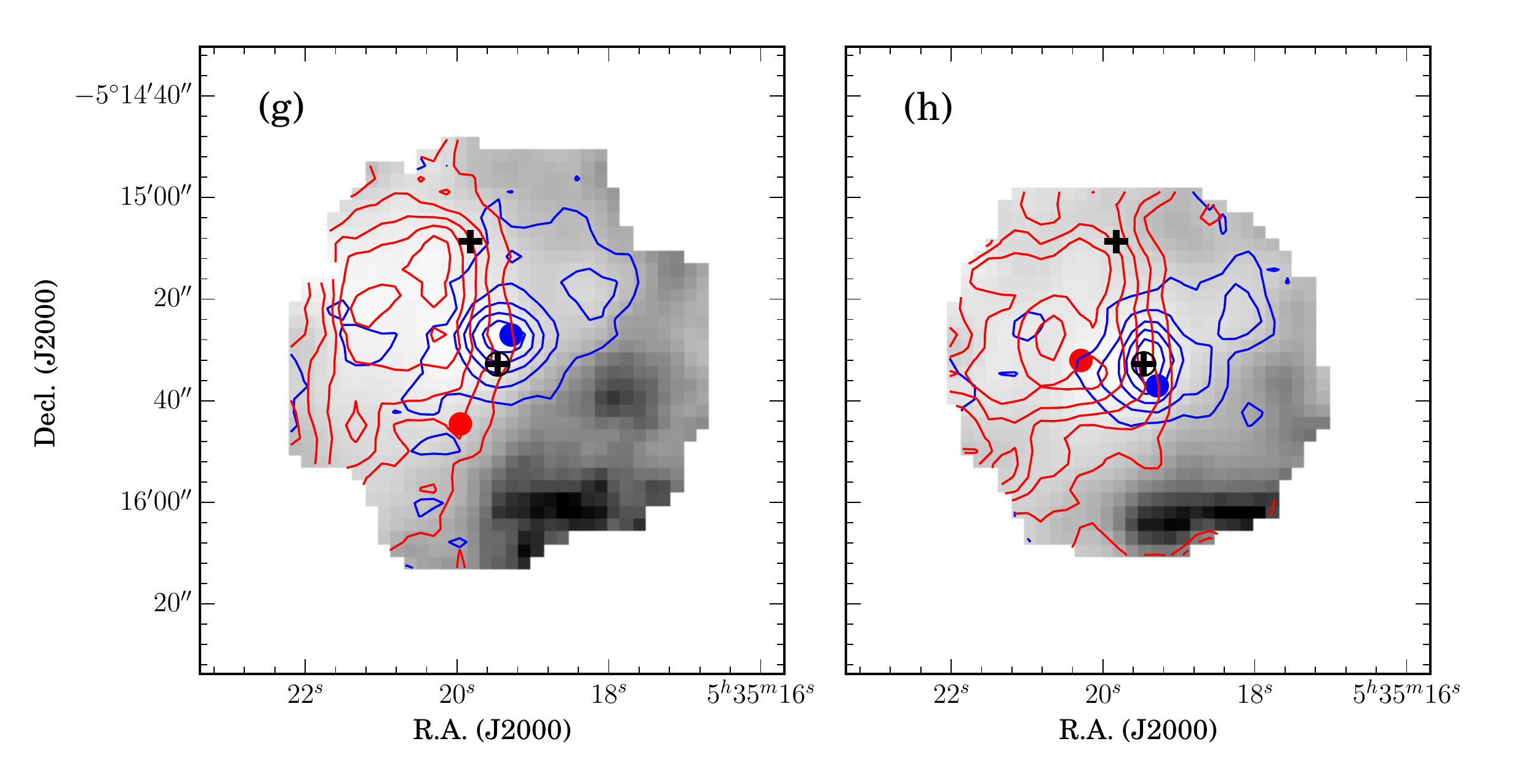}
\caption{
CO spectra and maps for HOPS 56 (OMC 2 CSO 33).
\label{fig_56_co}}
\end{figure*}

\begin{figure*}
\includegraphics[width=1.0\textwidth]{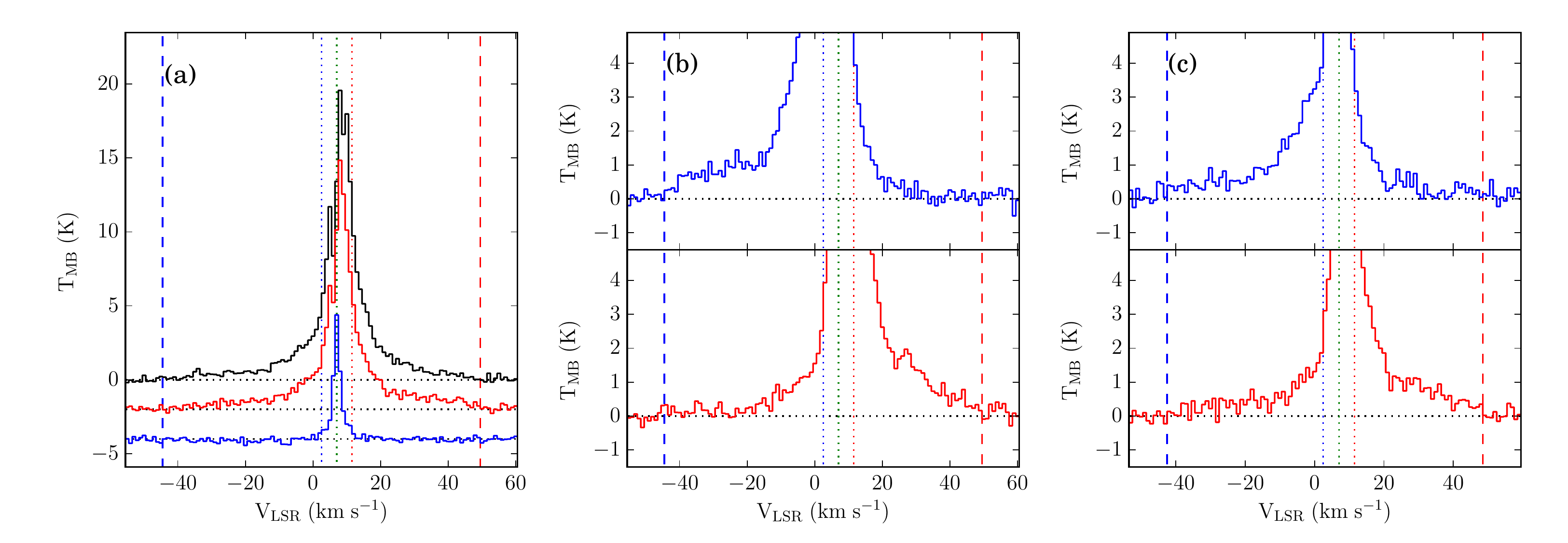}
\includegraphics[width=1.0\textwidth]{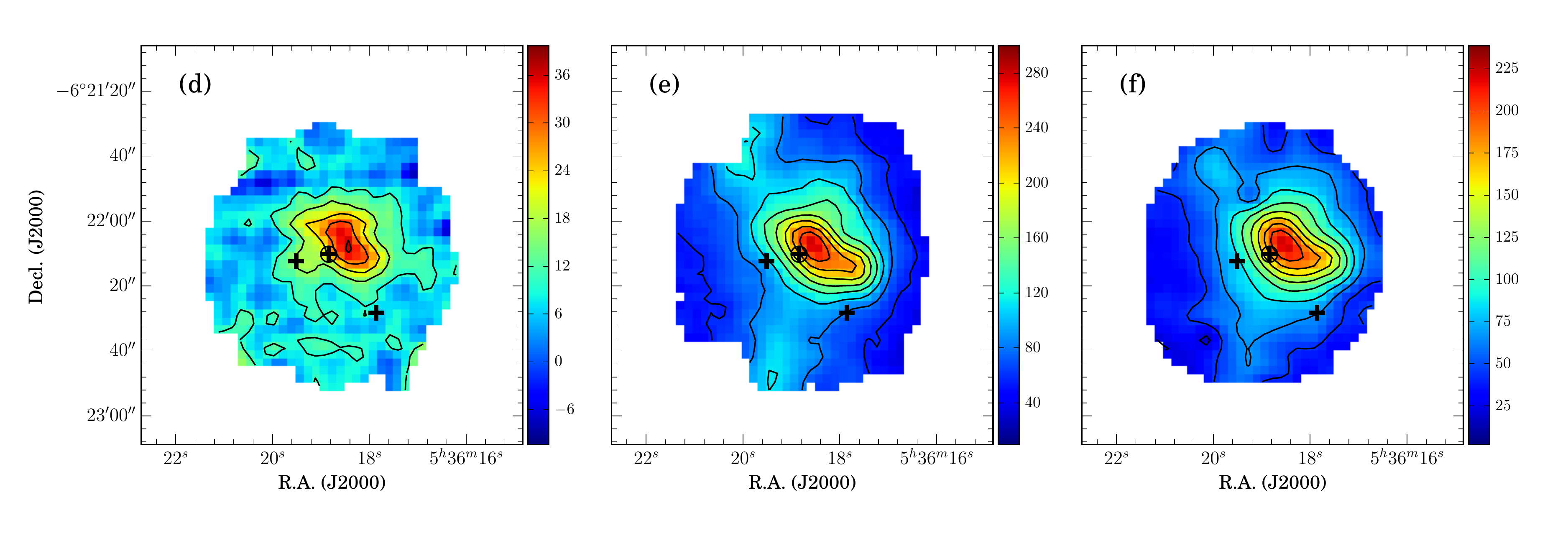}
\epsscale{0.9}
\plotone{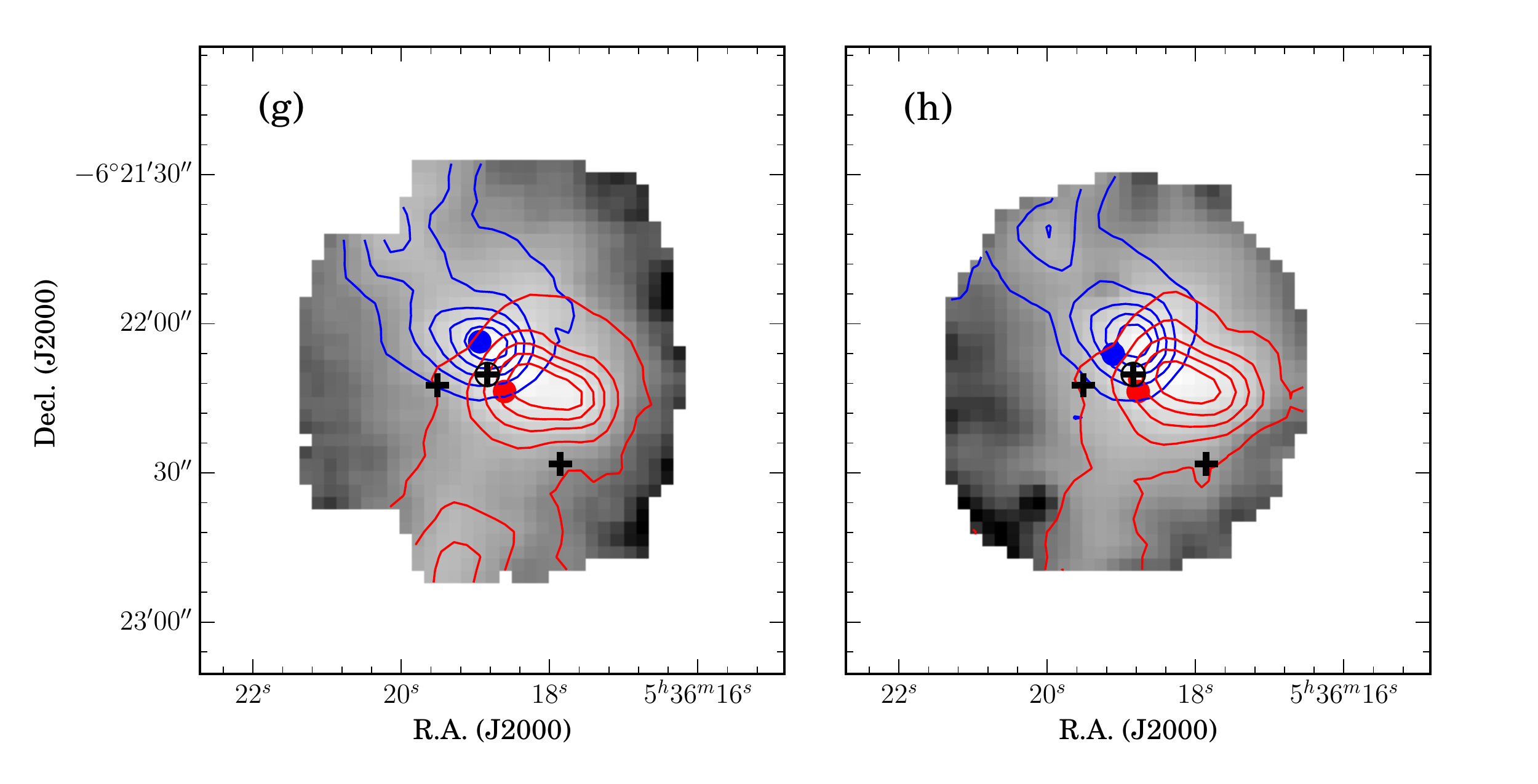}
\caption{
CO spectra and maps for HOPS 182 (L1641N MM1).
\label{fig_182_co}}
\end{figure*}

\begin{figure*}
\includegraphics[width=1.0\textwidth]{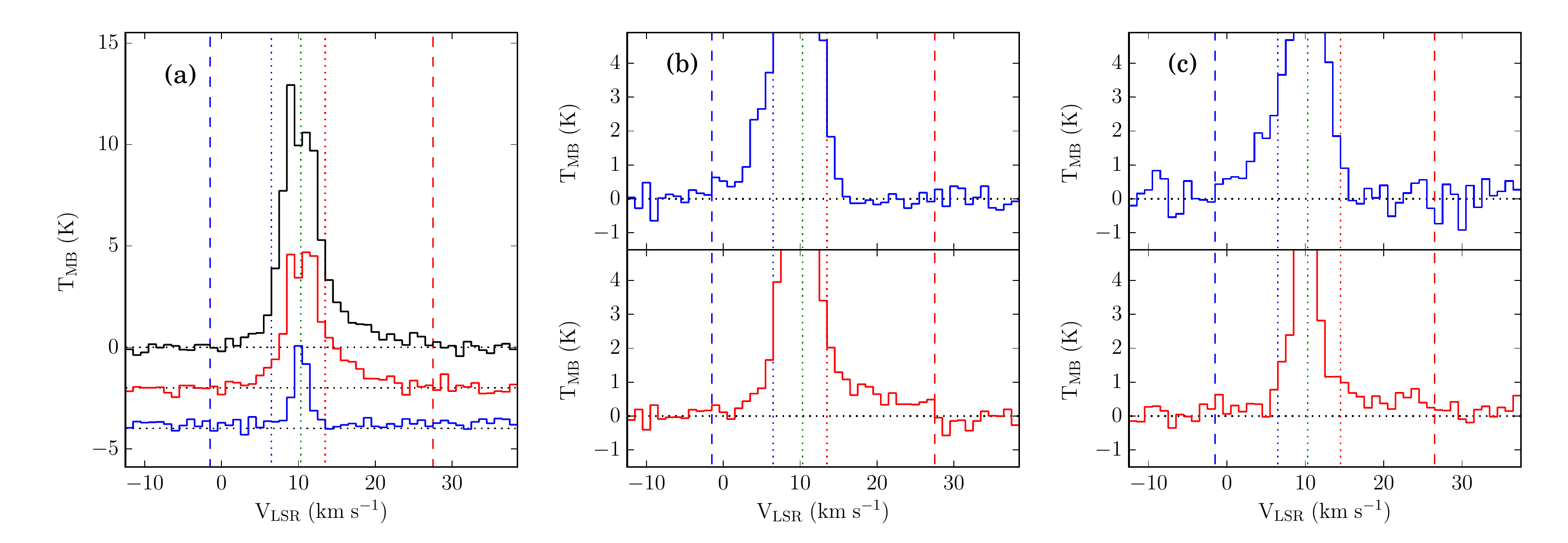}
\includegraphics[width=1.0\textwidth]{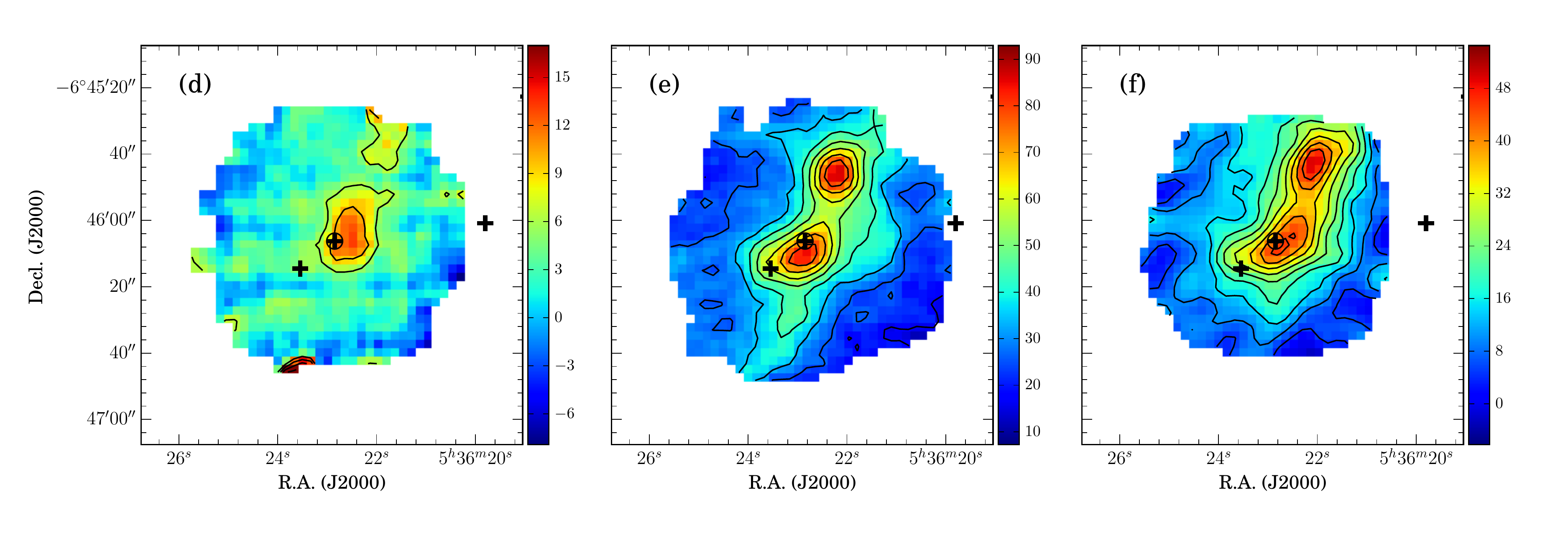}
\epsscale{0.9}
\plotone{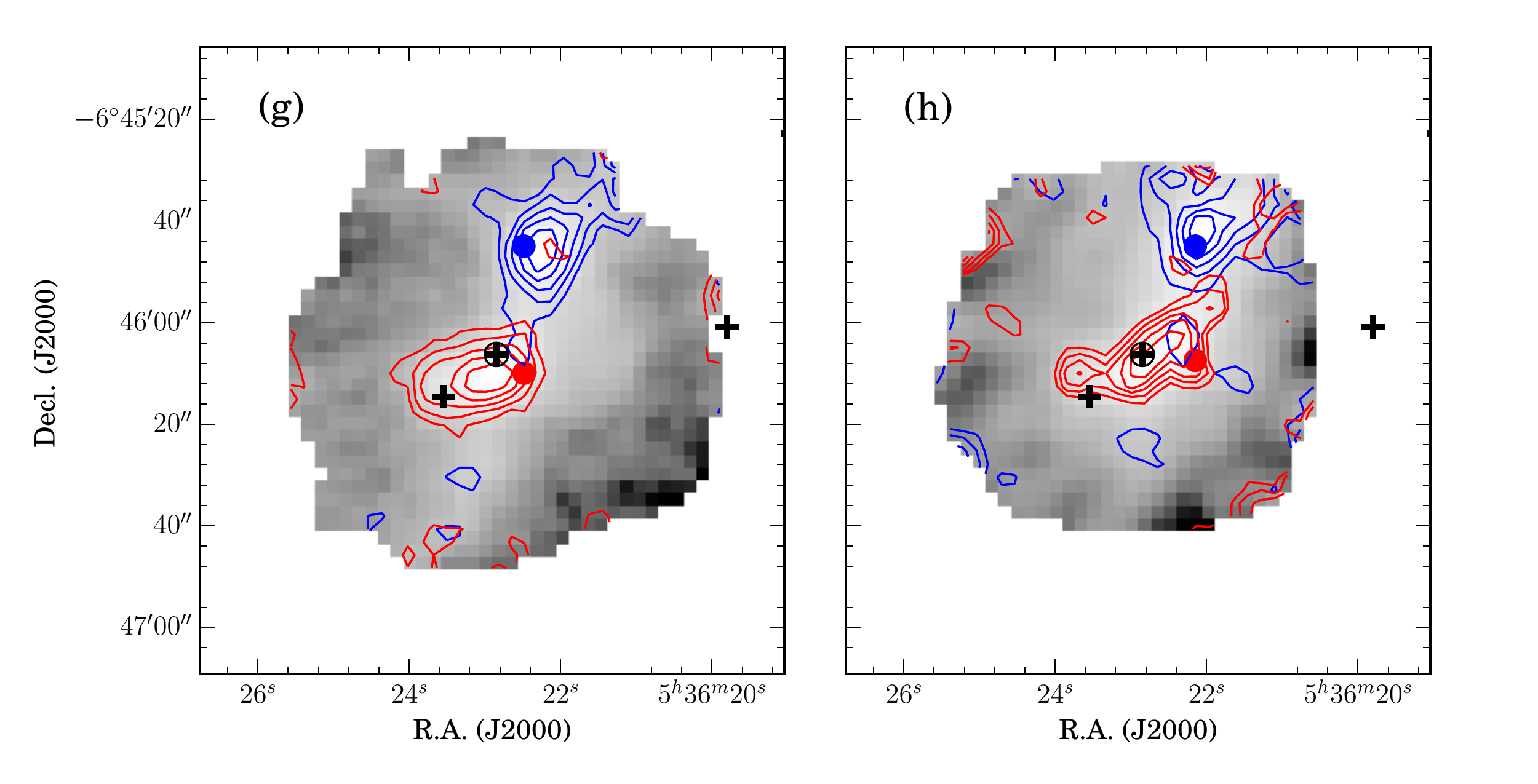}
\caption{
CO spectra and maps for HOPS 203 (HH 1/2 VLA 1).
\label{fig_203_co}}
\end{figure*}

\begin{figure*}
\includegraphics[width=1.0\textwidth]{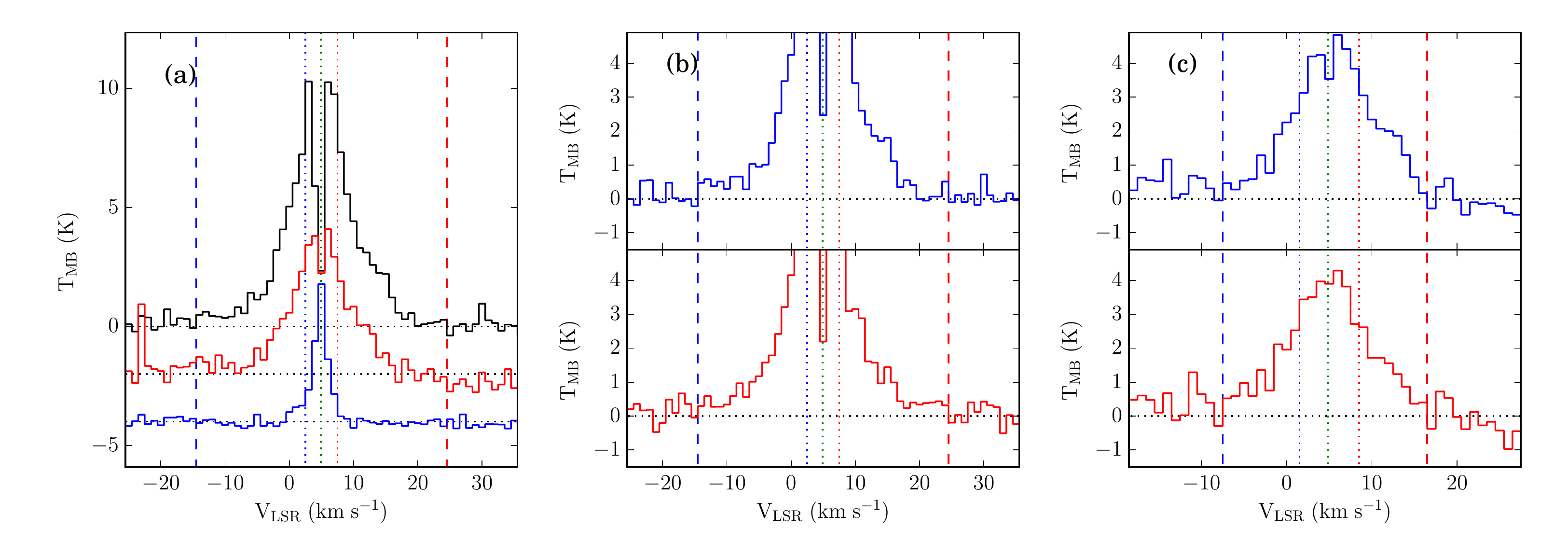}
\includegraphics[width=1.0\textwidth]{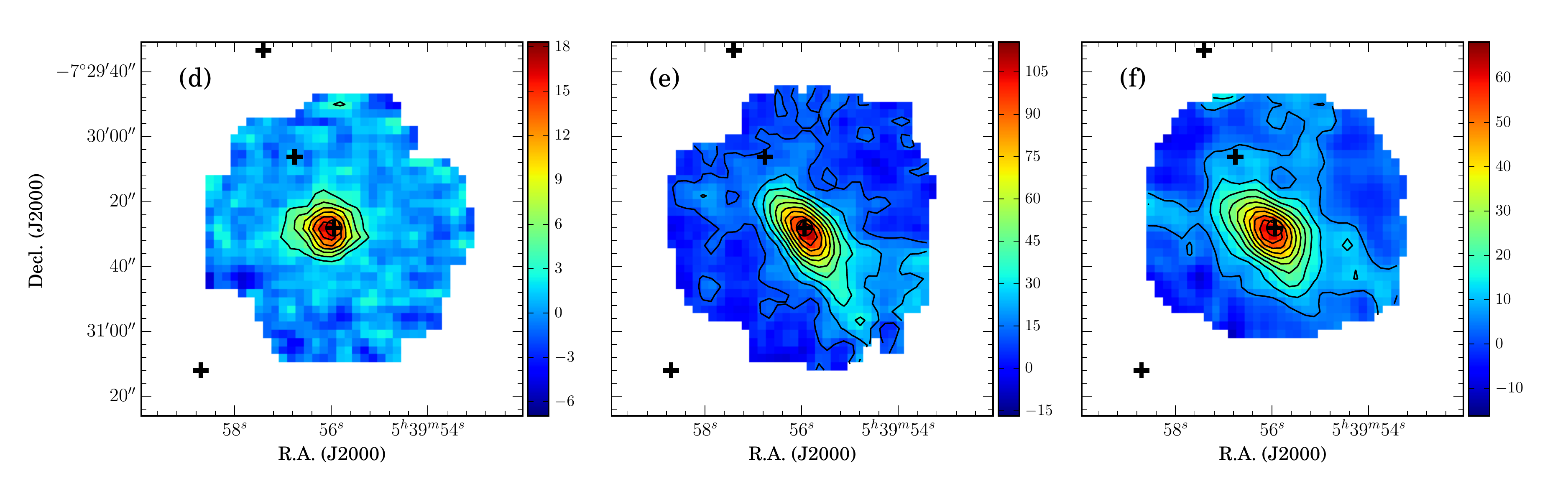}
\epsscale{0.9}
\plotone{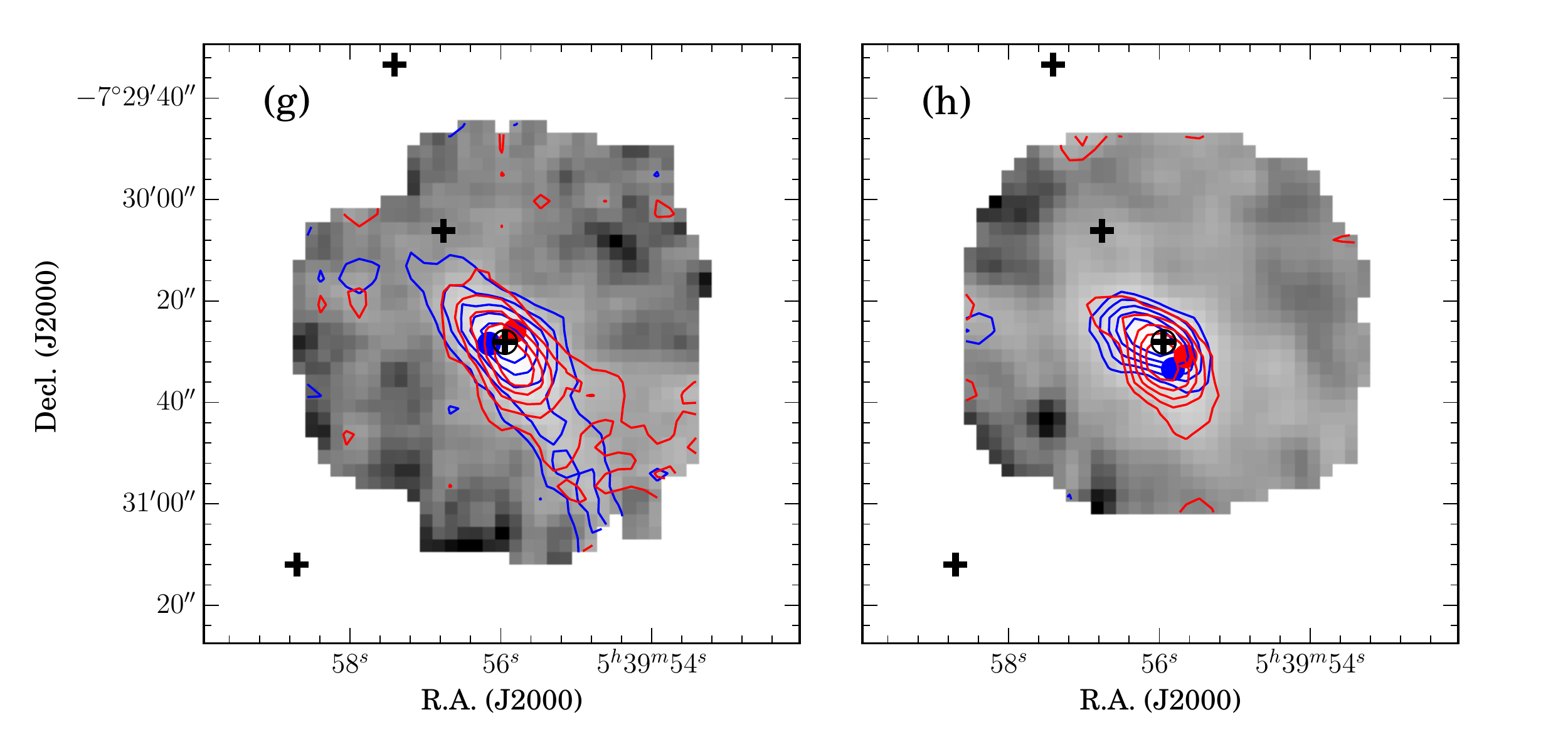}
\caption{
CO spectra and maps for HOPS 288 (L1641 S3 MMS 1).
\label{fig_288_co}}
\end{figure*}

\end{document}